\documentclass[final,5p,times,twocolumn]{elsarticle}
\usepackage{xcolor}
\usepackage{amsmath}
\usepackage{lineno,hyperref}

\usepackage{amssymb,graphicx}
\usepackage{float}
\usepackage{textcomp}
\usepackage{multirow}

\usepackage{tikz}
\usetikzlibrary{shapes,arrows,calc,decorations.pathmorphing,snakes}
\usepackage{esvect}
\usepackage{circuitikz}
\usepackage{subfig}

\journal{Journal of \LaTeX\ Templates}

\begin{document}

\begin{frontmatter}


\title{Systematic approach to measure the performance of microchannel-plate photomultipliers}

\author[a]{A.~Lehmann\corref{mycorrespondingauthor}}
\cortext[mycorrespondingauthor]{Corresponding author}
\ead{albert.lehmann@fau.de}
\author[a]{M.~B\"{o}hm}
\author[a]{M.~Götz}
\author[a]{K.~Gumbert}
\author[a]{S.~Krauss}
\author[a]{D.~Miehling}
\author[a]{M.~Pfaffinger}

\address[a]{Friedrich Alexander-University of Erlangen-Nuremberg, Erlangen, Germany}

\begin{abstract}
In this paper, we present our approach to systematically measure numerous performance parameters of microchannel plate photomultiplier tubes. The experimental setups, the analyses and selected results are discussed. Although the techniques used may be different in other locations, the document is intended as a guide for comparable measurements with other types of microchannel plate photomultiplier tubes. Measurements are shown for the following performance parameters: spectral and spatial quantum efficiency, collection efficiency, gain as a function of voltage, position and magnetic field, time resolution, rate capability and lifetime. By using a dedicated 3-axis stepper and an FPGA-based data acquisition system, also inner PMT parameters are measured as a function of the active area, such as relative detection efficiency, dark count rate, time resolution, recoil electron and afterpulse distributions, as well as charge sharing and electronic crosstalk. In addition, some of the parameters are investigated inside a strong magnetic field. For many of these measurements, the change of most setup parameters and the subsequent analysis can be controlled semi-automatically by software scripts.

\end{abstract}


\end{frontmatter}

\nolinenumbers

\section{Introduction}  \label{introduction}

\begin{figure}[!htbp]
	\centering
	\includegraphics[width=.98\columnwidth]{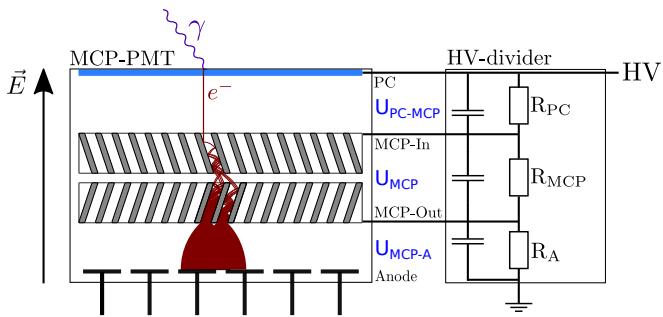}
	\caption{Diagram of an MCP-PMT with its high voltage (HV) divider \cite{gumbert}. Many of the terms and designations used in this document are defined here. The capacitances inside the HV divider are usually 4.7 nF.}
	\label{fig:mcp}
\end{figure}

Recently, microchannel-plate photomultiplier tubes (MCP-PMTs) have become very attractive sensor candidates for applications where ultrafast single photon detection is required in high magnetic fields, at very high photon rates and in harsh radiation environments \cite{iijima11,gys15,hink16,lehmann23}. This is due to the fact that their lifetime has increased considerably in the last $\gtrsim$10 years \cite{lehmann22}. Typical sizes today range from 1$\times$1 inch$^{2}$ and 2$\times$2 inch$^{2}$ active area offered by commercial suppliers such as Hamamatsu Photonics K.K. (HPK) \cite{hama}, Photek Ltd \cite{photek} and PHOTONIS \cite{photonis}, to 8$\times$8 inch$^{2}$ Large Area Picosecond Photodetectors (LAPPD) \cite{adams16} from Incom Inc \cite{incom}. These PMTs can be manufactured in various configurations depending on the intended application. A schematic representation of a typical MCP-PMT with a connected voltage divider is shown in Fig.~\ref{fig:mcp}. For detailed qualification tests of such tubes, reliable setups and measurement methods are required to measure and compare the performance parameters.

MCP-PMTs are the photon sensors of choice for the DIRC detectors \cite{bdtdr, carsten, eddtdr, ilknur} of the PANDA experiment \cite{pandatpr, pandapb} at the new FAIR accelerator complex at GSI in Darmstadt, Germany. In this context, various specialized measurement setups and analysis tools have been developed over the years for a thorough and comprehensive investigation of the performance characteristics of MCP-PMTs. Recently, the first series production MCP-PMTs for the PANDA Barrel DIRC \cite{bdtdr, carsten} were tested for their quality by measuring and comparing as many key performance parameters as possible. Here, we present a systematic approach to the detailed measurement of most MCP-PMT performance parameters within a manageable period of a few days. In particular, we focus on the description of the experimental setups and the procedures used to record the different parameters. Since the approaches are quite different, we describe them separately in each section. In addition, the analyses used and some representative results for each parameter are discussed.

In Section~\ref{sec:PDE} the spectral quantum efficiency (QE) between 200 and 850 nm and the spatial QE homogeneity are determined in specific scans. A setup for measuring the collection efficiency (CE) is also presented, which is applicable for almost all PMTs. The gain of the MCP-PMTs (Section~\ref{sec:gain}) is determined as a function of voltage and its homogeneity over the active area, for some of the tubes also within a magnetic field. Both the precise transit time spread (TTS) and a root-mean-square (RMS) time resolution are determined in Section~\ref{sec:timeres}, using a fast oscilloscope or a waveform digitizer. The advantages and disadvantages of two methods -- "current mode" and "pulse mode" -- for measuring rate capability are explained in Section~\ref{sec:rate}, followed by the presentation of an approach for measuring the MCP recharging time. The recently observed "escalation effect" (Section~\ref{sec:escal}) is briefly debated before our illumination setup and the measurement and analysis procedure for the aging studies are discussed in detail in Section~\ref{sec:lifetime}.

In the last two chapters, we present an efficient method for measuring some internal parameters of an MCP-PMT as a function of the active area. Position scans are performed with a self-built 3-axis stepper and an FPGA-based TRB3 data acquisition (DAQ) system developed at GSI. The setup, the scanning procedure and the analyses are discussed in detail in Section~\ref{sec:scans}, so that it should be possible to assemble a comparable system at other locations. The parameters accessible with this setup are the relative detection efficiency, the dark count rate (DCR), the time resolution (including electronics) and the distribution of recoil electrons and afterpulses. The investigation and quantification of different types of crosstalk is also possible with this system, e.g. charge sharing, electronic and "coherent" oscillations. Finally, the approach to perform such scans within the magnetic field is presented in Section~\ref{sec:scans_bfield}. The latter provides further insight into the behavior of primary and secondary electrons on the PMT performance when electromagnetic focusing is involved.

Although numerous other MCP-PMTs have also been investigated, in this article we focus on the performance measurements of the PHOTONIS Planacon 2-inch MCP-PMTs XP85112-S-BA and XP85132. These have a square shape with 2 MCP stages, 8$\times$8 or 100$\times$3 anode pixels, respectively, and 10 \textmu m pores coated with an ALD (Atomic Layer Deposition) technique \cite{arradiance} to extend the lifetime of the tube.

\section{Photon Detection Efficiency}  \label{sec:PDE}

A very important parameter of any photosensor is its photon detection efficiency (PDE), which is a measure of the probability that an incident photon at the photocathode will be converted into a detectable signal at the anode. The PDE is the product of the QE, the CE and the fraction of active area or form factor (FF) of the sensor:

\begin{equation}
	PDE = QE \cdot CE \cdot FF
\end{equation}

\subsection{Fraction of Active Area or Form Factor}

The form factor of an MCP-PMT is determined by the ratio between the active photocathode area and the total surface area of the PMT exposed to the incident photon flux. Obviously, the FF depends on the type and geometry of the PMT. It ranges from $\ll$50\% for small MCP-PMTs to 81\% for the 2-inch Planacon tubes used in the PANDA DIRCs. The value of the FF is determined exclusively by the layout and overall dimensions of the respective tube.

\subsection{Quantum Efficiency}

The QE describes the probability that a photon hitting the photocathode (PC) of the PMT is converted into a photoelectron (PE) that escapes into the vacuum. A typical peak QE of MCP-PMTs ranges from $\sim$20\% to $>$30\%, depending on the PC material. To qualify the detection efficiency of the PMT, we measure the QE as a function of the photon wavelength, but also its homogeneity across the photocathode at a fixed wavelength (in our case 372 nm or 633 nm, depending on the pulsed picosecond laser (PiLas from Advanced Laser Diode Systems \cite{pilas}) used).

\subsubsection{Spectral QE}  \label{spectralQE}

\begin{figure}[!htbp]
	\centering
	\includegraphics[width=.96\columnwidth]{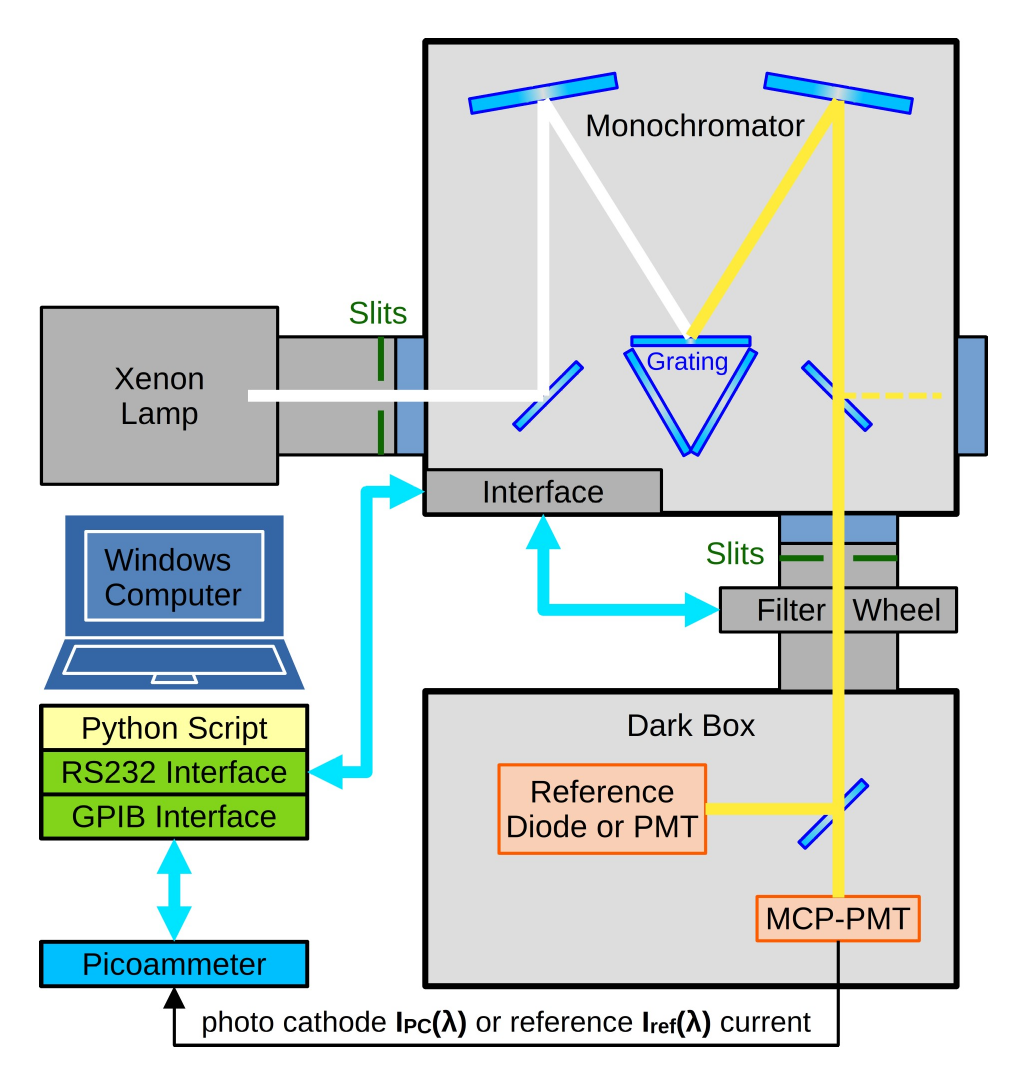}
	\caption{Schematic of the automated setup for measuring the QE as a function of the wavelength of the photons \cite{herold}.}
	\label{fig:QElambdasetup}
\end{figure}

The spectral QE is measured in a special in-house setup (see Fig.~\ref{fig:QElambdasetup}), which consists of a Xenon lamp as light source and a monochromator LOT Oriel MSH301 (later renamed Oriel Cornerstone 260 \cite{newport}) with different gratings and $<$1 nm wavelength resolution in a range from 200 to 850 nm. After passing through slits, a grating, several mirrors and, at the exit of the monochromator, another slit to eliminate second-order diffraction effects and a filter wheel to adjust the intensity, the wavelength-selected light is directed into a dark box containing the MCP-PMT and a reference diode [Hamamatsu S6337-01, $QE_{ref}(\lambda)$]. In the dark box the beam is pointed at either the reference sensors or the MCP-PMT by using a mirror. A PC current of the PMT of typically a few nA and/or the diode current are measured with a Keithley \cite{keithley} 487 picoammeter. The entire setup is controlled by a Python software script on a Windows computer, which communicates with the monochromator, the filter wheel and the picoammeter via a GPIB interface.

In practice, the Xenon lamp is warmed up for 2 to 3 hours before the actual measurement program is started to ensure a stable light intensity. The MCP-PMT is placed in the dark box well before the start of data collection in order to reduce the dark count rate. A voltage of 200 V is applied between the PC and MCP-In, and the photocurrent $I_{PC}(\lambda)$ directly at the PC is measured with the picoammeter. The other components (MCPs, anode) of the MCP-PMT are usually shorted. Before and after each wavelength scan, the dark current $I_{dark,PMT}$ is measured for a short time ($\sim$10 s) and subtracted in the QE analysis. The diode current $I_{ref}(\lambda)$ for each wavelength is measured before and sometimes after scanning one or more MCP-PMTs and then used to calculate the QE of the PMT at each wavelength (Eq. \ref{eq:2}). The scan over the entire wavelength range runs automatically.

\begin{equation} \label{eq:2}
	QE_{PC}(\lambda) = QE_{ref}(\lambda) \cdot \frac{I_{PC}(\lambda) - I_{dark,PMT}}{I_{ref}(\lambda)}
\end{equation}

\begin{figure}[!htbp]
	\begin{minipage}{.40\columnwidth}
		\centering
		\includegraphics[width=.98\columnwidth]{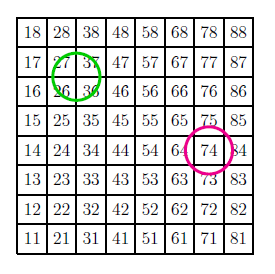}
	\end{minipage}
		\hspace{0.5pc}%
		\begin{minipage}{.56\columnwidth}
			\centering
			\includegraphics[width=.98\columnwidth]{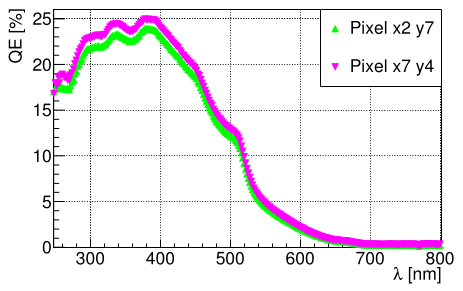}
		\end{minipage}
		\caption{Positions where the QE was measured (left) and spectral QE obtained at these two spots for the PHOTONIS XP85112 [SN 9002193] (right).}
		\label{fig:QEwave}
	\end{figure}

Normally, the QE is measured at two (or more) PC positions of the MCP-PMT. The illuminated light spots with a diameter of $\sim$5 mm are indicated by the green and magenta circles on the anode pixel map in Fig.~\ref{fig:QEwave} (left). As an example, the measured QE wavelength distributions are shown in Fig.~\ref{fig:QEwave} (right), which illustrate the slight position dependence of the QE over the PC.

\subsubsection{QE Homogeneity}  \label{positionQE}

\begin{figure}[!htbp]
	\centering
	\includegraphics[width=.95\columnwidth]{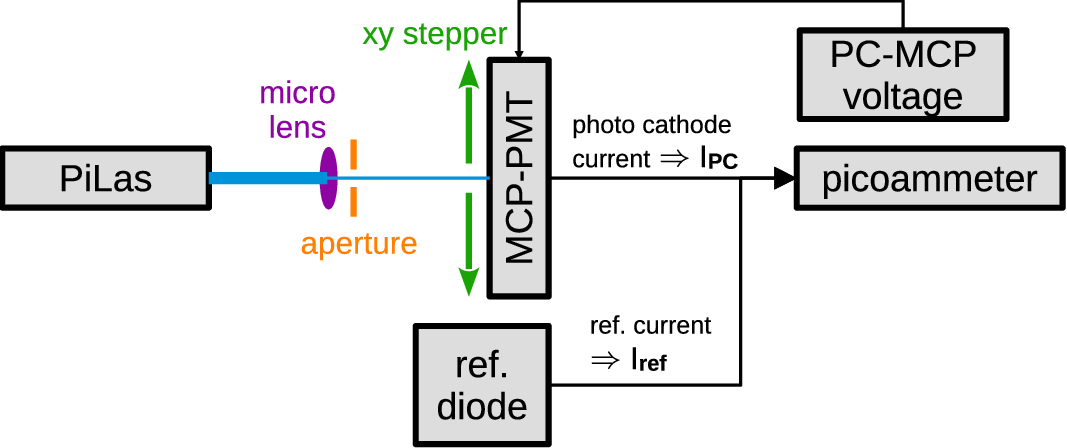}
	\caption{Block diagram of the setup for measuring the position-dependent QE.}
	\label{fig:QEsetup}
\end{figure}

In our quality control tests of the MCP-PMTs, we also measure the spatial dependence of the QE over the entire photocathode area. This is done inside a light-tight box (referred to as "dark box" in the following) with an automated 3-axis (xyz) stepper (see Section~\ref{sec:perfsetup}) using a PiLas at a wavelength $\lambda_{scan}$ of usually 372 nm (UV/blue) and sometimes also at 633 nm (red) with a timing resolution ($\sigma$) of 14 ps or 12 ps ($\sigma$), respectively. The light is guided from the laser head to the PC surface using single-mode optical fibers that end in a microfocus lens. The light intensity can be controlled and adjusted using neutral density (ND) filters and the laser rate. The final focus of the laser beam on the PC can be optimized by adjusting the distance between the lens and the MCP-PMT with the z-coordinate of the stepper. However, the light spot should not be too narrow in order to avoid saturation of the PC current. An additional aperture of about 1 mm diameter is located downstream of the lens to avoid stray light. The actual QE position scan is performed in steps of typically 0.5 mm in both the x and y directions. Our stepper can cover a range of $\sim$40 cm in both directions. This allows to measure four (and even more) different MCP-PMTs without having to change the setup, and we can perform long-term scans over nights and weekends. Here too, a voltage of 200 V is applied between the PC and MCP-In, and the other PMT electrodes are shorted. To obtain a measurable PC current of 5-10 nA, we usually do not use ND filters in the QE scans and apply a laser rate of $\sim$100 kHz. The local PC current $I_{PC}(xy)$ at each scanning position $xy$ is measured with a Keithley 6485 or 6487 picoammeter. A block diagram of this setup is shown in Fig.~\ref{fig:QEsetup}.

Before the actual measurement, the MCP-PMT is placed in the dark for some time to reduce the DCR. After typically four scanned x-rows, the laser spot is moved to the position of the reference diode S6337-01 with calibrated $QE_{ref}(\lambda_{scan})$ and a reference current $I_{ref}$ and the dark current $I_{dark,PMT}$ of the MCP-PMT are measured. $I_{dark,PMT}$ is interpolated for all points along the position scan. The local quantum efficiency $QE(xy)$ at the PC is calculated as follows:

\begin{equation}
	QE(xy) = QE_{ref}(\lambda_{scan}) \cdot \frac{I_{PC}(xy) - I_{dark,PMT}}{I_{ref}}
\end{equation}

\begin{figure}[!htbp]
	\centering
	\includegraphics[width=.98\columnwidth]{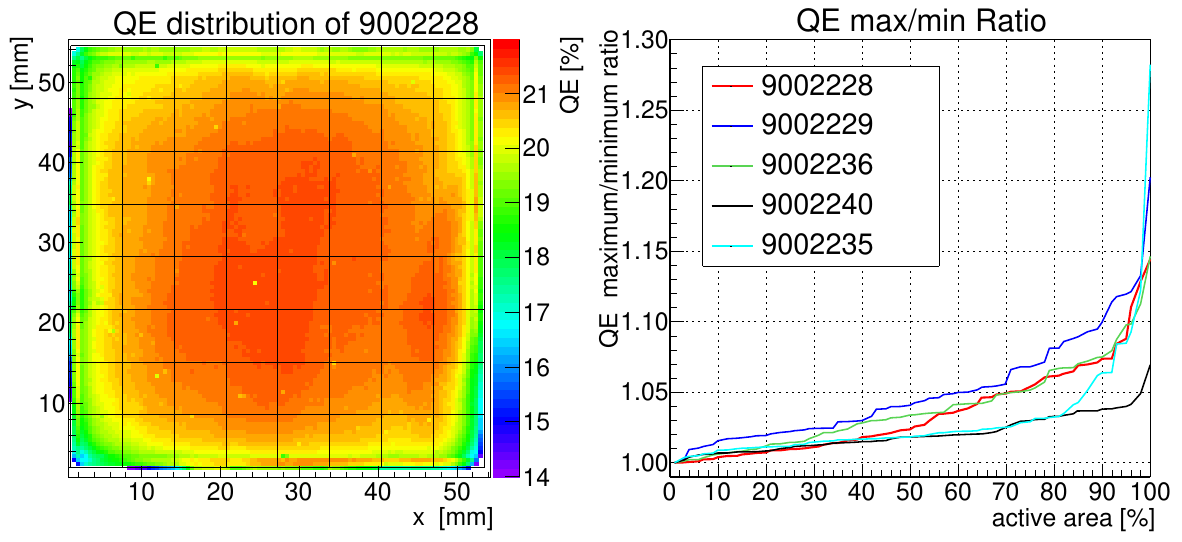}
	\caption{QE as a function of PC position (left) for the PHOTONIS SN 9002228, with the anode grid superimposed in black, and typical QE max/min ratios for a set of five PHOTONIS MCP-PMTs (right).}
	\label{fig:QEpos}
\end{figure}

Fig.~\ref{fig:QEpos} (left) shows the QE of the PHOTONIS SN 9002228 as a function of the PC area. Typically, the QE is fairly homogeneous, with slightly lower values towards the rims of the tube. The black grid in the graph shows the position of the PMT anode pixels. Using the measured local QE values at different ($>$100) illuminated $xy$ points, an average QE value can be determined for each pixel. The ratio between the pixel with maximum QE and the QEs of all other pixels (the QE max/min ratio) is calculated from the mean QEs. The resulting diagram is shown in Fig.~\ref{fig:QEpos} (right). This approach can be used to quantify how much of the PC area has a low QE, but for typically $\gg$90\% of the area the QE is very homogeneous with a max/min ratio of $<$1.2.

\subsection{Collection Efficiency}  \label{CE}

In order to check the CE values advertised by the various manufacturers, we have developed our own measurement method. Although the calculation of the CE appears simple at first glance, tiny PC currents in the picoampere range make the actual measurement considerably more difficult. The CE is given by

\begin{equation}
	CE = \frac{N_{pe@anode}}{N_{pe@PC}}
	= \frac{N_{pe@anode@LR} \cdot e \cdot LR}{I_{PC@HR}} \cdot \frac{I_{diode@HR}}{I_{diode@LR}}
\end{equation}

\noindent where $N_{pe@anode}$ is the number of photoelectrons (Npe) derived directly from the measured pulse height distribution at the anode, and $N_{pe@PC}$ is deduced from the PC current. $N_{pe@anode@LR}$ is measured at a low photon rate (LR, $\mathcal{O}$(kHz)), while $I_{PC@HR}$ for approximately the same Npe is the PC current obtained at a high photon rate (HR, $>$10 MHz). The currents $I_{diode@HR}$ and $I_{diode@LR}$ are determined by a reference diode to scale the low and high rate measurements. $e$ is the elementary charge.

\begin{figure}[!htbp]
	\centering
	\includegraphics[width=.9\columnwidth]{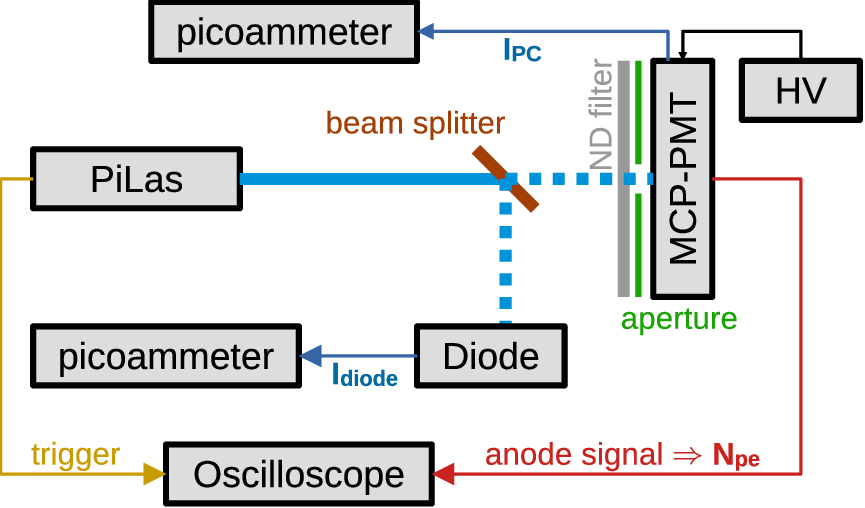}
	\caption{Block diagram of the setup for measuring collection efficiency.}
	\label{fig:CE}
\end{figure}

The CE setup is sketched in Fig.~\ref{fig:CE}. For the measurement of $I_{PC}$ the voltage $U_{PC-MCP}$ between PC and MCP-In is the same as in normal operation and ranges from $\sim$150 V in an $R_{PC}$:$R_{MCP}$:$R_A$ = 1:10:1 voltage divider configuration to $\sim$750 V in a 4:10:1 configuration. In this case, the electrode MCP-Out and the anode are usually shorted. It is important to note that the CE depends slightly on the configuration used. For the latest PHOTONIS MCP-PMTs with the 4:10:1 voltage divider configuration, a large CE close to 100\% was measured. The large CE is probably achieved by the deposition of a secondary emissive layer on top of the input electrode MCP-In \cite{orlov18}. However, the latter increases the tail of the time distribution (see Fig.~\ref{fig:tres}) and requires the use of a higher $U_{PC-MCP}$.

\begin{figure}[!htbp]
	\centering
	\includegraphics[width=.98\columnwidth]{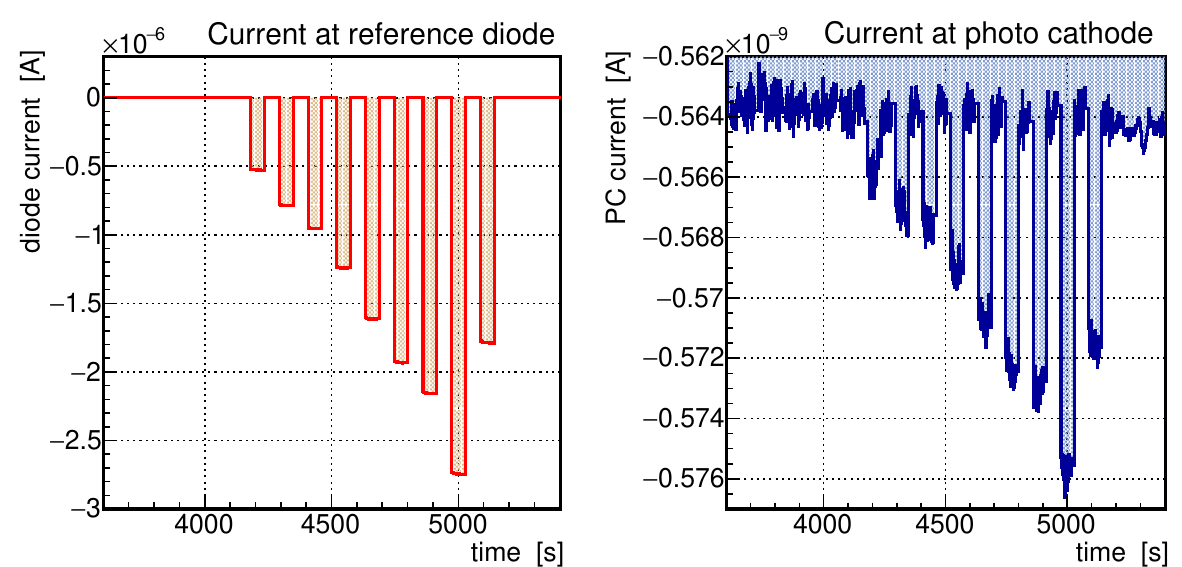}
	\caption{Comparison of the currents obtained during a CE measurement of a PHOTONIS (SN 9002230) MCP-PMT at the reference diode (left) and at the photocathode (right) with $U_{PC-MCP}$ = 190 V. The peaks are generated by laser on and laser off settings and correspond to a laser rate from 10 to 50 MHz in 5 MHz steps. After the tube was in the dark for $\sim$1 hour, the average dark current during this measurement was ~0.564 nA.}
	\label{fig:CEcurr}
\end{figure}

Before the actual measurements, the MCP-PMT is placed in the dark for several hours to reduce the dark currents. The light from a PiLas laser is passed through a beam splitter, with one part falling on the PC of the MCP-PMT and the other part on the reference diode. The current of the reference diode is measured with a Keithley 6485 picoammeter at both low ($I_{diode@LR}$) and high ($I_{diode@HR}$) illumination rates. This is necessary because the light intensity of the laser is not proportional to its pulse frequencies $LR$ and $HR$. The number of photoelectrons ($N_{pe@PC}$) emerging from the PC is determined by measuring its current ($I_{PC@HR}$) at high laser rates using a second Keithley picoammeter (6487). As an example, the time evolution of the raw PC current $I_{PC}$ and the diode current $I_{diode}$ is shown in Fig.~\ref{fig:CEcurr} for several applied laser rates (peaks) from 10 MHz to 50 MHz. The tiny photon-induced PC current of $\lesssim$12 pA with laser switched on indicates the difficulty of the measurement.

To measure $N_{pe@anode}$, the full voltage settings are applied to the MCP-PMT. The number of photoelectrons ($N_{pe@anode}$) that are amplified and reach the anode is derived from the pulse height distribution (see section \ref{sec:gain}) measured with a 3 GHz LeCroy WavePro 7300A oscilloscope with 20 GSa/s for a 2-channel (or 10 GSa/s for a 4-channel) operation at low light intensity (controlled by ND filters) and at low rates ($LR$). All these measurements are performed for different rate and thus intensity configurations, and the final CE is determined as the average of the obtained results.

\section{Gain}  \label{sec:gain}

The gain is a measure of the number of secondary electrons generated in the amplification process within the MCPs for each photoelectron that reaches the MCP-In. This is a very important performance parameter for an MCP-PMT and can range from $<$10$^{5}$ to $>$10$^{7}$. The secondary electrons are generated in a continuous amplification process in the MCP lead or borosilicate glass pores with a diameter of 3 to 25 \textmu m. The amplification is a function of the length-to-diameter aspect ratio of the MCP pores, the voltage (HV) applied across the MCPs and a potential magnetic field (magnitude and orientation). In this work, we show results of MCP-PMTs with 10 \textmu m pores coated by an ALD technique to extend the lifetime of the PMT.

\subsection{Gain vs High Voltage} \label{gain_vs_hv}

One of the first qualification steps is to measure the gain of each MCP-PMT as function of HV by illuminating the PC with laser light at a very low photon rate and intensity. The latter is controlled by neutral density filters. A primary goal of this step is to determine the voltage setting at which a gain of 10$^6$ is reached. This gain ensures efficient single photon detection and is the goal for operation at the PANDA DIRCs.

\begin{figure}[!htbp]
			\centering
			\includegraphics[width=.8\columnwidth]{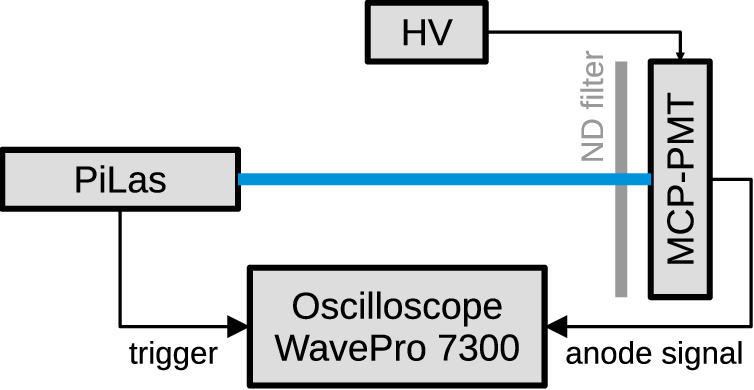}
		\caption{Block diagram of the setup for gain measurements at different HVs.}
		\label{fig:gain1}
\end{figure}

\begin{figure}[!htbp]
	\centering
	\includegraphics[width=.98\columnwidth]{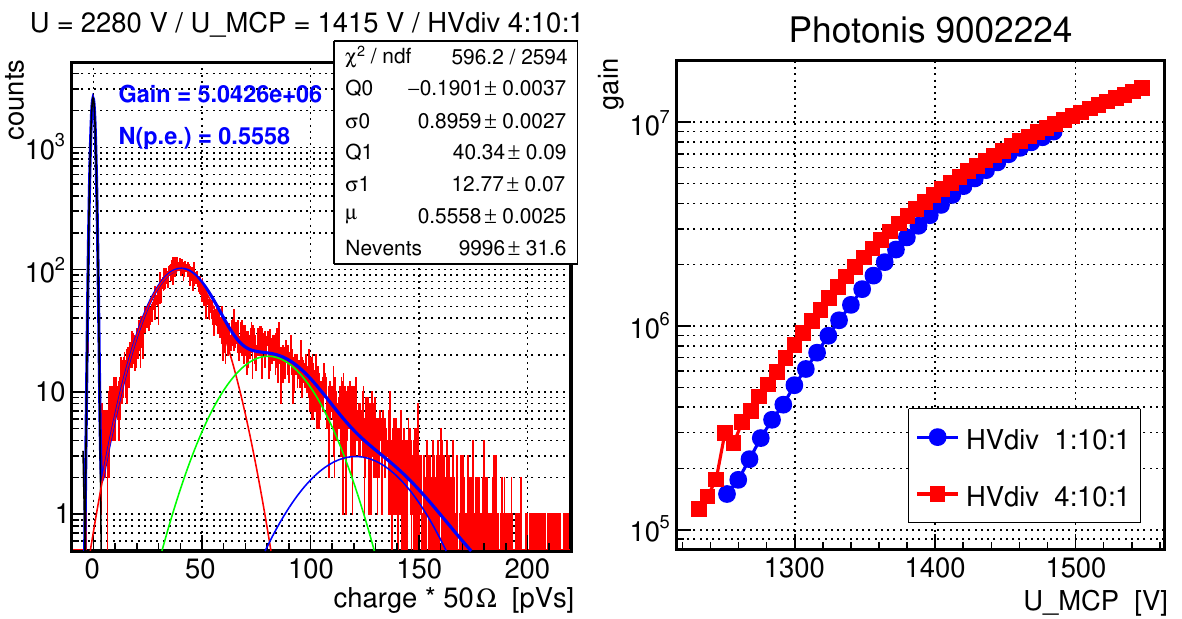}
	\caption{Pulse height distribution and response function fit (left) and gain curve (right) for different voltage divider ratios of the PHOTONIS MCP-PMT SN 9002224. In the fit, $w$ and $\alpha$ of equation \eqref{eq:response} were set to zero.}
	\label{fig:gain3}
\end{figure}

The anode signal of a central pixel (e.g. 44 or 45 in Fig.~\ref{fig:QEwave} (left)) is fed into the LeCroy WavePro 7300A oscilloscope \cite{lecroy} or a CAEN DT5742B desktop waveform digitizer \cite{caen}. The signal charge is integrated (usually directly in the oscilloscope) and the accumulated pulse height distributions are stored for further analysis. A schematic representation of our typical setup is shown in Fig.~\ref{fig:gain1}. The gain and the number of photoelectrons can be derived by fitting the pulse height distribution (Fig.~\ref{fig:gain3} (left)). If required, this plot can also be used to determine the peak-to-valley (P/V) ratio between the noise band (pedestal) and the single photon signal (first Gaussian of the fit). The measured pulse height distribution is deconvoluted using an approach described in \cite{bellamy}:

\small
\begin{multline} \label{eq:response}
	S(x)=\sum_{n=0}^{\infty} \frac{\mu^{n}\cdot e^{-\mu}}{n!} \cdot \Biggl((1-w)G_{n}(x-Q_{0})+w\cdot\frac{\alpha}{2}\exp\left[-\alpha(x-Q_{n}-\alpha\sigma_{n}^{2})\right] \\
	\cdot\biggl(\text{erf}\left[\frac{|Q_{0}-Q_{n}-\sigma_{n}^{2}\alpha|}{\sigma_{n}\sqrt{2}}\right]
	+\text{sign}(x-Q_{n}-\sigma_{n}^{2}\alpha)    \cdot\text{erf}\left[\frac{\left|x-Q_{n}-\sigma_{n}^{2}\alpha\right|}{\sigma_{n}\sqrt{2}}\right]\biggr)\Biggr)
\end{multline}

\normalsize

with

\begin{align}
	G_n(x-Q_0) &= \frac{1}{\sigma_n \sqrt{2 \pi}} \cdot
	\exp\left(-\frac{(x - Q_n)^2}{2 \sigma_n^2}\right) \\
Q_n &= Q_0 + nQ_1 \\
\sigma_n &= \sqrt{\sigma_0^2 + n\sigma_1^2}
\end{align}

The response function $S(x)$ has seven free parameters \eqref{eq:response}: $Q_0$ and $\sigma_0$ define the pedestal, $Q_1$ and $\sigma_1$ characterize the pulse height distribution of the first photoelectron. $\mu$ is the mean number of photoelectrons and $n$ the photoelectron index, which is usually limited to $n \le 24$ in our fits. $w$ and $\alpha$ describe a potential exponential background below the pedestal that extends into the photon peak. The most important parameters for the analyses are: (a) the mean number of photoelectrons $\mu$ that emerge from the PC and are amplified in the MCPs, and (b) the mean signal charge for a photoelectron $Q_1$. The PMT gain $G$ can be calculated directly from $Q_1$.

In our analyses, we generally use the simplified fit function $S_{MCP}(x)$ \eqref{eq:MCPresponse}. The background is neglected ($w = 0$ and $\alpha = 0$), as the accumulation of the pulse height distribution is only triggered if a photon pulse was emitted by the laser. This simplifies the response function to

\begin{equation} \label{eq:MCPresponse}
S_{MCP}(x)=\sum_{n=0}^{24} \left\lbrace \frac{\mu^n\cdot e^{-\mu}}{n!} \cdot 
\frac{1}{\sigma_n \sqrt{2 \pi}} \cdot
\exp\left[-\frac{(x - Q_n)^2}{2 \sigma_n^2}\right] \right\rbrace
\end{equation}

Typical gain curves are shown in Fig.~\ref{fig:gain3} (right) for different voltage divider configurations. The voltage between the PC and the MCP is varied in the same way as the voltage between the two MCPs. The curves show that the voltage between the PC and MCP also has a significant influence on the gain.

\subsection{Gain Homogeneity}

\begin{figure}[!htbp]
	\centering
	\includegraphics[width=.95\columnwidth]{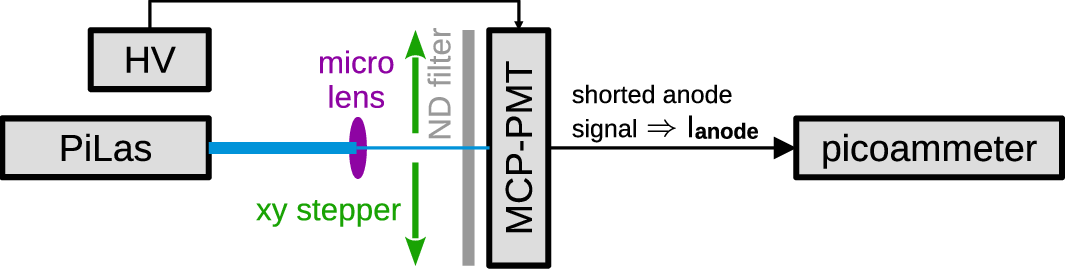}
	\caption{Block diagram for the setup of position-dependent gain measurements.}
	\label{fig:gainxy}
\end{figure}

Knowing the gain homogeneity across the active sensor area is important, as it affects the overall photon detection efficiency when the same signal threshold is applied to all pixels during PMT operation. The gain homogeneity measurement setup is based on the same principle as the QE position scan. A schematic representation of the measurement setup is shown in Fig.~\ref{fig:gainxy}. 

The MCP-PMT is operated with a voltage configuration that provides a gain of approximately $10^6$ and the sensor surface is automatically scanned in xy steps of typically 0.5 mm with an attenuated laser beam. At an intensity of $\sim$1 photoelectron, 50-100 kHz laser rate and $\sim$10$^6$ gain, an anode current of $I_A(xy) \sim 5 - 10$ nA is measured for about 0.2 s at each scan position. In this case, the laser spot is tightly focused, but before the scan, the anode current is checked at different laser rates to avoid charge saturation. In this measurement, all anode pixels are shorted, and we measure the anode current at each xy position. Similarly to the QE scans, after typically four scanned x-rows, the laser spot is placed far outside the active PC area and the dark current $I_{dark}$ of the MCP-PMT is measured. Each raw anode current $I_A(xy)$ is corrected for the dark current $I_{dark}$ and scaled with the current $I_{A,refpix}$ and the gain $G_{refpix}$ at the reference pixel (e.g. 44 or 45) measured in the HV scan (see Section~\ref{gain_vs_hv}). After the correction for $QE(xy)$ at each measured position (see Section~\ref{positionQE}), this results in a QE-corrected gain distribution $G(xy)$ over the entire active PC surface. 

\begin{equation}
	G(xy) = \frac{I_A(xy)-I_{dark}}{I_{A,refpix}-I_{dark}} \cdot G_{refpix} ~/~ QE(xy)
\end{equation}

\begin{figure}[!htbp]
	\centering
	\includegraphics[width=.98\columnwidth]{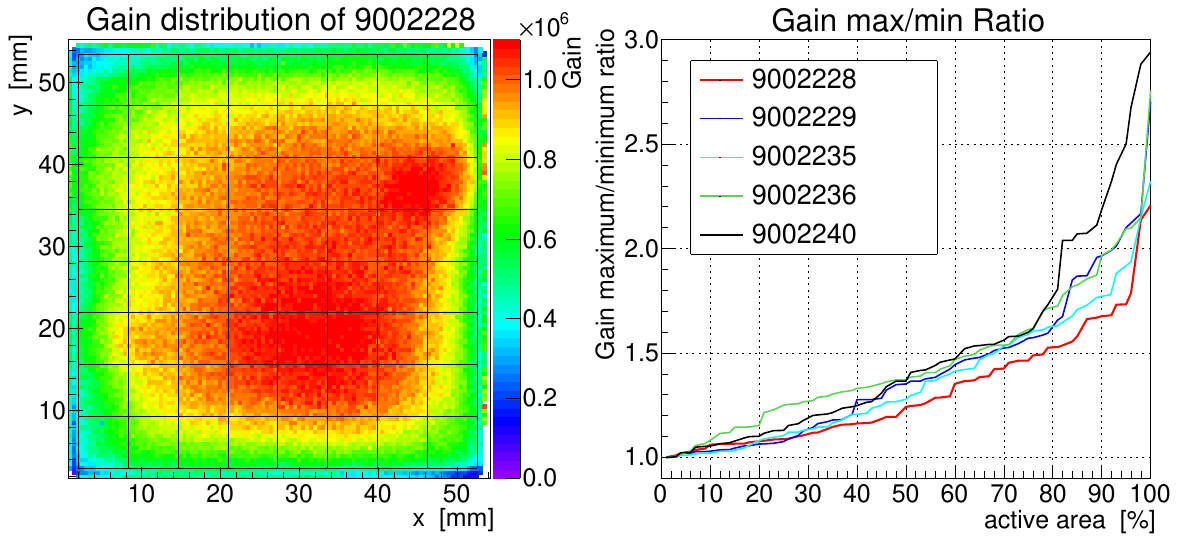}
	\caption{QE-corrected gain as a function of PC position (left) for the PHOTONIS SN 9002228 with the black overlaid anode grid and the ratio of maximum gain to minimum gain (max/min) for a set of five PHOTONIS MCP-PMTs (right).}
	\label{fig:GainPos}
\end{figure}

Fig.~\ref{fig:GainPos} (left) shows the gain of the PHOTONIS SN 9002228 as a function of the PC position. It is clear that the gain is less homogeneous than the QE. It is generally significantly lower towards the rims of the tube. Here too, the black grid in the plot illustrates the position of the anode pixels of the PMT. Similar to Section~\ref{positionQE}, an average gain is calculated for each anode pixel from the various measurement positions and a max/min ratio of the gain is determined from these values. The results are shown in Fig.~\ref{fig:GainPos} (right) for five example PMTs. This approach can be used to quantify the gain variation over the active area of the MCP-PMT. Typically, $>$80\% of the area has a gain max/min ratio of $<$2. This is also observed in many multianode dynode PMTs, and thus is not uncommon.

\subsection{Gain inside Magnetic Field}

The very compact PANDA detector leads to the requirement that the MCP-PMTs of the DIRCs must be placed inside a magnetic field of $\sim$1 Tesla, with the field lines crossing the PMT axis at an inclination angle $\theta$ of up to $\sim$20 degrees. Even under these conditions, the gain must be at least 10$^6$ to enable efficient single photon detection. This requires a measurement of the gain as a function of the magnitude and angle of the magnetic field.

\begin{figure}[!htbp]
	\centering
	\includegraphics[width=.99\columnwidth]{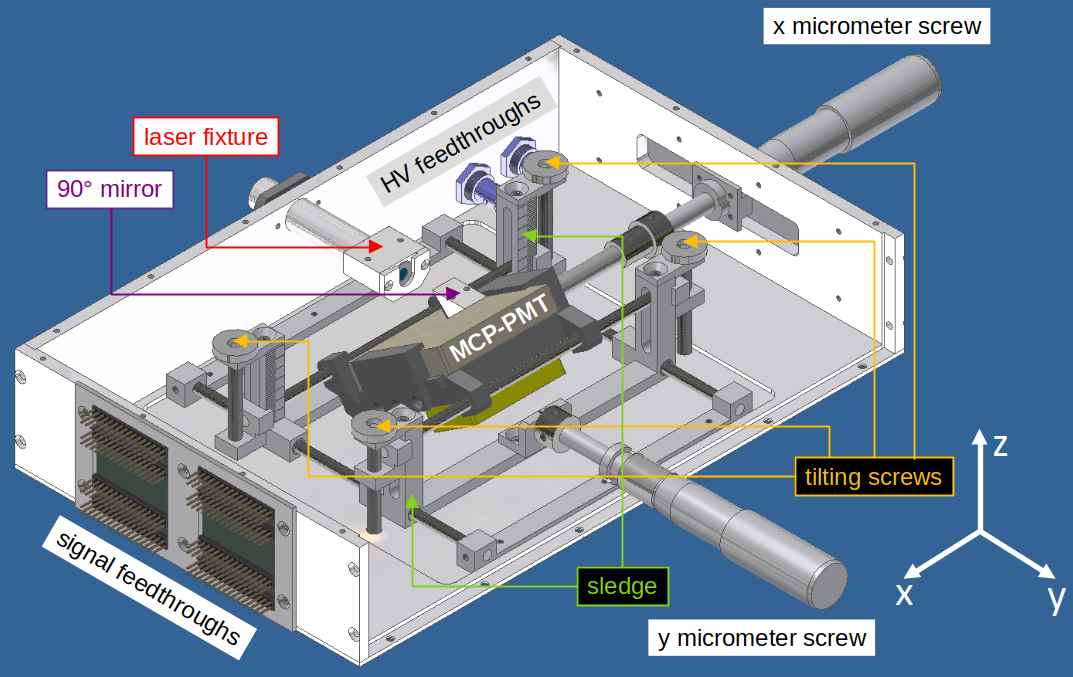}
	\caption{Dark box with the most important parts for B-field measurements.}
	\label{fig:Bfield1}
\end{figure}

The measurements were carried out on a dipole magnet with a field of up to 2 Tesla at the Julich Research Center in Germany. The B-field was monitored with a Hall probe. Unfortunately, the pole shoe gap of this dipole magnet is only 6 cm wide. To place the PMTs in this gap, we designed a special aluminum dark box with blackened side walls (Fig.~\ref{fig:Bfield1}). The total size of the box is 32$\times$19$\times$6 cm$^3$, and it consists of four main parts: (1) a top cover with a small mirror mounted on the bottom at a 45$^{\circ}$ angle, (2) a light-tight box with a holder for inserting a lens and several cable feed-throughs for the high voltage and the anode signals, (3) four special screws for adjusting the tilt angle $\theta$ and (4) an interior slide which holds the MCP-PMT in place. The slide can be moved in x and y directions using micrometer screws. The screws can also be moved with stepper motors to enable automatic position scans (Section~\ref{sec:scans_bfield}).

\begin{figure}[!htbp]
	\centering
	\includegraphics[width=.95\columnwidth]{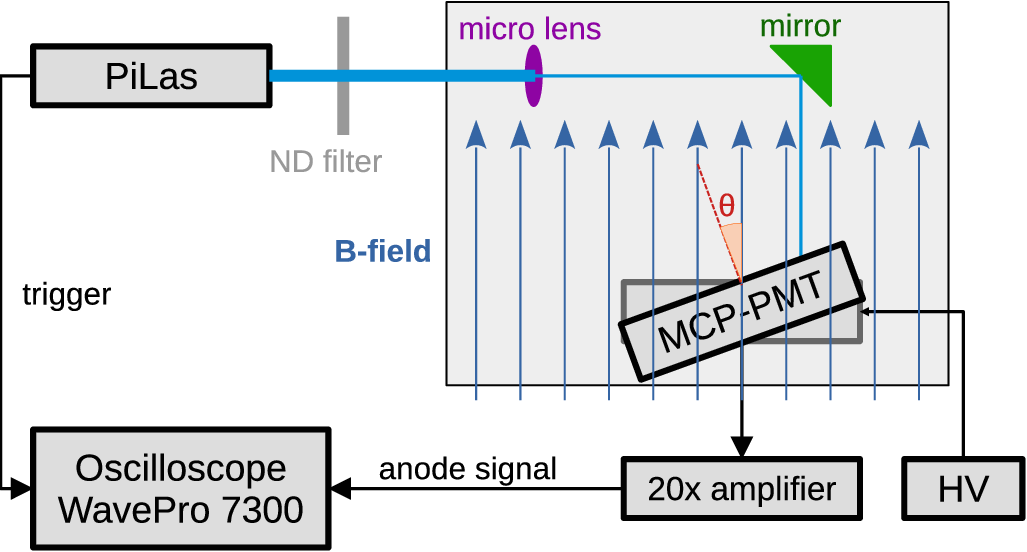}
	\caption{Setup for measuring the gain as a function of the magnetic field.}
	\label{fig:gainB}
\end{figure}

Fig.~\ref{fig:gainB} shows a schematic representation of the setup for the measurements within the B-field. The MCP-PMT is illuminated with a laser beam that is attenuated to single photon level with ND filters. The light is transported into the box through an optical fiber, which is connected to a microfocus lens at its downstream end. In a special tube, the lens is pushed into the dark box and the photons are deflected at the top cover mirror by 90$^{\circ}$. By varying the lens position, the light is focused to a point on the PC of the MCP-PMT that corresponds to the center of one of the anode pixels, except for scans. The high voltages for the tube are fed into the box via SHV feed-throughs, while the voltage divider is placed outside the box. The measurements are performed at low laser rates ($\le$5 kHz) to avoid gain saturation due to high local intensities. The anode signals from one or more pixels are read out via Samtec LSHM 140 (or 150) 50 $\Omega$ high-speed microcoaxial cables and amplified with a $\times$20 amplifier (Phillips Scientific PS775) before being passed to a LeCroy WavePro oscilloscope which accumulates pulse height distributions with typically 50k - 100k events for each selected setting [HV, tilt angle, B-field]. The data sets are stored and analyzed in the same way as described in an earlier subchapter.

\begin{figure}[!htbp]
	\centering
	\includegraphics[width=.98\columnwidth]{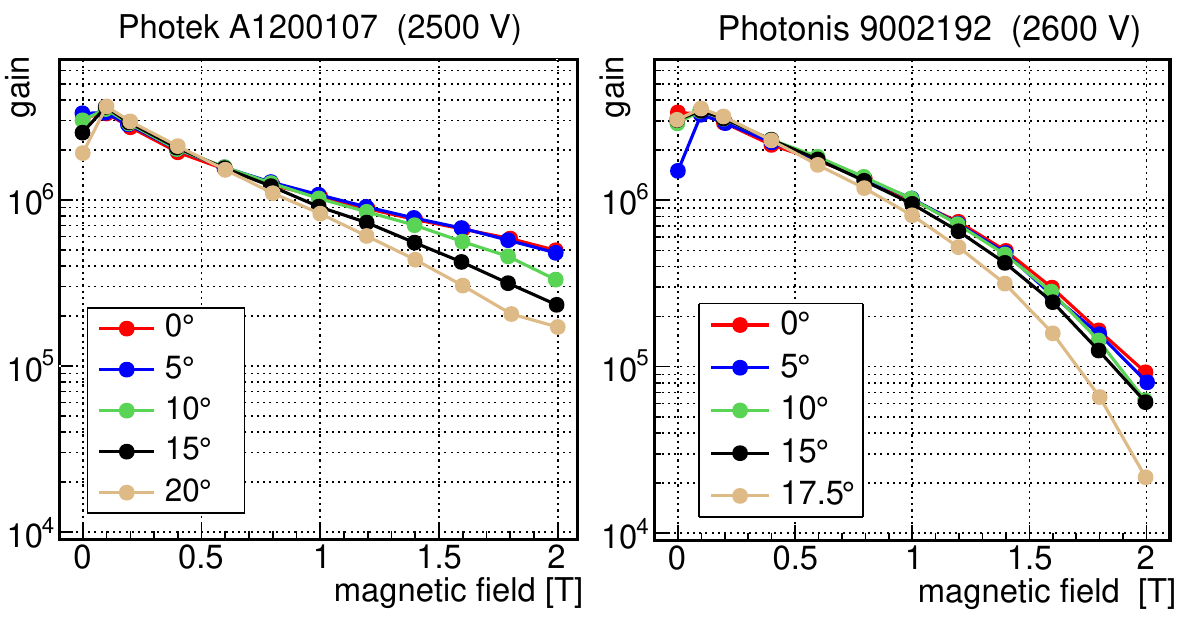}
	\caption{Gain as a function of the magnetic field for different tilt angles $\theta$ for the Photek A1200107 and the PHOTONIS SN 9002192 MCP-PMT with 6 \textmu m and 10 \textmu m pores, respectively.}
	\label{fig:gainB_K}
\end{figure}

For an MCP-PMT with 6 \textmu{m} pores from Photek and one with 10 \textmu{m} pores from PHOTONIS, the resulting gains for different B-fields and tilt angles are shown in the plots of Fig.~\ref{fig:gainB_K}. It can be clearly seen that the decrease in gain by a factor $\sim$3 between 0 and 1 T is similar in both PMTs. Beyond 1 T, the Photek PMT performs significantly better than the PHOTONIS tube, which is to be expected due to the smaller MCP pore diameter.


\section{Time Resolution}  \label{sec:timeres}

A unique feature of MCP-PMTs is their excellent time resolution for single photons. Due to the compact dimensions and small path variations of the electrons on their way from the PC to the anode, the transit time spread ($\sigma_{TTS}$) is typically $\ll$50 ps. However, this is a lower limit as the true time resolution is impaired by some primary electrons bouncing back from the MCP-In electrode, which becomes visible as an additional time component arriving later than the main TTS peak (as shown in Fig.~\ref{fig:tres}). For experimental applications, it is therefore better to also determine the mean time resolution $\sigma_{RMS}$ within a certain time interval, in our case from -0.5 to 2.0 ns around the TTS peak. This is actually the most important time resolution for many applications. Another approach has been chosen in \cite{vavra2007} by fitting two Gaussians with $\sigma_{narrow}$ and $\sigma_{wide}$ to the time distribution.

\begin{figure}[!htbp]
	\centering
	\includegraphics[width=.95\columnwidth]{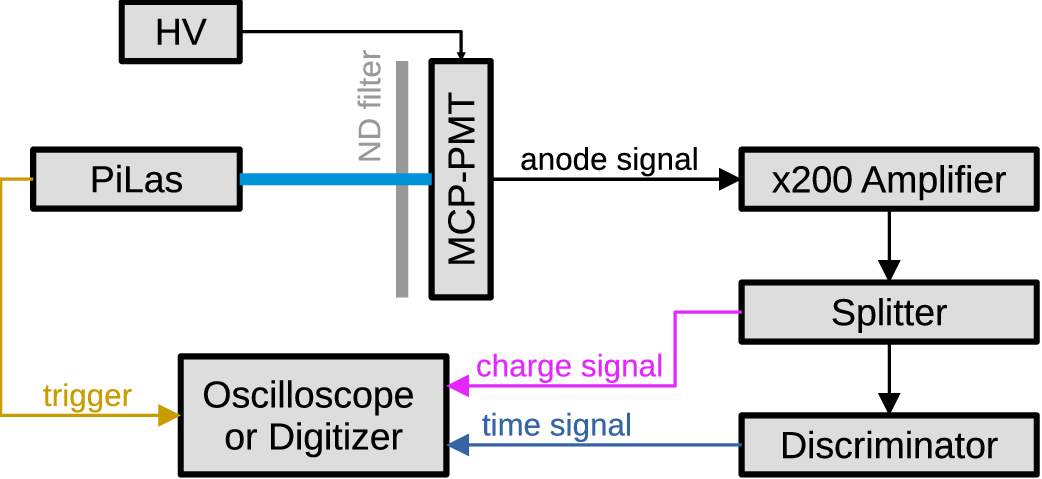}
	\caption{Block diagram of the setup for the time resolution measurements.}
	\label{fig:timeres}
\end{figure}

\begin{figure}[!htbp]
	\centering
	\includegraphics[width=.9\columnwidth]{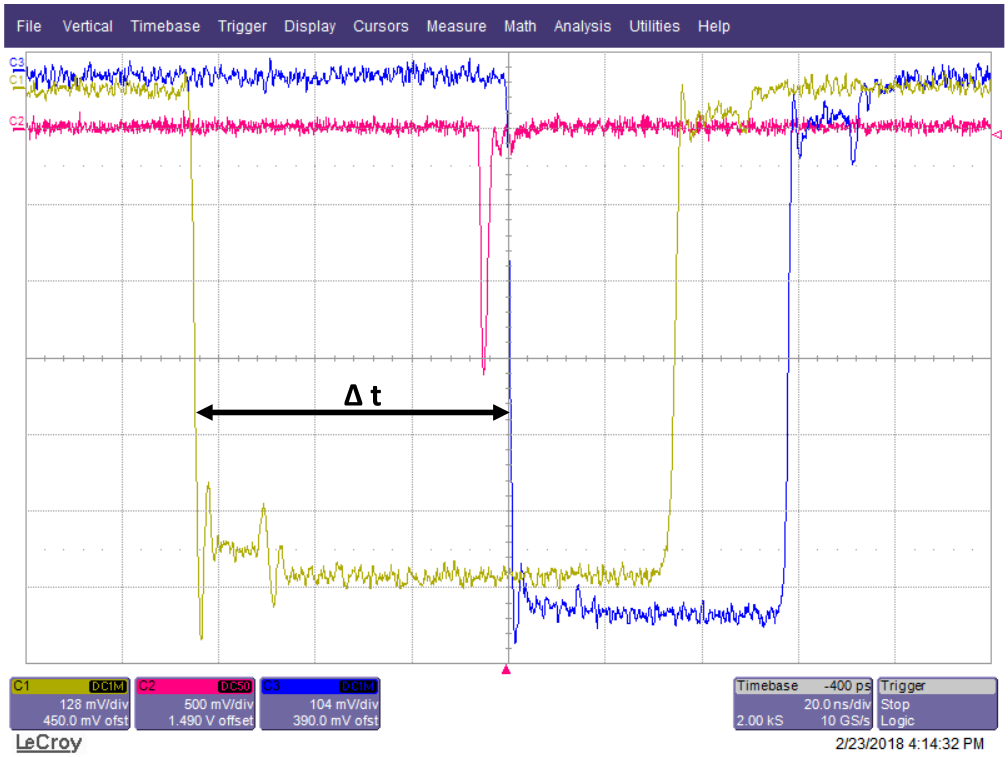}
	\caption{Oscilloscope screenshot of the signals used to determine the time resolution: amplified anode charge signal (red), laser trigger signal (yellow) and discriminated anode signal (blue). The time resolution is the jitter in the delay $\Delta t$ between the yellow and blue signal. The time resolution of the laser trigger signal is $<$3 ps (RMS), and the internal time resolution of an oscilloscope channel was measured to be $\sim$5 ps ($\sigma$).}
	\label{fig:tscreenshot}
\end{figure}

\begin{figure}[!htbp]
	\centering
	\includegraphics[width=.98\columnwidth]{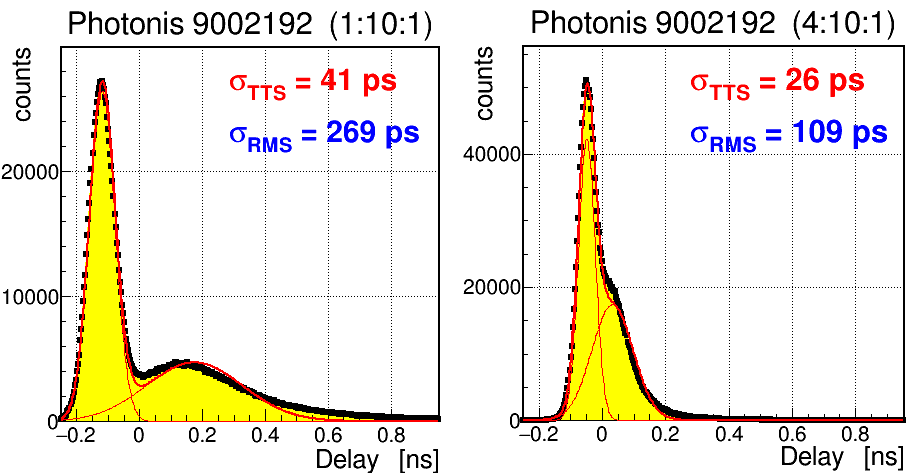}
	\caption{Time distributions ($\sigma_{TTS}$, $\sigma_{RMS}$) for a PHOTONIS MCP-PMT with different voltage divider configurations \cite{lehmann22}: $R_{PC}$:$R_{MCP}$:$R_{A}$ = 1:10:1 (left) and 4:10:1 (right). The right-hand plot shows a clear improvement in $\sigma_{RMS}$.}
	\label{fig:tres}
\end{figure}

The time resolution can be measured by illuminating a single pixel in the center of the MCP-PMT with laser light attenuated to single photon level ($N_{pe@anode}$ $\approx$ 0.5 - 2). The anode signal is amplified 200-fold with an Ortec FTA820 device of 350 MHz bandwidth before being split into two parts with a simple passive 50 $\Omega$ splitter circuit. One signal branch is fed directly into the LeCroy WavePro 7300A oscilloscope (3 GHz; 20 GSa/s) to measure the integrated anode charge of the pulse, the other signal branch is passed through a leading edge discriminator (LeCroy 821, Phillips Scientific PS705, or PS707) which provides the timing signal at the oscilloscope. Also, the CAEN DT5742B digitizer is sometimes used to measure many waveform samples of the anode signals, which are later analyzed to obtain the time resolution. Due to the limited bandwidth (500 MHz) and sampling frequency (5 GSa/s) of the digitizer, the time resolution obtained is slightly worse than that determined with the oscilloscope. Data acquisition and analysis are automatic and controlled by Python scripts. A schematic representation of the time resolution setup is shown in Fig.~\ref{fig:timeres}, while a screenshot in Fig.~\ref{fig:tscreenshot} illustrates the most significant signals at the oscilloscope. 

By analyzing the jitter of the time delay between the anode time signal and the trigger signal of the laser controller, we can measure both $\sigma_{TTS}$ and $\sigma_{RMS}$. The charge and delay of each photoelectron are stored in files, with one measurement containing $\sim$100k events. In the analysis, these charge and delay values are used for a time-walk correction, resulting in distributions as shown in Fig.~\ref{fig:tres}. The obtained histograms are fitted with two Gaussian curves to roughly unfold the main TTS peak (left) and the recoil electron peak (right). The two distributions shown were measured with different HV divider configurations, where $U_{PC-MCP}$ differs by a factor of 4. This leads to a shift of the recoil electron peak, which shows that a larger voltage between PC and MCP-In reduces $\sigma_{RMS}$. 

\section{Rate Capability}  \label{sec:rate}

\begin{figure}[!htbp]
	\centering
	\includegraphics[width=.98\columnwidth]{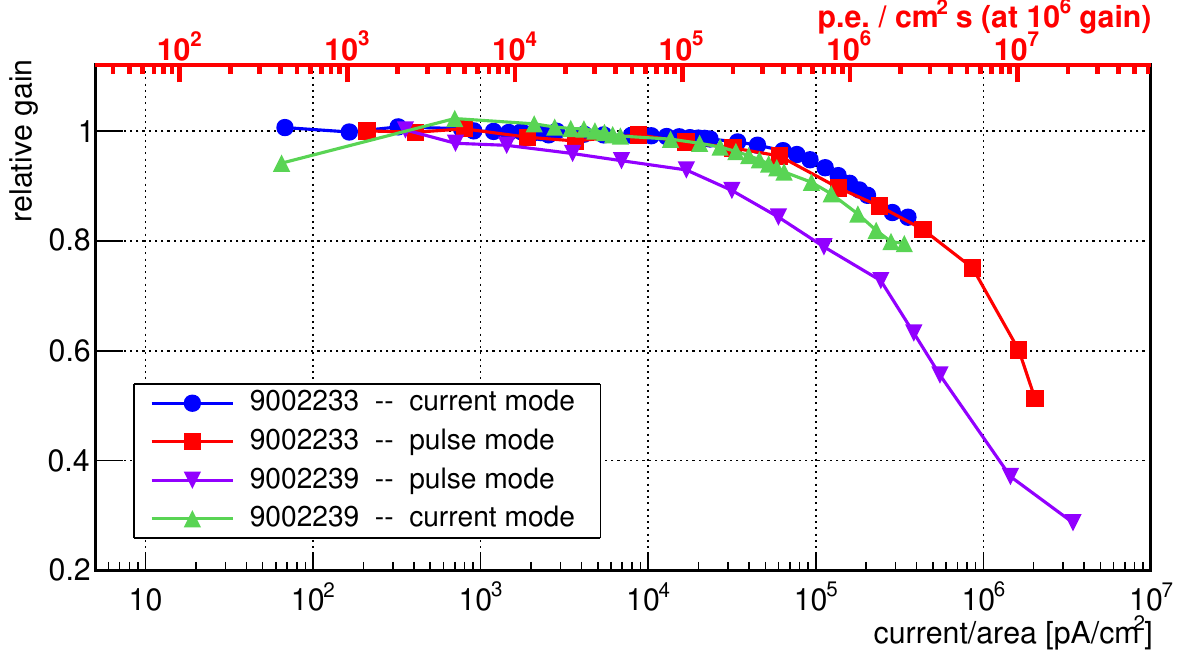}
	\caption{Rate capability of the latest 2-inch PHOTONIS MCP-PMTs (SN 9002233 and SN 9002239) with two ALD layers. The results obtained with the "current mode" and the "pulse mode" are compared.}
	\label{fig:ratedep}
\end{figure}

The photoelectrons emerging from the PC are accelerated towards the MCP, where secondary electrons are generated during the amplification process in the MCP pores. The missing charge of these electrons must be restored in the MCP walls, which defines a MCP recovery or recharge time $\tau = RC$. The capacitance [$\mathcal{O}$(100 pC)] depends on the MCP area and the resistance $R$ (tens of M$\Omega$) is rather large. Consequently, the recharge current is limited. If the amount of emitted secondary electrons is large, which is particularly the case in the second MCP, the gain can be reduced at high photon rates. This effect limits the rate capability of MCP-PMTs and depends on the photon intensity and the illuminated area. We measure the rate capability as a function of the anode current per area in two different ways: (1) illuminating the entire active PC area and directly reading the current with all anode pixels shorted ("current mode"); (2) illuminating only one anode pixel and analyzing the corresponding pulse height distribution ("pulse mode"). When comparing the results of the two modes, the CE is not taken into account, as it is $>$90\% for the PHOTONIS MCP-PMTs examined here. Exemplary results of the rate capability obtained with both methods are shown in Fig.~\ref{fig:ratedep} for two recent 2-inch PHOTONIS MCP-PMTs with 2 ALD layers. The anode current $I_{anode}$ per area $A$ is plotted against the gain relative to low rates. The quantity on the upper axis is the corresponding photoelectron rate $R$~$\cdot$~cm$^{-2}$ assuming a gain of $G = 10^6$. The rate capability for each other gain $G$ can be found by applying the following trivial formula which is used to calculate $R$
 
\begin{equation}
	R = \frac{I_{anode}}{A \cdot G \cdot e}
\end{equation}

In most cases, both measurement techniques provide the same result, with the gain reduction becoming relevant ($<$90\%) at a photoelectron rate of $\lesssim$1 MHz/cm$^2$. Some 1-inch MCP-PMTs show a rate capability of up to 10 MHz/cm$^2$ \cite{britting2011,lehmann14,inami15}. Sometimes we also observe different results with both measurement methods. This could be due to an inhomogeneity of the resistivity across the active MCP-PMT area.

\subsection{Measurement with Anode Current}

\begin{figure}[!htbp]
	\centering
	\includegraphics[width=.98\columnwidth]{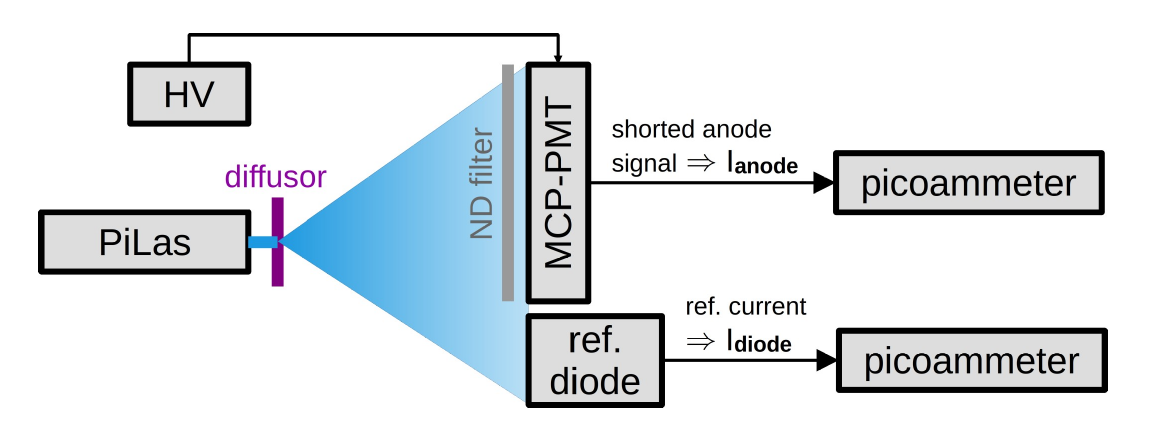}
	\caption{Block diagram of the setup for rate capability measurements with a shorted anode current at different laser frequencies.}
	\label{fig:ratestab}
\end{figure}

Measuring the shorted anode current $I_{anode}$ as a function of photon rate simulates the expectation in a real experiment where the entire PC surface is illuminated with photons. To measure this, we broaden the laser light with a square diffuser ED1-S20-MD from Thorlabs to cover the entire active sensor area. With an ND filter, the light intensity is reduced to 5 - 20 photons per laser pulse and entire PC area. In this type of measurement, all anode pixels are shorted and $I_{anode@f}$ is measured with a picoammeter at different laser frequencies $f$. In parallel, the current of a reference diode $I_{diode@f}$ is measured with a second picoammeter. This is necessary because the laser intensity does not increase proportionally to the laser frequency and must be corrected. The MCP-PMT is operated with a voltage that provides $\sim$10$^6$ amplification. The setup is shown schematically in Fig.~\ref{fig:ratestab}.

A correction for the dark current is made for each measured frequency $f$. For example, after measuring of the anode current at a certain laser frequency, the dark currents of the PMT $I_{anode,dark}$ and the reference diode $I_{diode,dark}$ are measured. In relation to a reference point at a low laser frequency $f_{ref}$, at which the MCP-PMT does not yet saturate, the relative gain $G_{rel}$ can be calculated for each laser frequency $f$ according to \eqref{ratecurrent}.

\begin{equation} \label{ratecurrent}
	G_{rel} 
	= \frac{I_{anode@f} - I_{anode,dark}}{I_{anode@f_{ref}} - I_{anode,dark}}  \cdot \frac{I_{diode@f_{ref}} - I_{diode,dark}}{I_{diode@f} - I_{diode,dark}}
\end{equation}

\subsection{Measurement with Pulse Height Distribution}

Analyzing the pulse height distribution is another method for measuring the rate capability of an MCP-PMT. The photocathode is illuminated with a laser light spot that is attenuated to single photon level. An aperture of $\sim$5 mm diameter limits the spot size to one anode pixel, usually at position 44 (see Fig.~\ref{fig:QEwave} [left]) or elsewhere. All anode pixels except the illuminated one are shorted, and the charge spectrum is recorded for different laser frequencies $f$ and intensities $N_{pe}$. The pulse height distributions are analyzed in the same way as described in Section~\ref{gain_vs_hv}. The fit provides the gain $G$ and the number of photoelectrons $N_{pe}$, which can be used to determine the anode current per area $A$ by

\begin{equation}
	\frac{I_{anode}}{A} 
	= \frac{f \cdot e \cdot G \cdot N_{pe}}{A}
\end{equation}

\noindent where $f$ and $e$ are the applied laser frequency and the elementary charge, respectively. In contrast to the "current mode", this "pulse mode" method only illuminates a relatively small area of the MCP-PMT to determine the rate capability, which can lead to slightly different results. In principle, the "pulse mode" method can also be used to determine the rate capability as a function of the active surface. This has been done for a few tubes.

\subsection{MCP Recharge Time}

\begin{figure}[!htbp]
	\centering
	\includegraphics[width=.9\columnwidth]{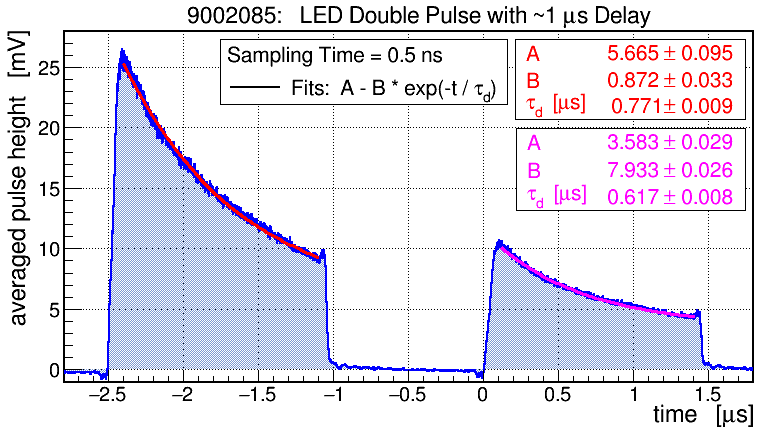}
	\includegraphics[width=.9\columnwidth]{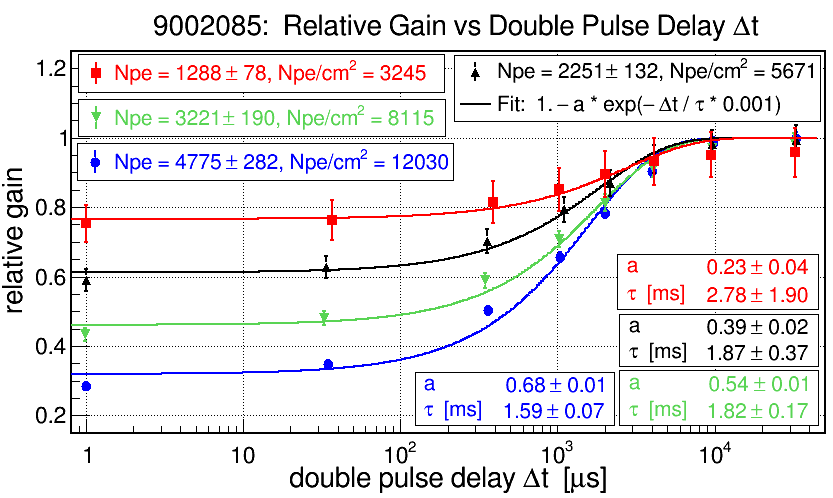}
	\caption{PHOTONIS SN 9002085: mean waveform (averaged over 1000 pulses) of an LED double pulse with 1.5 \textmu{}s width and $\sim$1 \textmu{s} delay (top); measured relative gain for different double pulse delays (bottom), fitted by the function $[1 - a\cdot exp(\Delta t/\tau)]$. It shows the MCP recovery behavior for different photon densities ($N_{pe}/cm^2$), resulting in an average recharge time of $\tau \approx$ 1.6 ms.}
	\label{fig:RecovTime}
\end{figure}

It is generally agreed that the rate capability depends on the charge recovery time $\tau$ of the MCP pores. Illuminating the photocathode with LED double pulses with different time delays is a simple way to measure this time. The electrical double signals from a Philips PM-5770 pulse generator are used to drive a blue LED and trigger an oscilloscope. The LED is placed near the input window of the MCP-PMT, attenuated by an ND filter to a moderate light intensity of a few hundred or thousand photons per pulse and "focused" by an aperture directly in front of the photocathode. The photons in each pulse are distributed over a time interval of typically 1.5 \textmu{s}. Since the PMT response to each photon is very fast ($\sim$1 ns width), at low intensity many of the individual photons can even be visualized with a fast oscilloscope. In Fig.~\ref{fig:RecovTime} (top) the averaged waveform is shown with two consecutive anode signals for a delay of $\sim$1 \textmu{s}. Normally $\sim$1000 of these waveforms are stored, averaged, and further analyzed. 

A first and surprising observation is the decay of the pulse heights with a time constant of $\tau_d < 1$ \textmu{s}. Since the pulser signal has a rectangular shape with a width of 1.5 \textmu{s}, this effect must be due to saturation of either the LED or the MCP-PMT. LED saturation was ruled out by performing the same measurement with a Philips XP2262B dynode PMT, which shows neither signal decay nor gain reduction at high rates. On the other hand, the signal decay time $\tau_d$ must have a different origin than the MCP-PMT charging time $\tau$ as it is too short. We assume that this decay is caused by space charges in the MCP pores when many photons hit the same pores almost simultaneously. The transit time in the MCPs and from the MCP-Out to the anode is $\mathcal{O}$(few ns) for an electron avalanche. Considering that for each LED pulse several thousand Npe/cm$^{2}$ ($\approx$ $\mathcal{O}$(100 Npe/pore)) are evenly distributed over a width of $\sim$1.5 \textmu s (see Fig. ~\ref{fig:RecovTime}), this multiplies to a time of $\mathcal{O}$(few hundred ns), in which the MCP pores and the region after the MCP-Out are filled with many electrons from the multiplication process and could lead to a lower gain due to a reduced electric field. This would mean that the measured decay constant $\tau_d$ is roughly comparable to the transit time of the secondary electrons multiplied by the number of photo electrons (Npe). However, more studies are needed to fully understand this signal decay. On the other hand, the rather small increase in charge between the end of the first and the beginning of the second pulse is due to the MCP charging time.

The waveforms are used to integrate the anode charge contained in each of the two pulses. Their charge ratio for a $\sim$10$^6$ single photon MCP gain is plotted against the time delay of the double pulse. The results for different photon intensities ($N_{pe}/cm^2$) are shown in Fig.~\ref{fig:RecovTime} (bottom). The fitted function $[1 - a\cdot exp(\Delta t/\tau)]$ provides the pore recharge time $\tau$, which does not seem to depend on the intensities within the given uncertainties. The weighted mean value for the PHOTONIS MCP-PMT SN 9002085 with 25 \textmu{m} pores is $\tau = (1.63 \pm 0.06)$ ms, which is smaller than the expected value $\tau_{exp} = R\cdot C$ = 7.5 ms from the MCP resistance $R$ and the capacitance $C$. This roughly agrees with measurements of the recharge time in other tubes where $\tau = k\cdot \tau_{exp}$ with $k$ = 0.4 - 2 \cite{kobayashi23} was observed, where $k$ should represent the properties of the unknown recharge circuit.

The data demonstrate that the relative gain at 1 \textmu{s} delay drops more sharply for larger photon densities. In addition, measurements at similar photon densities show that the relative gain decreases with increasing illumination area. It confirms the observation that the rate capability and thus the charging behavior depend on both the photon density and the overall photon intensity (= illuminated area). The more the capacitor charge is emptied when a new photon arrives, the longer it takes for the pores to fully recover, which can lead to lower amplification. The same measurements were repeated with the PHOTONIS MCP-PMT SN 9002228 with 10 \textmu{m} pores. The measured recovery time in this tube was $\tau = (2.67 \pm 0.16)$ ms with an expected value of $\tau_{exp}$ = 8.7 ms. This tube also shows a pulse decay time of $\tau_d \sim$ 1 \textmu{s}. The data indicate that one of the dominating factors for the rate capability is the initial charge of the "MCP-PMT capacitor" when a new photon is detected.

\section{Escalation}   \label{sec:escal}

\begin{figure}[!htbp]
	\centering
	\includegraphics[width=.98\columnwidth]{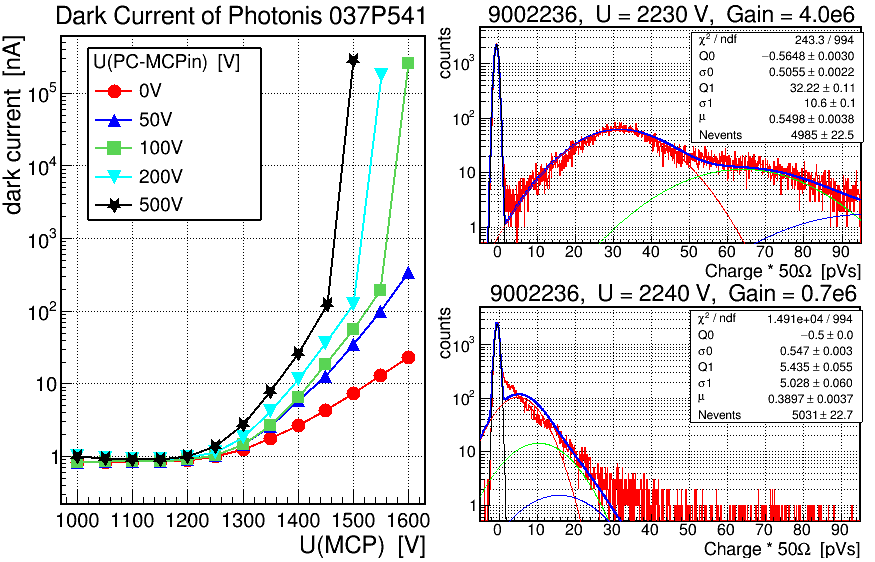}
	\caption{PHOTONIS 037P541: dark currents for different voltages at the two MCPs and between PC and MCP-In (left). PHOTONIS SN 9002236: fitted pulse height distributions for total PMT voltages $U_{PC-A}$ = 2230 V (top right) and $U_{PC-A}$ = 2240 V (bottom right). Both the steep rise in dark current by $\sim$3 orders of magnitude and the sudden drop in gain indicate the onset of "escalation".}
	\label{fig:Escalation}
\end{figure}

Some of the recent MCP-PMTs from PHOTONIS with two ALD-layers sometimes enter a strange and not yet understood state that was never observed in former Planacon-type tubes without or with only one ALD-layer. This "escalation" state \cite{daniel2022,steffen2023} can already start at quite low (few kHz single photons) or even without photon illumination when the MCP-PMT is operated at high gains ($\gtrsim$5$\cdot$10$^6$). The latter indicates that the effect is not correlated to the rate capability of the tube. At high photon intensities, the effect can already trigger at an amplification just above 10$^6$. The most important observations at the onset of escalation are:

\vspace*{-2mm}
\begin{itemize}
	\item   massive rise of count rate and dark current (Fig.~\ref{fig:Escalation} left)
\vspace*{-2mm}
	\item	significant gain drop due to the high rate (Fig.~\ref{fig:Escalation} right)
\vspace*{-2mm}
	\item   drop in MCP resistance
\vspace*{-2mm}
	\item   significant production of photons with a white spectrum
\vspace*{-2mm}
	\item   suppression of the effect already in moderate B-fields 
\end{itemize}

Close to the onset of "escalation", even a small increase of the MCP voltage by $\sim$10 V can cause a massive rise in the anode and MCP currents. When a digital camera or another PMT is placed opposite the "escalating" MCP-PMT, it can be clearly seen that a huge number of photons are generated in the sensor, which explains the increased count rate. Further tests have shown that these photons are generated at or in the MCP layers or at the electrodes, but do not originate from the PC. At this stage, it is not clear whether the observed "escalation" condition has a similar origin as the signal-induced noise reported in \cite{andreotti21} and other references. By optimizing their production processes, PHOTONIS was able to shift the onset of escalation to non-critical operation voltages. So far, we have not come across an MCP-PMT from Hamamatsu and Photek that also shows this effect, but their available tubes are still without two ALD-layers.

In our detailed performance screenings of the latest 2-inch PHOTONIS MCP-PMTs, we check each tube for "escalation". If it does not escalate to a gain of $\sim$10$^7$, we assume that the PMT is safe for operation in the PANDA environment. Nevertheless, it would be highly desirable for the cause of the "escalation" condition to be further investigated and understood.

\section{Lifetime}  \label{sec:lifetime}

The lifetime of MCP-PMTs is one of the most important issues for many recent applications. Until around 2010 \cite{britting2011,kishimoto2006}, MCP-PMTs were not an option for long-term experiments due to aging of the photocathode caused by feedback ions from the residual gas in the PMT. The prospects have changed since the MCP pores are coated with an ultra-thin layer of aluminum oxide and/or magnesium oxide using an ALD (atomic layer deposition) technique \cite{arradiance}. This drastically reduces PC aging and extends the lifetime of some MCP-PMTs by a factor $>$100 \cite{steffen2023}. Meanwhile, MCP-PMTs are an attractive option for many high photon rate applications that require immunity to magnetic fields, very fast timing and high radiation tolerance. For this reason, it is important to measure the lifetime of MCP-PMTs. In our approach, this is achieved by illuminating the photocathode with constant low-intensity light and measuring the QE (and other observables) at regular time intervals. The QE is plotted against the integrated anode charge (IAC), and a decrease in QE indicates the onset of aging of the MCP-PMT. Data examples are presented in several publications  \cite{lehmann22,conneely2013,matsuoka2017,moruyama2019}. In this paper, we will focus on the description of our approach to measure the MCP-PMT lifetime.

\subsection{Illumination Setup}

In 2011, the Erlangen group launched an extensive illumination campaign with lifetime-enhanced MCP-PMTs. The aim was to measure the aging behavior of tubes of various designs and from different manufacturers in the same environment. In order to enable the fairest possible comparison of the different PMT types, we decided to set up a test station where the aging behavior of all available lifetime-enhanced MCP-PMTs could be long-term illuminated and measured in parallel \cite{britting2011,lehmann14} and under similar conditions. This setup has already been briefly described several times in other papers \cite{britting2011, lehmann13,pfaffinger18}, but will be presented in more detail here. 

All included MCP-PMTs are permanently illuminated with a common LED light source at 460 nm. The special LED driver consists of a combination of an old Anritsu bit-generator and a self-built pulse generator, which generates square-wave pulses of $\sim$10 ns width and $\sim$70 V height. These electrical pulse signals are sufficient to trigger a very short LED flash. Unfortunately, the rather large and fast pulser signals also cause a considerable amount of stray electromagnetic radiation, which can interfere with all other ongoing measurements. For this reason, double-shielded Aircell7 cables had to be used to conduct the control signals from the pulser to the LED.

\begin{figure}[!htbp]
	\centering
	\includegraphics[width=.99\columnwidth]{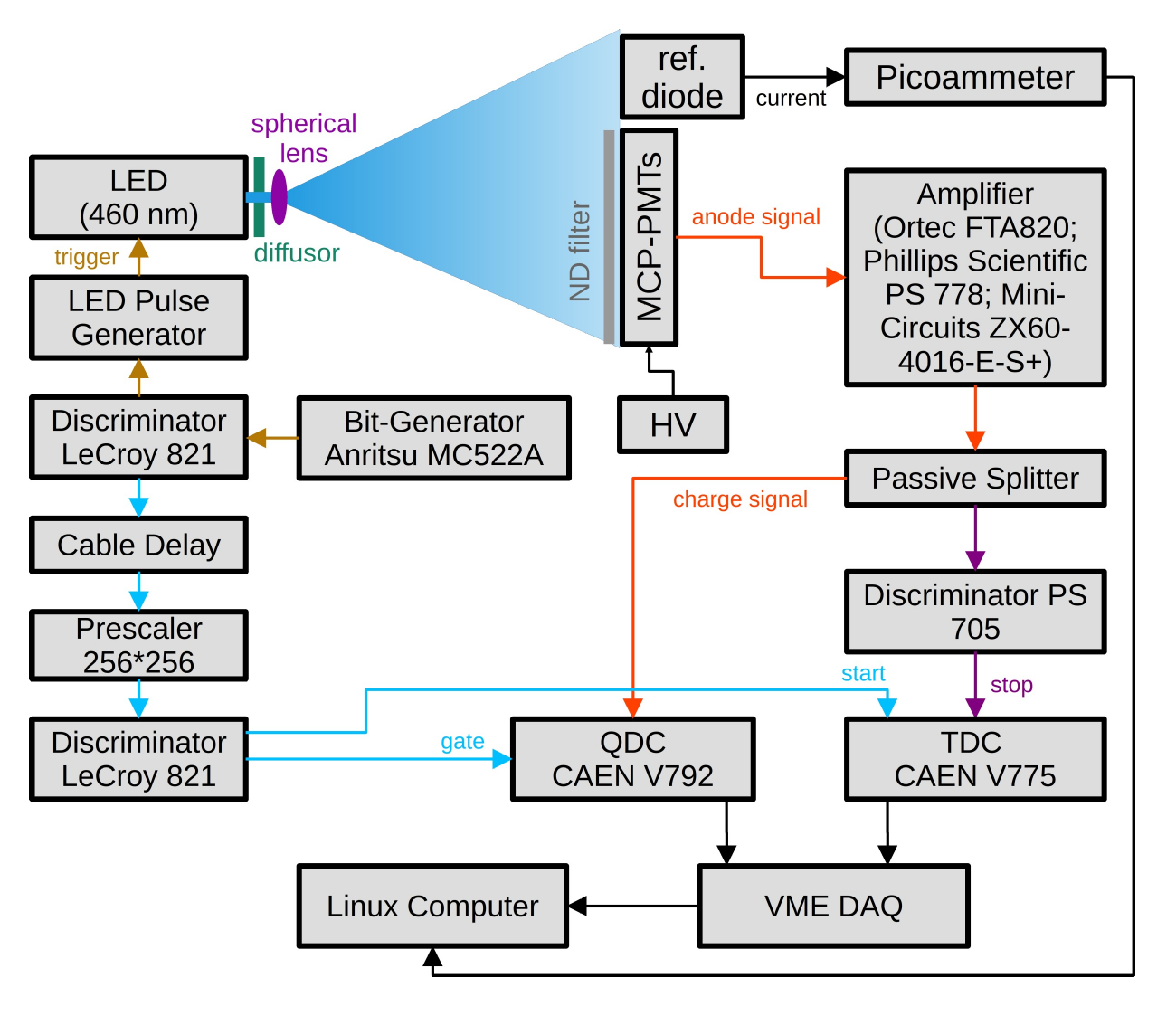}
	\caption{Block diagram with LED driver and readout as well as the DAQ electronics chain of the lifetime setup.}
	\label{fig:lifetime_block}
\end{figure}

\begin{figure}[!htbp]
	\begin{minipage}{.76\columnwidth}
		\centering
		\includegraphics[width=.98\columnwidth]{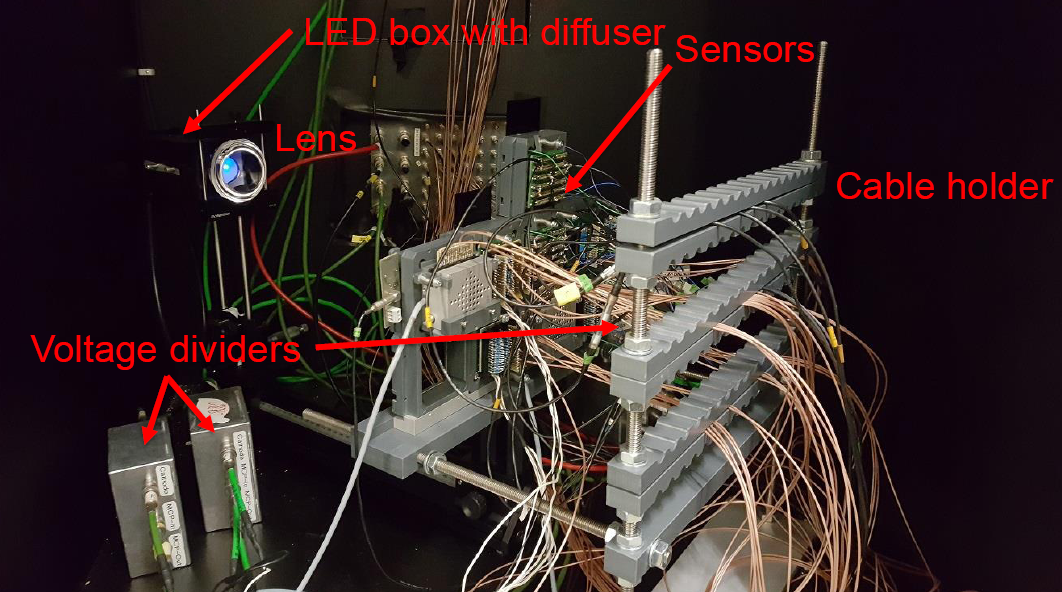}
	\end{minipage}
		\hspace{0.pc}%
		\begin{minipage}{.23\columnwidth}
			\centering
			\includegraphics[width=.98\columnwidth]{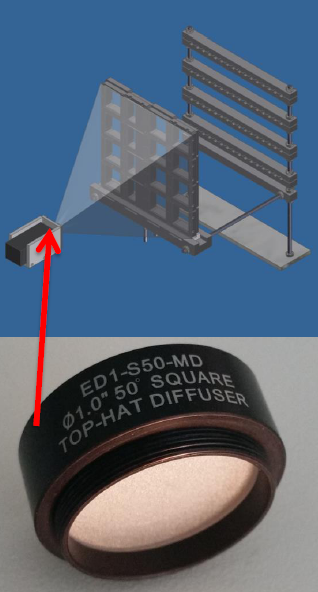}
		\end{minipage}
		\caption{Image (left) and schematic representation (right) of the illumination setup \cite{markusdiss} with the ED1-S50-MD square diffuser from Thorlabs.}
		\label{fig:lifetime_setup}
	\end{figure}

Under normal illumination conditions, the blue LED is driven at a frequency of 1 MHz and its intensity is attenuated to single photon level by neutral density filters. This is comparable to the expected boundary conditions at the PANDA DIRCs. At these illumination parameters and 10$^{6}$ MCP-PMT gain, each MCP-PMT accumulates $\sim$15 mC/cm$^{2}$ IAC per day. The stability of the LED light intensity is monitored with a photodiode near the MCP-PMTs during the entire illumination period. In addition, a VME data acquisition system permanently reads out the pulse heights of some selected anode pixels of each PMT and records them to hard disk at a highly prescaled rate ($2^{-16}$). This allows a straightforward determination of the integrated anode charge. A block diagram of the lifetime setup is shown in Fig.~\ref{fig:lifetime_block}. For some MCP-PMTs, one half of the PC is masked to allow a direct side-by-side comparison of the aging progress on the same tube. A homogeneously illuminated area of about 30 $\times$ 30 cm$^{2}$ at a distance of $\sim$50 cm from the blue LED is achieved with a square diffuser ED1-S50-MD from Thorlabs, followed by a spherical lens to further broaden the LED light. This currently allows us to illuminate up to sixteen 2-inch MCP-PMTs in parallel within the same setup (see Fig.~\ref{fig:lifetime_setup}). 

\subsection{Measurement Procedure, Analysis, and Results}

After the MCP-PMTs have been illuminated for several weeks (or months), measurements of the gain, DCR and QE are performed for each tube. First, the photon rate is reduced to a few kHz, the prescaling in the readout electronics is removed and the distribution of the height of the single photon pulses for some selected anode pixels of all MCP-PMTs in the illumination setup is recorded. This results in the gain estimate of the MCP-PMTs. The DCR is also determined for each tube after the LED has been switched off. In the next step, all MCP-PMTs are transferred to the in-house monochromator \cite{herold11}, which is described in Section~\ref{spectralQE} and where a QE wavelength scan is performed in steps of $\lambda$ = 2 nm from 200 to 850 nm. If no further QE surface scans are planned, the MCP-PMTs are reinstalled in the illumination setup and the pulse height spectra are checked for consistency. After increasing the LED rate to 1 MHz and resetting the prescaling factor, the aging process is restarted by permanently illuminating the MCP-PMTs for several more weeks.

\begin{figure}[!htbp]
	\centering
	\includegraphics[width=.98\columnwidth]{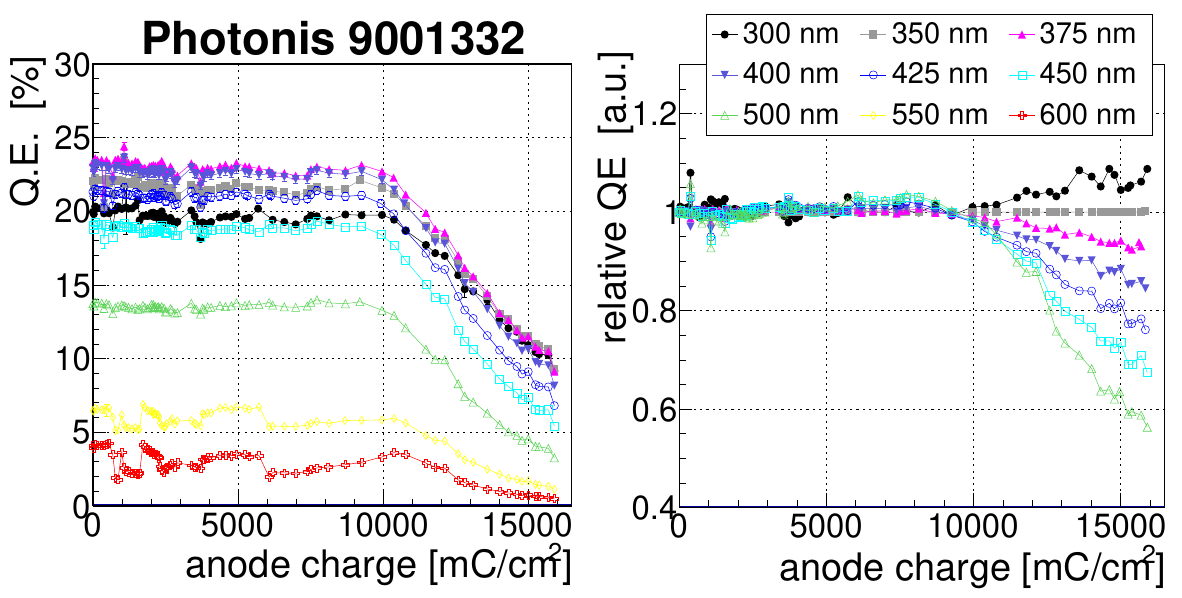}
	\caption{QE and relative QE (normalized to 350 nm) as a function of IAC for the ALD-coated 2-inch MCP-PMT PHOTONIS XP85112 (SN 9001332). The QE is stable up to an IAC of about 10 C/cm$^2$ and then decreases continuously. Obviously, the QE for red photons drops faster than for blue photons. Note that the relative QE for 550 nm and 600 nm is not plotted due to excessive fluctuations (see left graph).}
	\label{fig:lifetime_results}
\end{figure}

The charge and time distributions of the signals recorded continuously during the illumination are analyzed to determine the integrated anode charge for each MCP-PMT. The additional non-prescaled pulse height spectra are used to determine the gain and the DCR before and after each illumination period. Finally, QE, gain and DCR are plotted against the IAC, which provides the final results. In addition, measuring the entire QE spectrum between 200 and 850 nm allows an investigation of the evolution of the QE relative to a reference wavelength. As an example, the QE aging results are shown in Fig.~\ref{fig:lifetime_results} for the PHOTONIS MCP-PMT SN 9001332. The relative QE clearly shows that PC aging, once it has started, is more severe for red light than for blue light. Although it is not obvious for this MCP-PMT, we have also observed decreasing gain and DCR with progressive aging for some other tubes \cite{lehmann14}.

Every few months, a QE surface scan is performed for each MCP-PMT in small steps of 0.5 mm over the entire PC area to identify potential regions on the PC where the QE may decrease faster. The used setup is the same as that described for the spatial QE scans in Section~\ref{positionQE}. The measurements are conducted at a wavelength of 372 nm or (in rare cases) 633 nm.

\begin{figure}[!htbp]
	\centering
	\includegraphics[width=.98\columnwidth]{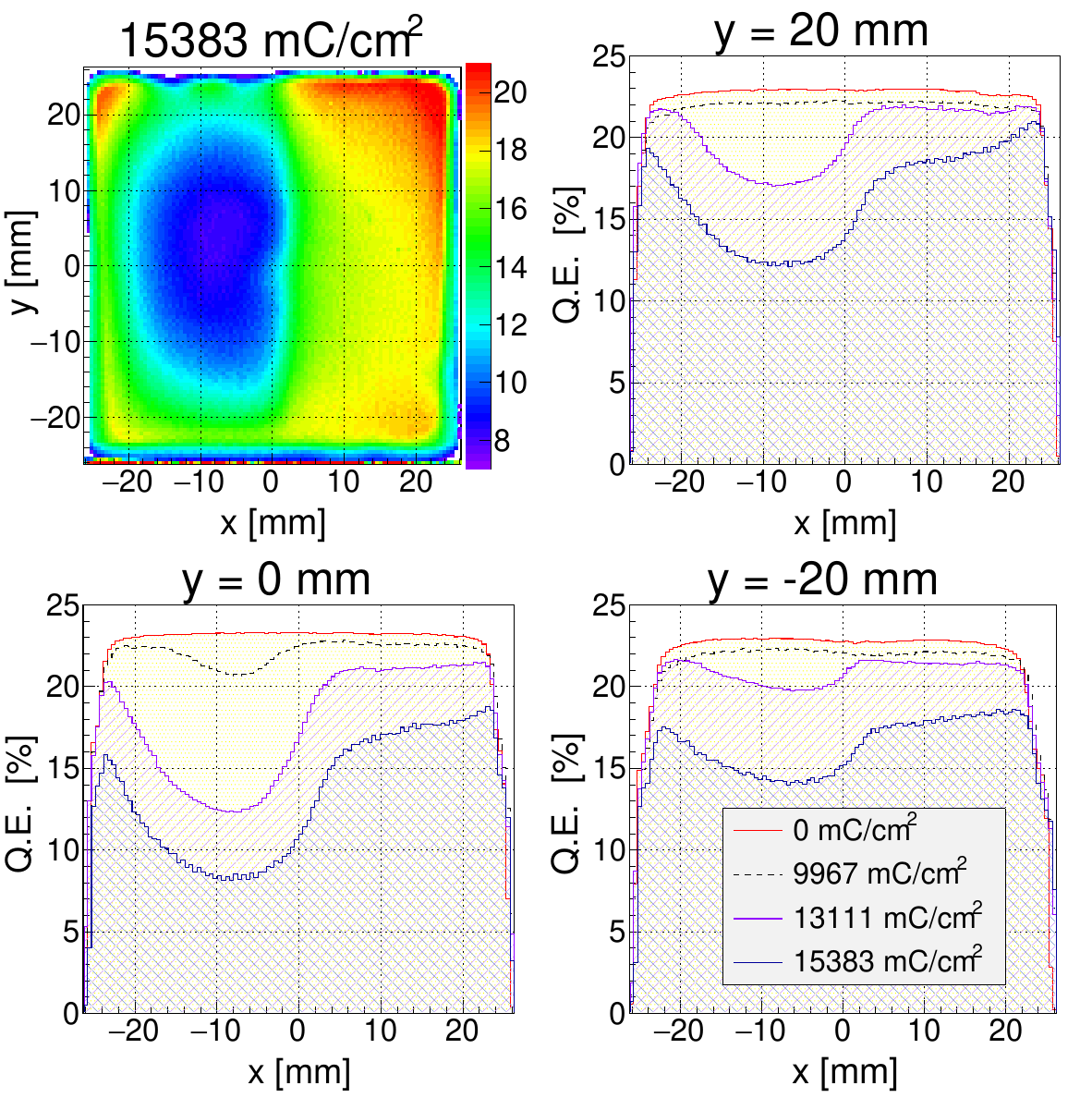}
	\caption{QE map (top left) after 15.4 C/cm$^2$ IAC for the ALD-coated MCP-PMT PHOTONIS XP85112 (SN 9001332). The right half of the PC was covered during illumination. The QE x-projections of such maps along three discrete y-positions for different IACs show the evolution of PC aging.}
	\label{fig:lifetime_scans}
\end{figure}

These QE scans provide quite interesting insights. We often observe (see Fig.~\ref{fig:lifetime_scans}) that the degradation of the QE starts at a certain PC position and grows from that point. The plots show the results of a QE scan and some x-projections at different IACs for the ALD-coated PHOTONIS SN 9001332. The right half of the PC was always covered during illumination. The QE degradation with increasing IAC is much more pronounced in the illuminated half, indicating that ion feedback is the main cause of PC aging. 

\begin{figure}[!htbp]
	\centering
	\includegraphics[width=.98\columnwidth]{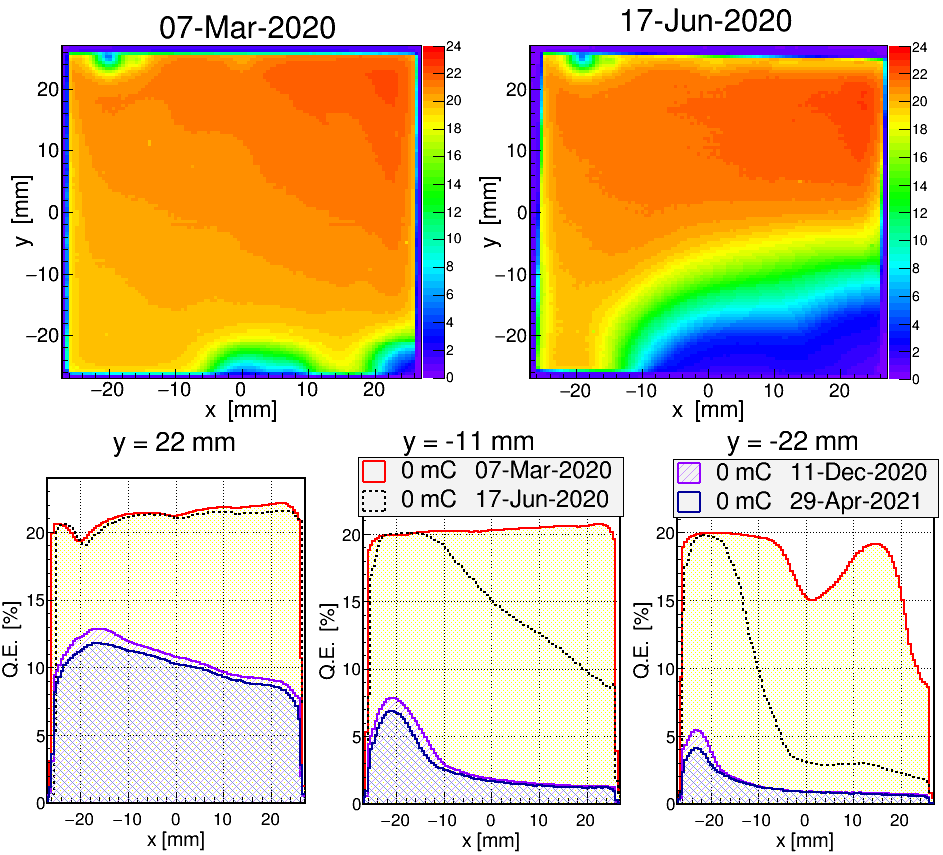}
	\caption{QE as a function of the PC position (top), recorded at different time points for the 2-inch Photek MAPMT253 MCP-PMT A1200116. The x-projections along different y-positions (bottom) show the evolution of the QE degradation over time, even when the PMT was not illuminated.}
	\label{fig:QEhole}
\end{figure}

In rare cases, significant "QE holes" were observed immediately after delivery of the tube, usually on the PC rims and/or corners, which deteriorated within weeks or months, even when the tube was not illuminated. This indicates vacuum micro-leaks in the housing, probably caused by a faulty sealing, and these tubes had to be rejected. An example is shown in Fig.~\ref{fig:QEhole} with the Photek MAPMT253 MCP-PMT (SN A1200116), where the first QE scan was made immediately upon arrival of the tube, followed by others several months later. It can be clearly seen that the QE degradation develops from these "QE holes". This shows that the rather simple QE scans are quite a powerful tool to detect tiny vacuum leaks in the MCP-PMT housing.

\section{Position Scans with PADIWA/DiRICH/TRB DAQ}  \label{sec:scans}

The conventionally assessed PMT performance parameters are QE, gain, time resolution, total DCR and sometimes the rate capability. Only rarely, the CE and lifetime of these tubes, for example, are measured. Especially in multianode MCP-PMTs, it is also very informative to investigate the less easily accessible internal parameters, such as the dark count rate per anode pixel, the spatial and temporal distribution of afterpulses triggered by feedback ions, the contribution of electrons bouncing back from the MCP input and the crosstalk (charge-sharing and electronic) between all anode channels. A more detailed explanation of the latter parameters can be found in the subsections~\ref{sec:perfDC} to \ref{sec:perfCrosstalk}. After setting up a sophisticated scanning device in combination with a suitable data acquisition system, we were able to investigate these parameters in comprehensive detail.

\subsection{Setup}
\label{sec:perfsetup}

\begin{figure}[!htbp]
	\begin{minipage}{.35\columnwidth}
		\centering
		\includegraphics[width=.98\columnwidth]{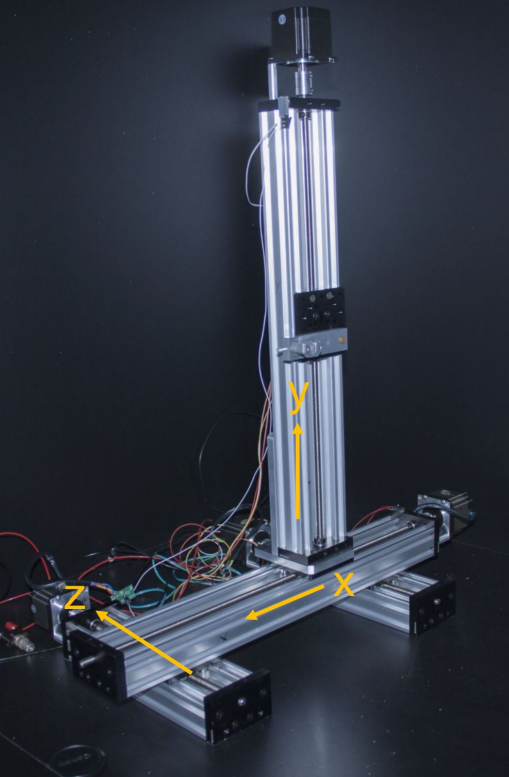}
	\end{minipage}
	\hspace{0.pc}%
	\begin{minipage}{.65\columnwidth}
		\centering
		\includegraphics[width=.98\columnwidth]{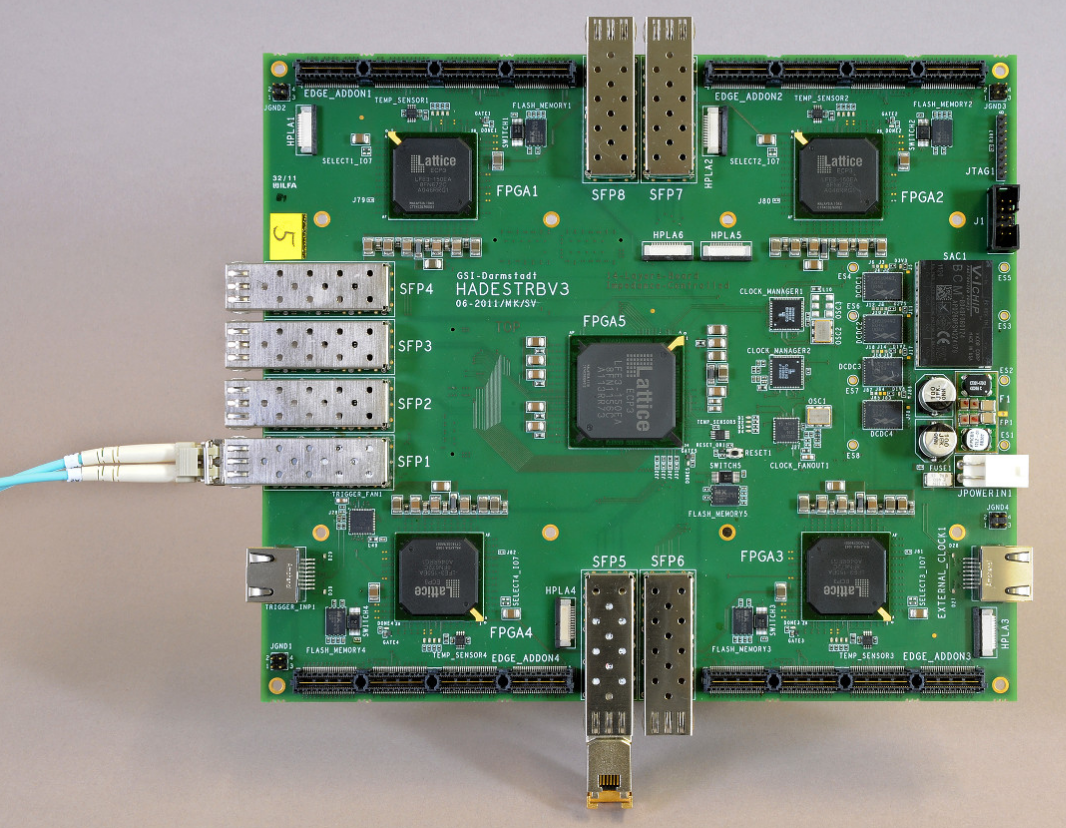}
	\end{minipage}
	\begin{minipage}{.99\columnwidth}
	\centering
	\includegraphics[width=.98\columnwidth]{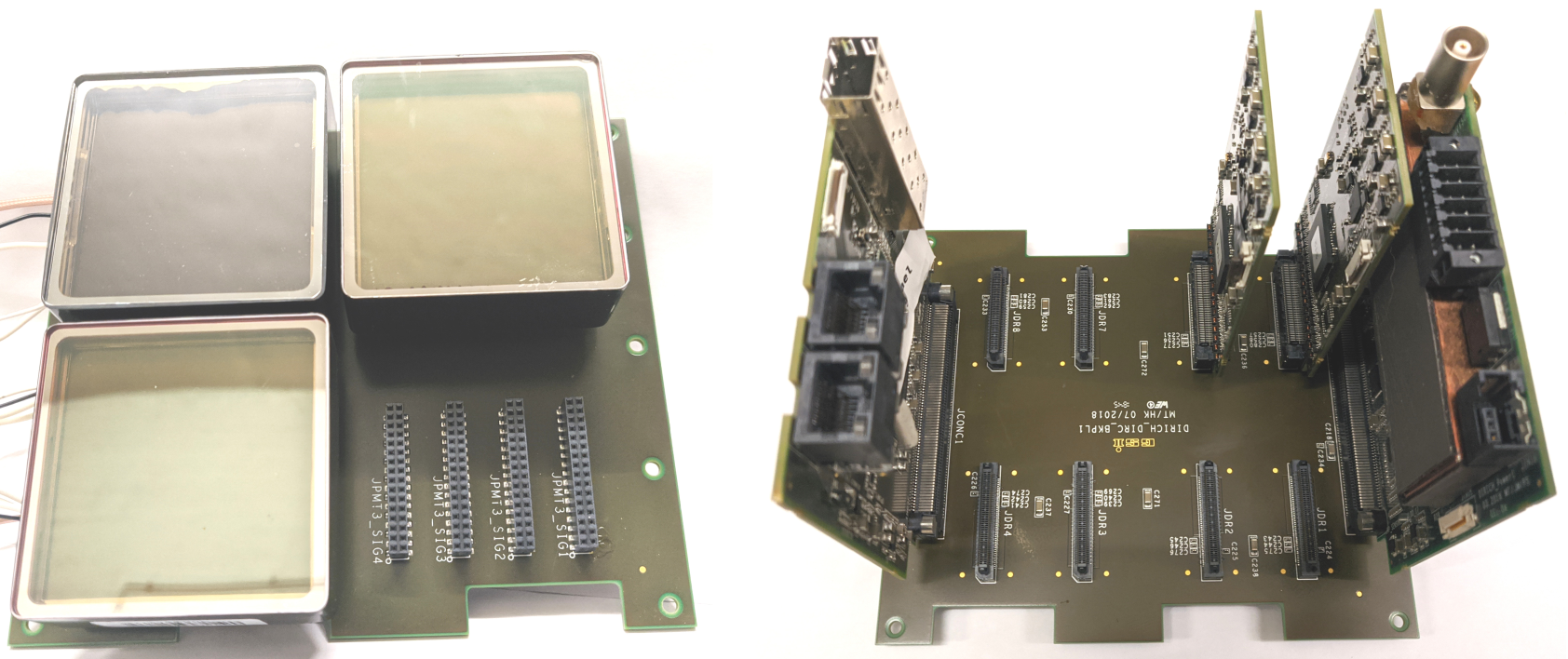}
	\end{minipage}
	\caption{Important parts of the setup for the position scans: 3-axis stepper (top left) and TRB3 board \cite{trb3} (top right). The backplane of the DiRICH system shows the front with 3 MCP-PMTs attached (bottom left) and the rear with (from left to right) data concentrator, two DiRICH3 boards and power module (bottom right) \cite{merlindiss}.}
	\label{fig:stepper_setup}
\end{figure}

The MCP-PMTs are surveyed in large light-tight wooden boxes whose outer walls are covered with thin copper sheets to shield the interior from electromagnetic radiation. This is essential for clean anode signals from the PMTs, but also because the investigation of some performance parameters requires the measurement of currents in the sub-nanoampere range. A self-built 3-axis stepper (Fig.~\ref{fig:stepper_setup}, top left) is installed in one of the dark boxes, which can be moved $\sim$40 cm in the x and y directions and $\sim$15 cm in the z direction. The large xy range allows us to measure four (and more) 2-inch MCP-PMTs in succession without having to change the setup. The stepper consists of 4 "C-Beam\textsuperscript{\textregistered}" linear actuators made of extruded aluminum, each equipped with a motor controlled by a "TMC2130 Trinamic" stepper motor driver. The advantage of this driver is that it is equipped with sensorless motor load detection, which recognizes when the stepper motor hits a limit or other obstacle. The driver is controlled by a "Teensy 3.5" microcontroller, which is programmed in an "Arduino" development environment \cite{arduino}. A stepper motor has 200 steps/revolution, while the lead screw used has a feed rate of 8 mm/revolution. By additionally activating a 16-fold microstep option, a theoretical distance of 2.5 \textmu m/step can be achieved. When approaching a position always from the same side, a real positioning accuracy of $\le$10 \textmu m was measured with a CMOS camera. This accuracy is sufficient to deliver high-quality scans in our applications.

A microfocus lens system is attached to the 3-axis stepper, which is connected to a PiLas laser head outside of the dark box by a single-mode optical fiber. The photocathode of an MCP-PMT is illuminated with ultrafast ($\sigma_{t} \lesssim$ 15 ps) blue or red laser pulses, and the light intensity is reduced to single photon level by neutral density filters. By adjusting the lens position along the z-axis of the stepper, the light spot can be focused to $\lesssim$10 \textmu{m} FWHM peak width at the PC. In practice, a slightly worse focus is used to avoid gain saturation of the MCP-PMTs at the typically applied photon rates of 5 - 50 kHz.

The signal of each anode pad of a multianode MCP-PMT is read out directly and fed into FPGA-based PADIWA \cite{ugur,trb} or DiRICH \cite{michel2017} frontend boards with 16 or 32 channels, respectively, which are located inside the dark box. The signals from the frontend boards are forwarded to a TRB3 (Trigger and Readout Board version 3) \cite{trb,trb3} board, as shown in Fig.~\ref{fig:stepper_setup} (top right), outside the dark box, which takes over the further data management. In our applications, a PADIWA/TRB3 setup is used for reading out MCP-PMTs with a high anode segmentation (e.g., 100$\times$3 pixels) or in spatially limited environments such as measurements inside the Julich dipole magnet, while the standard quality test scans for the 8$\times$8 anode pixel PMTs are performed with a DiRICH/TRB3 setup.

Depending on the setup, the PADIWA boards can either be connected directly to the backplane of the MCP-PMTs or via Samtec HLCD high-speed coaxial cables. Each PMT signal is 7$\times$ amplified and processed by an FPGA-based threshold discriminator, which switches an LVDS output as long as the signal is above the threshold. The LVDS signal of each channel is routed to a TRB3 board via simple ribbon cables. In addition to a central FPGA chip, each TRB3 board is equipped with four peripheral FPGAs which, in combination with the PADIWAs, can act as TDC to provide time and time-over-threshold (ToT) signals. In a single central FPGA, the trigger signal is generated with a Central Trigger System (CTS) and then distributed to the central FPGAs of any additional TRB3 boards. The clock signal is either provided by an external source or each TRB3 board generates its own clock, with the latter usually being applied. A TRB3 board has 256 usable data channels if no ToT information is required, and 4$\times$48 channels otherwise. It can process a maximum data readout trigger rate of up to 700 kHz. In principle, by combining many PADIWAs with enough TRB3 boards, it would be possible to read out an "infinite" number of PMT channels.

In order to significantly reduce the cable clutter caused by the use of PADIWAs, DiRICH frontend boards were developed that can be connected directly to the PMTs via intermediate adapter backplanes (Fig.~\ref{fig:stepper_setup}, bottom). One backplane can accommodate up to 4 MCP-PMTs with 64 channels each. A DiRICH channel is equipped with a $\sim$18 - 28$\times$ (dependent on the pulse height) amplifier, a discriminator and a multihit TDC and provides coarse pulse height information through a ToT technique, all of which is implemented in an FPGA. The intermediate backplane is additionally equipped with a data concentrator and a power supply board. The data concentrator is connected to a TRB3 board outside the dark box via a fiber optic cable for the data and one or two Ethernet cables for the distribution of the trigger (and possibly the clock) signal to the DiRICH boards. The latter signals are generated in the CTS of the TRB3. The power module is connected to a 32 V power supply outside the dark box.

\begin{figure}[!htbp]
	\centering
	\includegraphics[width=.98\columnwidth]{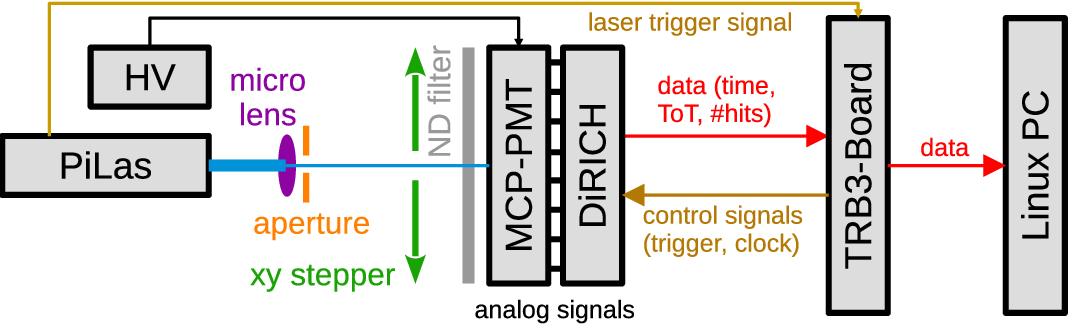}
	\caption{Block diagram of the setup for the measurements and xy-scans with the DiRICH/TRB3 DAQ system.}
	\label{fig:trbscan}
\end{figure}

A block diagram of the basic components of the scanning setup is shown in Fig.~\ref{fig:trbscan}. The DiRICH/TRB3 DAQ system has a flexible design and offers a timing accuracy of up to $\sim$10 ps per single TDC channel. An internal ring buffer for each channel can (theoretically) store the timing information of up to 32 hits per trigger and handle burst hit rates of $\sim$30 MHz. During xy-scans the DiRICH/TRB3 system permanently analyzes the data stream provided by all anode pixels of the MCP-PMT. The lead time (= timing of the pulse leading edge) and the ToT information of each hit and TDC channel (corresponding to an anode pad) are continuously queued in the ring buffer and can be read out and stored within an adjustable time window (e.g. -10 to +10 \textmu{s}) around a certain trigger time. In our applications, this is usually the trigger pulse of a PiLas laser, which for practical reasons is artificially shifted to t$_{0}$ = 100 ns for each channel in the analysis. 


\subsection{Data Taking and Analysis}
\label{sec:trbanalysis}

\begin{figure}[!htbpp]
	\centering
	\includegraphics[width=.98\columnwidth]{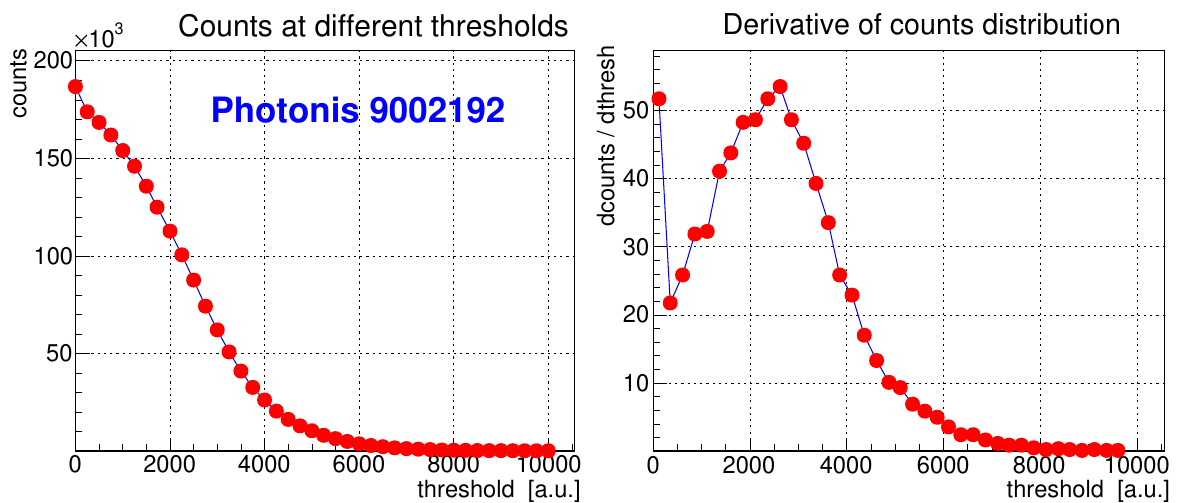}
	\caption{PHOTONIS SN 9002192 MCP-PMT. Left: measured counts for a fixed anode channel at different thresholds set in the DiRICH board. A threshold of 1000 digits corresponds to $\sim$3.5 mV pulse height. Right: derivative of this curve reflecting the pulse height distribution; the noise is clearly separated from the single photon peak \cite{merlindiss}.}
	\label{fig:threshscan}
\end{figure}

Before any accurate and rigorous measurement is possible with the DiRICH/TRB3 data acquisition system, the fine time delays of the individual TDC channels must be calibrated. Unfortunately, this calibration is temperature sensitive and needs to be repeated occasionally, especially if the ambient temperature has changed. Otherwise, the time resolution and linearity of the DAQ system can deteriorate considerably. Calibration is performed by triggering each channel with internal random time pulses generated by the CTS. This creates a statistical sample from which the fine time delay is calculated from longer delays and used to calibrate the sampling distances of the time axis. The next step is to set the thresholds for the incoming analog signals on each DiRICH channel. With the PMTs turned off, the offset (or threshold) voltages of the DiRICH channels are varied close to the edge of the noise band until a moderate count rate is measured for each channel. An interesting option in this context is the possibility to measure the pulse height distribution for each anode pixel by simply varying the signal threshold. Counting the hits per pixel for different thresholds gives the distribution shown in Fig.~\ref{fig:threshscan} (left). The derivative of this curve provides the analog pulse height distribution measured at each anode pixel (Fig.~\ref{fig:threshscan}, right). This allows us to examine the average pulse heights arriving at each DiRICH channel. In most of our scans, we apply a common threshold for all channels. In standard measurements, this is typically set to 10 to 25\% of the pulse height peak of the single photons in a central anode pixel.

The entire stepper/DAQ setup allows xy-scanning over the active surface of the MCP-PMT in steps down to $\sim$20 \textmu{m}. The laser beam is always directed exactly perpendicular to the PC of the PMT in order to suppress internal reflections in the optical entrance window. Any effects of this kind are not taken into account in the analyses. At each xy position, all data collected for a certain number (usually between 10k and 100k) of laser triggers are read out. The data of each measured position is first stored in files with a special format (.hld) and in a further step converted into root files \cite{root} for the final analysis. In combination with the DiRICH/TRB3 or the PADIWA/TRB3 DAQ system \cite{michel2017,trb,ugur}, we can read up to 256 anode pixels simultaneously with the DiRICH and more than 300 channels (192 per TRB3) with the PADIWA boards. In practice, the laser positioning along x and y with the stepper, data acquisition for a given number of triggers, file conversion and data analysis at each measured xy point is automated and can be controlled with appropriate Python scripts. The data obtained allows a view “inside the PMT” \cite{AL} in 3D (x,y,t) and many performance parameters can be measured as a function of the incident photon position. By performing an xy-scan (for the 8$\times$8 anode MCP-PMTs in steps of 0.5 or 1.0 mm) over the active surface of the MCP-PMT, the information x- and y-position of the laser spot and the lead time, ToT and channel number of each hit within a given trigger time window for each anode pad (channel) and each laser pulse is stored to disk. The spatial dark count distributions, the afterpulse distributions as a function of the xy position and the time after the trigger pulse (quasi the TOF of the feedback ions) and the spatial and temporal recoil electron distributions can be derived from this raw data. By counting the hits in narrow and wide time windows, the charge-sharing and electronic crosstalk behavior can be analyzed. These TRB scans make it possible to separate crosstalk events from those of recoil electrons and afterpulse hits.

\begin{figure}[!htbp]
	\centering
	\includegraphics[width=.98\columnwidth]{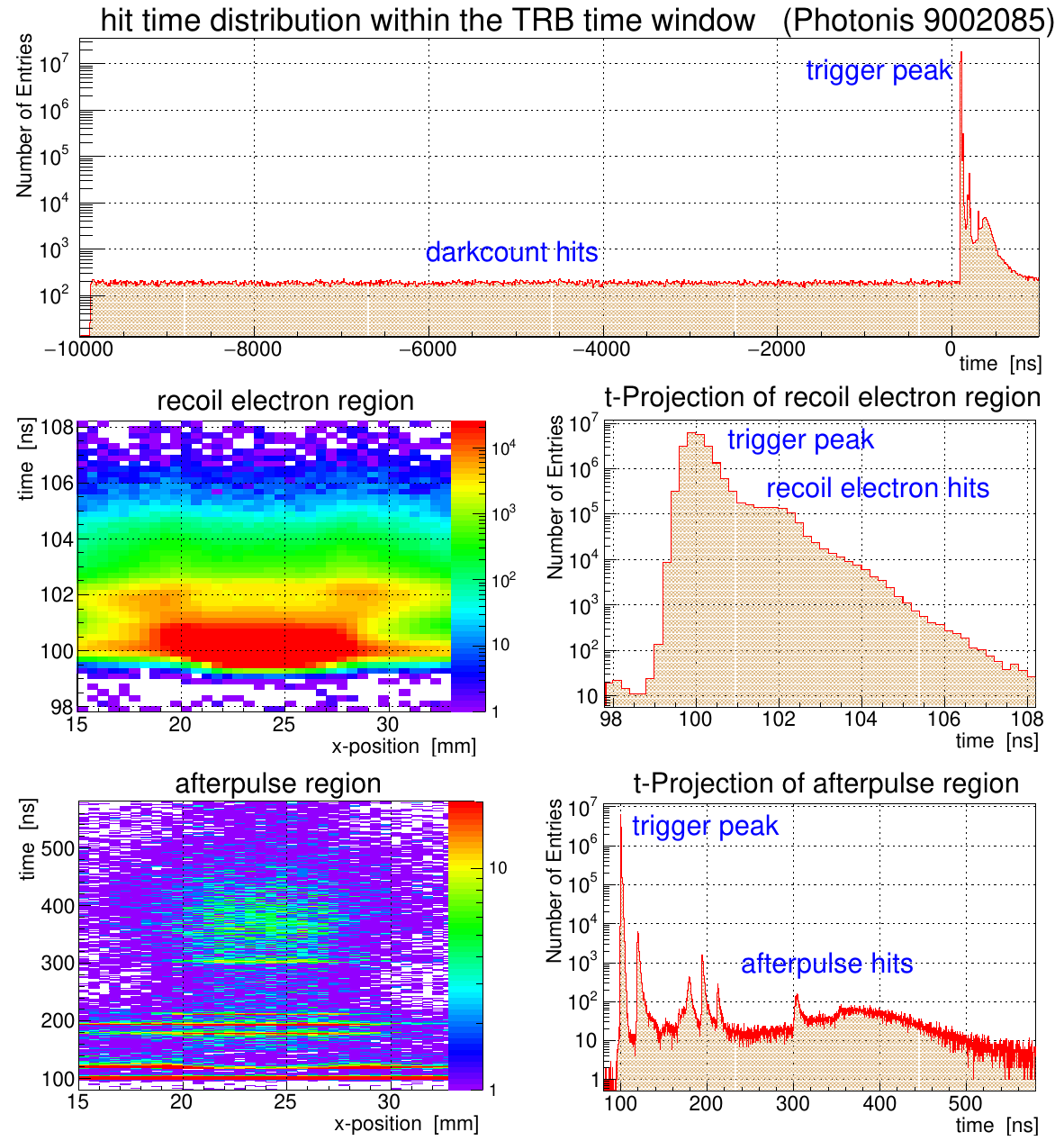}
	\caption{Information obtainable by PADIWA/TRB3 scans for the PHOTONIS XP85012 (SN 9002085) HiQE MCP-PMT \cite{AL}. Top: time distribution of all hits within a time window of [-10,1] \textmu{s} for any anode pixel x4-y6. This time interval contains all hits of interest, e.g. the dark count region at [-10,0] \textmu{s}. The trigger peak at 100 ns corresponds to photons of the laser pulse. Center: same time distribution, but zoomed to the time window [98,108] ns, which contains the trigger peak and the recoil electron hits just above 100 ns. Bottom: time distribution zoomed to the time range [80,580] ns with afterpulse hits that actually start at t$_{0}$ $\gtrsim$ 105 ns (see also Figs.~\ref{fig:recoilE} and \ref{fig:afterpulse}).}
	\label{fig:trbinfo}
\end{figure}

Figure~\ref{fig:trbinfo} (top) shows the lead time distribution of the exemplary anode pad x4-y6 within a [-10,1] \textmu{s} window for a HiQE PHOTONIS XP85012 (SN 9002085) MCP-PMT summed over all xy illumination points. A laser rate of 10 kHz was used, which corresponds to a delay of 100 \textmu s between two laser pulses. Two distinct regions can be identified: the long flat distribution before the trigger peak at 100 ns contains pure dark count hits, while clearer discrete structures appear later. The DCR is determined by integrating the number of hits between -10 and 0 \textmu{s}. Two enlarged time window regions are shown in the other plots. In these distributions, the number of hits measured in the anode pad x4-y6 is plotted for laser x-positions between 15 and 34 mm, which also cover the left and right adjacent anode pads. In the middle plots of Fig.~\ref{fig:trbinfo}, the time window of [98,108] ns is zoomed very close around the main trigger peak, which contains recoil electrons and charge-sharing events. At the bottom plots, the time window is zoomed to [80,580] ns, showing that the afterpulse hits arrive much later than the main trigger pulse at 100 ns.

\begin{figure}[!htbp]
	\centering
	\includegraphics[width=.98\columnwidth]{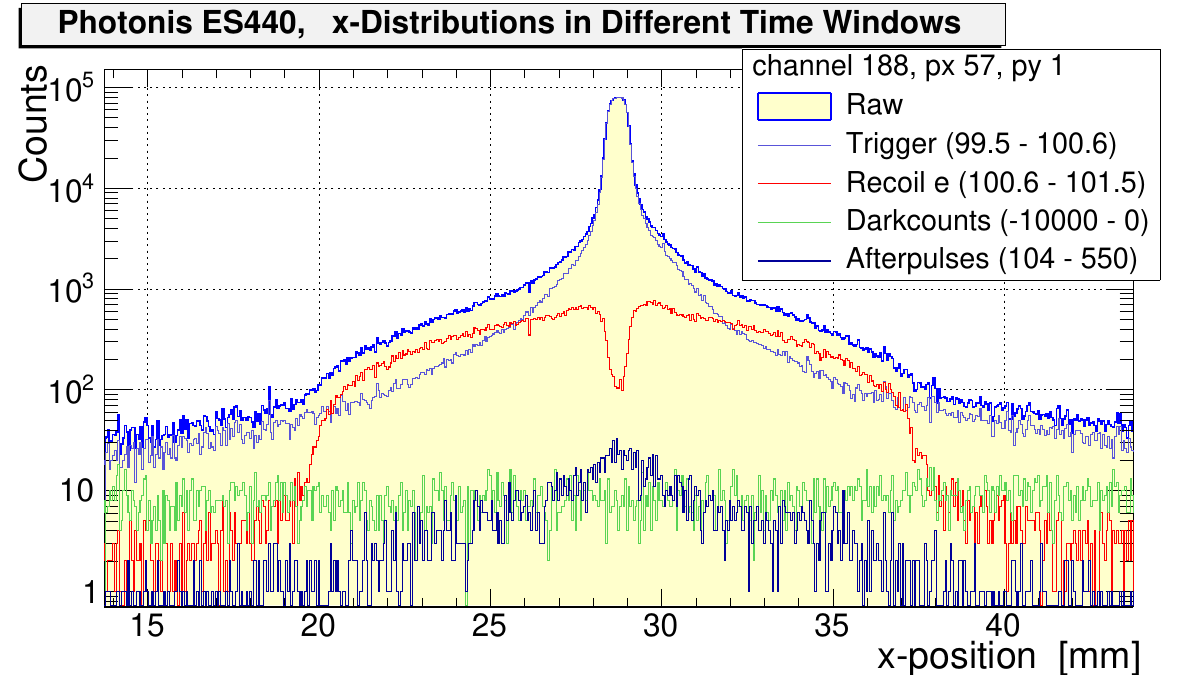}
	\caption{Spatial distributions (dark count, trigger, recoil, afterpulse) along x for different time windows for the PHOTONIS ES440 MCP-PMT with 100$\times$3 anode pixels. The read-out pixel was x57-y1 while the laser spot was moved along x in 25 \textmu m steps.}
	\label{fig:xdistr}
\end{figure}

The DAQ system also gives us access to the spatial distribution of different hit classes defined by specific time windows around the trigger peak. In Fig.~\ref{fig:xdistr} such distributions are shown for a high-resolution PHOTONIS MCP-PMT (ES440) with a 100$\times$3 anode grid. Shown is the spatial distribution of four classes of hits: (1) the dark count hits between [-10, 0] \textmu s show a flat distribution in space;  (2) around the main trigger time ([99.5, 100.6] ns) we obtain, as expected, a narrow peak at the pixel position of direct photoelectrons; (3) in the recoil electron time region ([100.6, 101.5] ns) a well-defined hump about 15 mm wide appears with a central hole due to dead time; (4) events in the afterpulse time region ([104, 500] ns) show a broad scatter with a maximum at the position of the read-out pixel. Please note that in this plot the anode pixel x57-y1 was read out and data for many x-positions from 14 to 44 mm were recorded in steps of 25 \textmu m.

\subsection{Relative Detection Efficiency}

\begin{figure}[!htbp]
	\centering
	\includegraphics[width=.99\columnwidth]{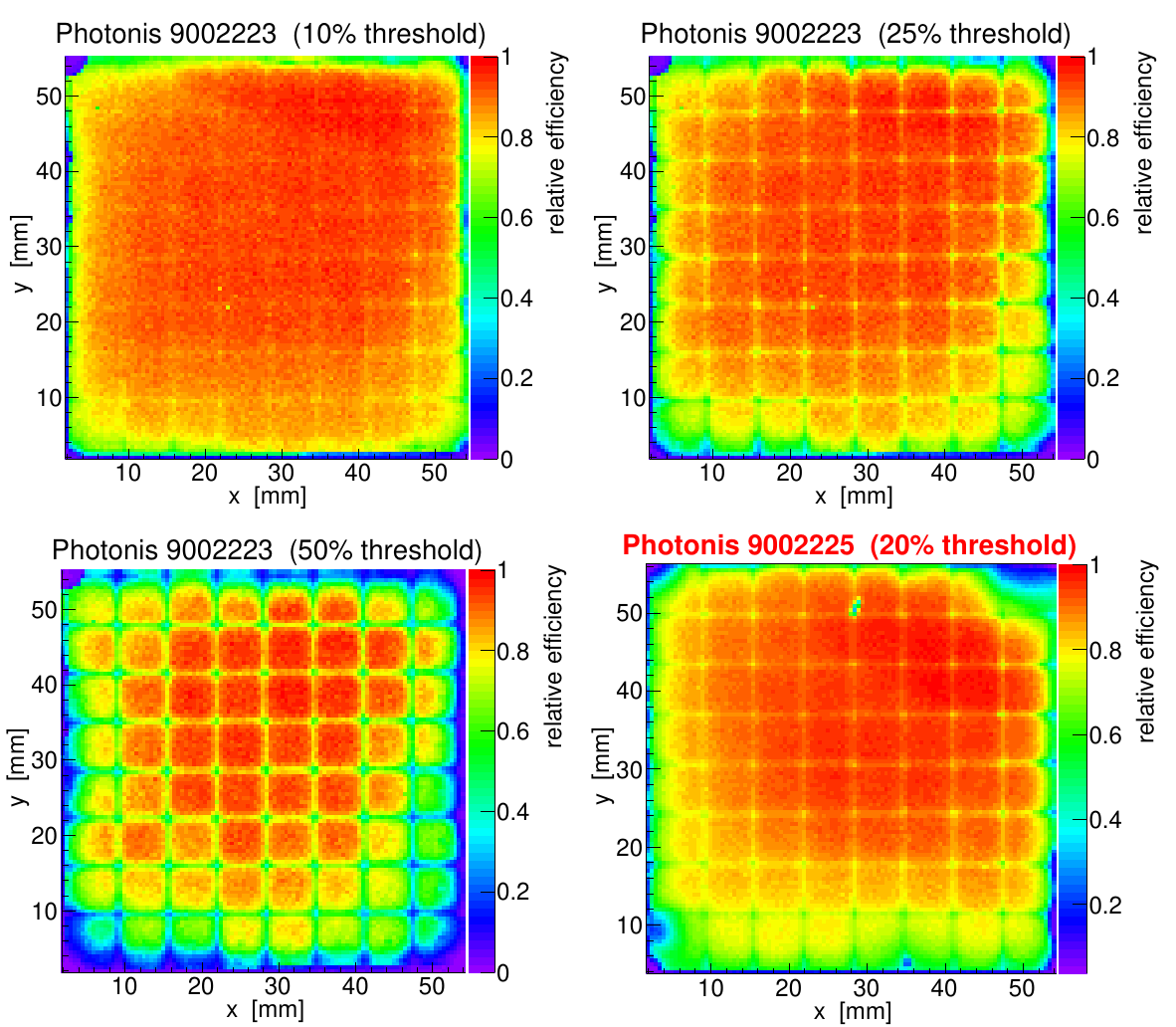}
	\caption{Relative efficiency as a function of position and signal threshold for two PHOTONIS XP85112 MCP-PMTs: SN 9002223 (top left, top right, bottom left) at different single photon peak height thresholds and SN 9002225 (bottom right) with QE holes in the corners.}
	\label{fig:releff}
\end{figure}

The detection efficiency over the entire active area of an MCP-PMT depends on the QE (and CE), the gain and the threshold applied to the anode signals. To estimate the efficiency that would be obtained in a real experimental setup, we compare the number of expected hits with the measured ones. For each xy illumination point, the number of measured signals within a narrow time window [99.5,100.5] ns around the main peak at 100 ns in the time distribution is divided by the number of expected laser triggers. The ratio obtained for each position is normalized to the highest value within the entire active area, which provides the spatial "relative efficiency plots" as shown in Fig.~\ref{fig:releff}. It can be seen that the relative efficiency is lower towards the PMT rims and between the anode pixels because (1) the gain is generally smaller at the rims of the tube, probably due to distorted electric fields, and (2) the signal between the anodes is lower due to the sharing of charges. Typically, the same absolute threshold is set to each channel of a tube, and the relative efficiency is highest somewhere in the center of the PMT. At higher threshold values, the efficiency between the pixels and at the rims decreases more. The net effect of different thresholds (10\%, 25\%, and 50\% of the mean single photon pulse height at a central pixel) is shown for the PHOTONIS MCP-PMT SN 9002223 in the first three plots of Fig.~\ref{fig:releff}. The relative efficiency map of a PMT (SN 9002225) with a low QE in some areas is also shown in Fig.~\ref{fig:releff} (bottom right). A low QE naturally leads to a significantly reduced detection efficiency (see top right and bottom left corners), as fewer photoelectrons per laser trigger exit the PC in this area.

\subsection{Leakage Current and Dark Count Rate}
\label{sec:perfDC}

In addition to dark count events (DCR) caused by thermally emitted electrons at the photocathode, we usually measure a typical DC leakage current of a few nA even when the MCP-PMT is not illuminated. The leakage current comes from a non-infinite resistance between the PC and the MCP-In electrode when a voltage is applied between them. Ideally, the leakage current is in the sub-nA range, because if it is $\gg$10 nA, this current can become problematic for measuring some parameters such as CE and spectral and spatial QE, where a very small photon-induced current at the MCP-In electrode needs to be accurately determined. Fortunately, the leakage current is not a problem under normal experimental conditions as it does not reach the anode and therefore does not affect the pulse detection of the MCP-PMT.

\begin{figure}[!htbpp]
	\centering
	\includegraphics[width=.99\columnwidth]{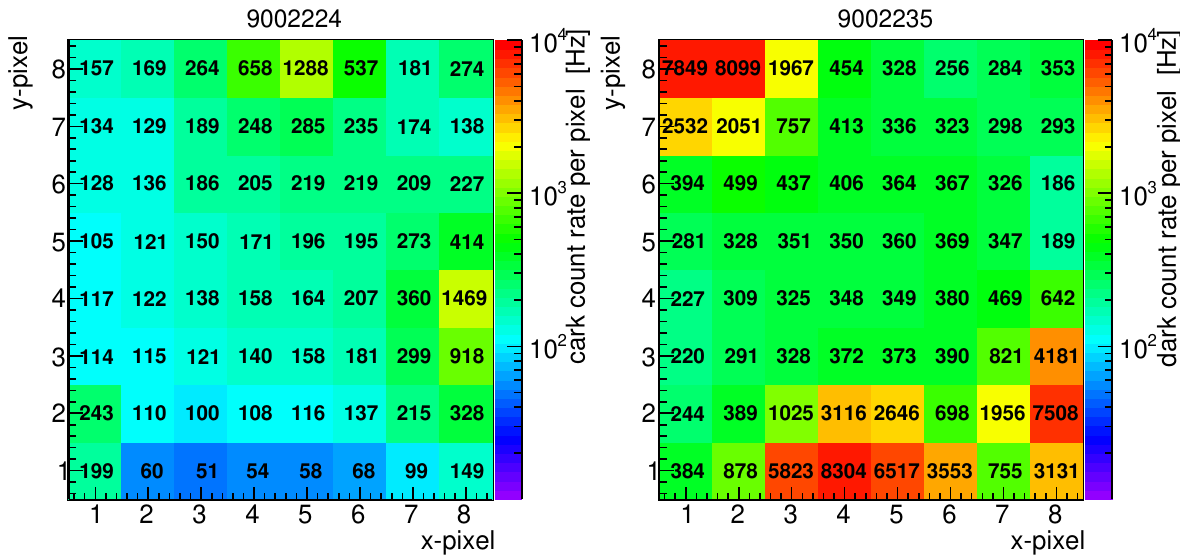}
	\caption{DCR maps per anode pixel for the 2-inch MCP-PMTs PHOTONIS XP85112-S-BA SN 9002224 (left) and SN 9002235 (right). The mean PMT gain was set to $\sim$10$^{6}$, with a threshold well below the typical single photon pulse height. The color-coded scale (z-axis) corresponds to the DCR/pixel in Hz.}
	\label{fig:DCR}
\end{figure}

The thermally induced DCR decreases after the PMT has been placed in the dark. For the series production PHOTONIS MCP-PMTs, the stabilization time is typically 30 to 60 minutes. For this reason, we wait about 1 hour after closing the dark box before starting any measurement. After this period the DCR has usually stabilized at $\lesssim$1 kHz/cm$^{2}$. A by-product of the xy-scans described above is a dark count map for all anode pixels of each investigated MCP-PMT. The DCR for a pixel is derived from the integrated number of hits in the time window from -10 to 0 \textmu{s} before the actual trigger peak, scaled by 10$^{5}$ and finally divided by the laser rate at each laser position to obtain the DCR in Hz. At a typical laser trigger rate of $<$10 kHz, this time window is free of other background hits such as afterpulses and recoil electrons. As expected and shown in Fig.~\ref{fig:trbinfo} (top) and Fig.~\ref{fig:xdistr}, the DCR distribution is uniform in time and space. Fig.~\ref{fig:DCR} shows the measured DCR/pixel for two recent MCP-PMTs from PHOTONIS with 8$\times$8 anodes. It varies between $<$100 Hz and $\sim$10 kHz, and we observe that pixels with high DCR are usually found at the rims or corners and not in the center of the anode region. This is the case for most of the scanned MCP-PMTs; a possible explanation could be a distorted electric field due to residues of the sealing material at the PMT rims. With a mean gain of $\sim$10$^6$ and a signal threshold of $\sim$20\% of the single photon pulse height at a central anode pixel, the typical DCR, averaged over the entire active area, is between 100 Hz/cm$^2$ and several kHz/cm$^2$ for recently measured PHOTONIS MCP-PMTs.

\subsection{Time Resolution}

\subsubsection{Time-over-Threshold (ToT)}

\begin{figure}[!htbpp]
	\centering
	\includegraphics[width=.99\columnwidth]{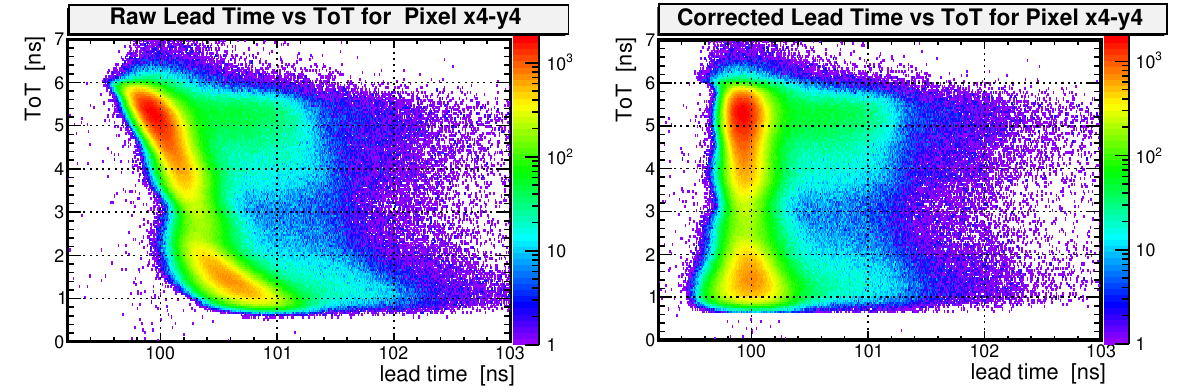}
	\caption{Typical raw (left) and timewalk-corrected (right) distributions of lead time versus ToT in the region of the main laser peak for signals from a recent PHOTONIS SN 9002223 MCP-PMT. The two visible structures correspond to two classes of hits: The upper bump originates from standard unipolar pulses, the lower bump from bipolar signals, which are mainly caused by hits in close proximity to the edge of the read-out pixel. The latter seems to be a kind of electronic crosstalk between neighboring anode pixels, but the origin is not fully understood. On the other hand, real charge-sharing hits are found at the lower edge of the upper bump, which is reasonable because of their lower pulse heights.}
	\label{fig:tot}
\end{figure}

As the time stamp for each analog signal is determined using a leading edge discrimination technique in the DiRICH boards, the measured lead time depends on the pulse height of the analog signal and has to be corrected for timewalk. Rough pulse height information is available with the Time-over-Threshold information, which is determined for each signal. The ToT is defined as the time interval in which a signal is above a fixed threshold value. This information is used in the analysis to apply a coarse timewalk correction and derive the true time resolution for an anode pixel. A typical ToT distribution before and after the timewalk correction is shown in Fig.~\ref{fig:tot}. Technically, for each bin along the "ToT" direction, the lead time distribution is fitted by a Gaussian distribution, which provides the correction for the "lead time". The corrected time peak is artificially shifted to 100 ns.

\subsubsection{TTS and RMS Time Resolution}

\begin{figure}[!htbp]
	\centering
	\includegraphics[width=.99\columnwidth]{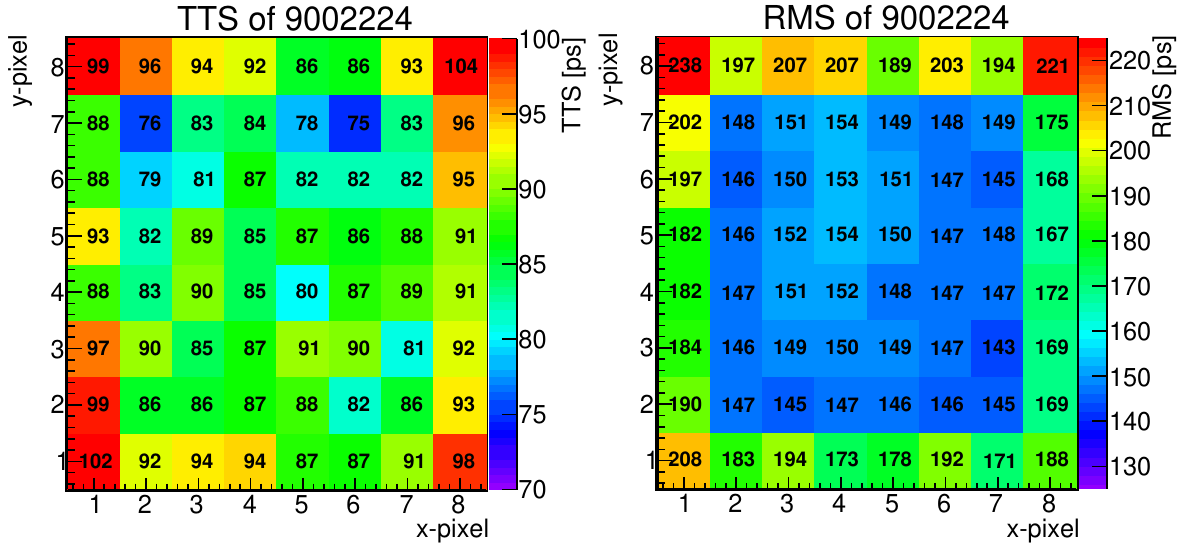}
	\caption{$\sigma_{TTS}$ (left) and $\sigma_{RMS}$ (right) time resolutions for each pixel of the PHOTONIS SN 9002224. The applied voltage divider ratio was $R_{PC}$:$R_{MCP}$:$R_A$ = 4:10:1. With a divider ratio of 1:10:1, $\sigma_{RMS}$ would be about 100 ps worse, while $\sigma_{TTS}$ would remain essentially unchanged.}
	\label{fig:timeres2d}
\end{figure}

It is known that in most MCP-PMTs the transit time spread $\sigma_{TTS}$ is well below 50 ps (see also Section~\ref{sec:timeres}). However, the more important parameter for real experiments is the RMS time resolution $\sigma_{RMS}$ within a certain time window around the TTS peak. For PANDA, $\sigma_{RMS}$ in a time window from -0.5 to 2.0 ns around the TTS peak was chosen as a performance criterion. The RMS time resolution depends on both the internal structure of the PMT and the applied voltage divider ratio. Fig.~\ref{fig:timeres2d} shows pixel maps with the measured time resolutions $\sigma_{TTS}$ and $\sigma_{RMS}$ for one of the latest PHOTONIS XP85112-S-BA (SN 9002224) MCP-PMTs. The values shown are the time resolutions at the corresponding anode pixel, convolved with the resolution of the related DiRICH/TRB3 electronic channel. The electron recoil bump after the TTS peak (Section~\ref{sec:timeres}) is mainly responsible for the significantly poorer RMS time resolution. These plots also clearly show that the time resolution is worse for pixels at the rim of the active area. The given time resolutions are the mean of the entire anode pixel area. $\sigma_{TTS}$ is much better (50 - 60 ps) if only a local position (e.g., one laser spot) in the center of the pixel is analyzed.


\subsection{Recoil Electrons}
\label{sec:perfErecoil}

\begin{figure}[H]
		\begin{minipage}{.42\columnwidth}
			\centering
			\begin{tikzpicture}[scale=0.85]
				\hspace{-0.3pc}
				
	            \fill[ultra thick,cyan] (1,2) rectangle (5.12,2.08)node[at start, above, black]{PC};
	\draw[blue,very thick](2.8,2) parabola (3.08,0.4); 
	\node[blue] at (2.5,1.7) {$e^-$};
	\fill[red](4.26,0) -- (4.4,0) -- (4.46,0.4) -- (4.32,0.4) -- cycle;
	
	\fill[red](4.06,-0.2) -- (4.2,-0.2) -- (4.26,-0.6) -- (4.12,-0.6) -- cycle;
	\fill[red](4.26,-0.2) -- (4.4,-0.2) -- (4.46,-0.6) -- (4.32,-0.6) -- cycle;
	\fill[red](4.46,-0.2) -- (4.6,-0.2) -- (4.66,-0.6) -- (4.52,-0.6) -- cycle;
	
	\foreach \i in {1,1.2,1.4,...,5}{
		\filldraw[thick,fill=black!30!white] (\i,0) -- (\i+0.06,0) -- (\i+0.12,0.4) -- (\i+0.06,0.4) -- cycle;
	}
	\draw[thick](1,0) rectangle (5.12,0.4);
	\foreach \i in {1,1.2,1.4,...,5}{
		\filldraw[thick,fill=black!30!white] (\i+0.06,-0.6) -- (\i+0.12,-0.6) -- (\i+0.06,-0.2) -- (\i,-0.2) -- cycle;
	}
	\draw[thick](1,-0.6) rectangle (5.12,-0.2);
	
	\draw[red,very thick](3.08,0.4) arc(140:40:0.86cm);
	\node[red] at (4.1,1.1){Recoil $e^-$}; 
	
	\foreach \l in {4.13,4.21,...,4.54}{
		\draw[red](4.33,0) parabola (\l,-0.2);
	}
	
	\foreach \k in {3.9,4.02,...,4.5}{
		\draw[red](4.2,-0.6) parabola (\k,-1.2);
	}
	\foreach \k in {4.1,4.22,...,4.7}{
		\draw[red](4.4,-0.6) parabola (\k,-1.2);
	}
	\foreach \k in {4.3,4.42,...,4.9}{
		\draw[red](4.6,-0.6) parabola (\k,-1.2);
	}

	\draw[very thick](1,-1.2) -- (2.24,-1.2);
	\draw[very thick](2.44,-1.2) -- (3.68,-1.2);
	\draw[very thick](3.88,-1.2) -- (5.12,-1.2); 
	
	\foreach \j in {1.62,3.04,4.5}{
		\draw[very thick](\j,-1.2) -- (\j,-1.4);						
	}
	\draw[very thick, ->](5.3,-1.4) -- (5.3,2)node[pos=0.5,right]{$\vv{E}$};
				
			\end{tikzpicture}
		\end{minipage}
		%
		\hspace{1.5pc}%
		\begin{minipage}{.42\columnwidth}
			\centering
			\begin{tikzpicture}[scale=0.85]
				
	            \draw[blue,very thick](2.7,2.02) parabola (3.0,0.4); 
	\node[blue] at (2.4,1.7) {$e^-$};
	
	\fill[green!70!black](3.86,0) -- (4,0) -- (4.06,0.4) -- (3.92,0.4) -- cycle;
	\fill[green!70!black](3.66,-0.2) -- (3.8,-0.2) -- (3.86,-0.6) -- (3.72,-0.6) -- cycle;
	\fill[green!70!black](3.86,-0.2) -- (4.0,-0.2) -- (4.06,-0.6) -- (3.92,-0.6) -- cycle;
	\fill[green!70!black](4.06,-0.2) -- (4.2,-0.2) -- (4.26,-0.6) -- (4.12,-0.6) -- cycle;
	
	\fill[blue](2.86,0) -- (3,0) -- (3.06,0.4) -- (2.92,0.4) -- cycle; 
	\fill[blue](2.66,-0.2) -- (2.8,-0.2) -- (2.86,-0.6) -- (2.72,-0.6) -- cycle;
	\fill[blue](2.86,-0.2) -- (3.0,-0.2) -- (3.06,-0.6) -- (2.92,-0.6) -- cycle;
	\fill[blue](3.06,-0.2) -- (3.2,-0.2) -- (3.26,-0.6) -- (3.12,-0.6) -- cycle;
	
	\foreach \i in {1,1.2,1.4,...,5}{
		\filldraw[thick,fill=black!30!white] (\i,0) -- (\i+0.06,0) -- (\i+0.12,0.4) -- (\i+0.06,0.4) -- cycle;
	}
	\draw[thick](1,0) rectangle (5.12,0.4);
	\foreach \i in {1,1.2,1.4,...,5}{
		\filldraw[thick,fill=black!30!white] (\i+0.06,-0.6) -- (\i+0.12,-0.6) -- (\i+0.06,-0.2) -- (\i,-0.2) -- cycle;
	}
	\draw[thick](1,-0.6) rectangle (5.12,-0.2);
	
	\draw[very thick] (3.48,2.02) -- (3.02,0.4); 
	\draw[green!70!black,very thick] (3.45,2.02) parabola (4,0.38); 
	\node at (3.5,0.66){Ion};
	\node[green!70!black] at (4.6,0.7){AP $e^-$};
	
	\fill[ultra thick,cyan] (1,2) rectangle (5.12,2.08)node[at start, above, black]{PC};
	
	\foreach \l in {3.73,3.81,...,4.14}{
		\draw[green!70!black](3.93,0) parabola (\l,-0.2);
	}
	
	\foreach \k in {3.5,3.62,...,4.12}{
		\draw[green!70!black](3.8,-0.6) parabola (\k,-1.2);
	}
	\foreach \k in {3.7,3.82,...,4.32}{
		\draw[green!70!black](4,-0.6) parabola (\k,-1.2);
	}
	\foreach \k in {3.9,4.02,...,4.52}{
		\draw[green!70!black](4.2,-0.6) parabola (\k,-1.2);
	}

	\foreach \l in {2.73,2.81,...,3.14}{
		\draw[blue](2.93,0) parabola (\l,-0.2);
	}
	
	\foreach \k in {2.5,2.62,...,3.12}{
		\draw[blue](2.8,-0.6) parabola (\k,-1.2);
	}
	\foreach \k in {2.7,2.82,...,3.32}{
		\draw[blue](3,-0.6) parabola (\k,-1.2);
	}
	\foreach \k in {2.9,3.02,...,3.52}{
		\draw[blue](3.2,-0.6) parabola (\k,-1.2);
	}

	\draw[very thick](1,-1.2) -- (2.24,-1.2);
	\draw[very thick](2.44,-1.2) -- (3.68,-1.2);
	\draw[very thick](3.88,-1.2) -- (5.12,-1.2); 
	
	\foreach \j in {1.62,3.04,4.5}{
		\draw[very thick](\j,-1.2) -- (\j,-1.4);						
	}
	\draw[very thick, ->](5.3,-1.4) -- (5.3,2)node[pos=0.5,right]{$\vv{E}$};

			\end{tikzpicture}
		\end{minipage}
		\caption{Schematic representation of signal development by recoil electrons (left) and afterpulses from feedback ions (right) \cite{steffen}.}
		\label{fig:RecoilAP}
	\end{figure}
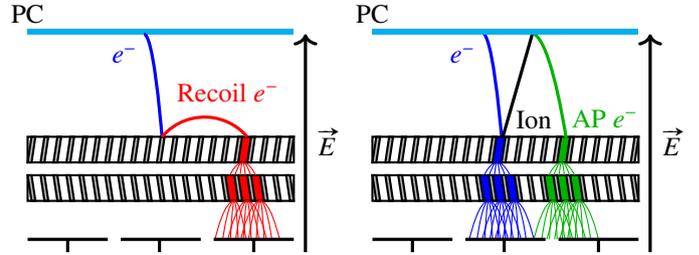

There is a certain probability that the photoelectron will not enter a pore directly, but will be scattered back at the MCP-In surface. The recoil electron can then reach the MCP at another location and trigger a delayed electron avalanche. The principle of this effect is sketched in the left part of figure~\ref{fig:RecoilAP}.

\begin{figure}[!htbp]
	\centering
	\includegraphics[width=.99\columnwidth]{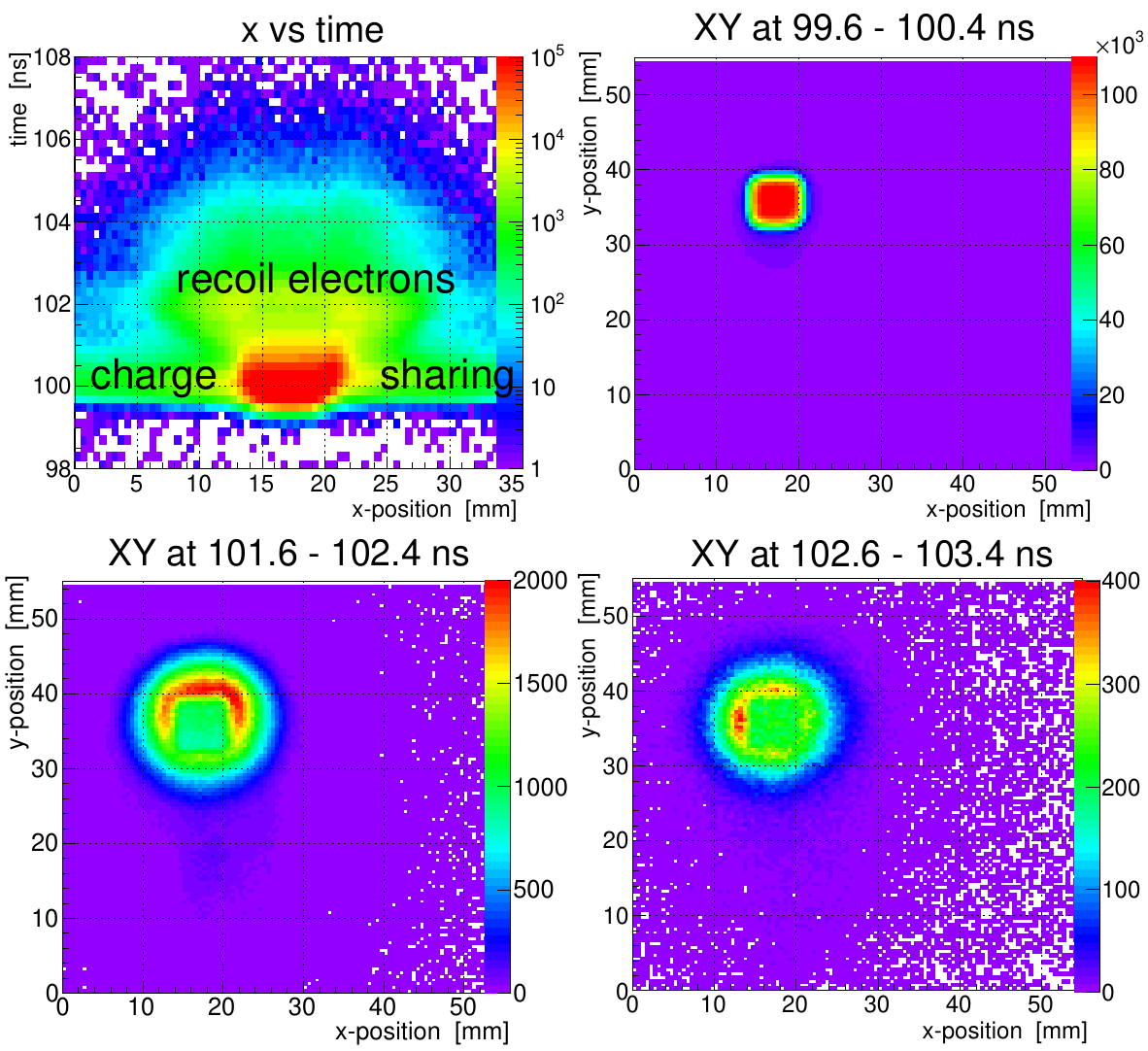}
	\caption{Temporal and spatial distribution of recoil electrons at the read-out pixel x3-y6 for the PHOTONIS XP85012 (SN 9002085). Top left: x-position vs time, regions where hits of recoil electrons and charge-sharing are expected are marked; top right: xy distribution for hits very close (99.6 to 100.4 ns) to the main trigger peak, corresponding to direct photoelectrons; bottom: xy distributions for temporally shifted events (101.6 to 102.4 ns [left] and 102.6 to 103.4 ns [right]), corresponding to recoil electrons.}
	\label{fig:recoilE}
\end{figure}

The signals from recoil electrons become visible in the time region directly after the main trigger peak at 100 ns, where a clear structure can be seen in the center-right plot of Fig.~\ref{fig:trbinfo}. The number of entries is approximately constant from 101 to 102.5 ns. The latter time corresponds to photoelectrons elastically scattered normal to the surface. In addition, the number of hits decreases over several orders of magnitude up to $\sim$108 ns. In Fig.~\ref{fig:recoilE} (top left), the distribution x-position is plotted against time. It shows that the recoil electrons are scattered over an area around the read-out anode pixel. It is instructive to examine the spatial distribution of these hits by shifting the time window. In the other plots of Fig.~\ref{fig:recoilE} it can be observed that at a narrow time slice around the trigger peak only the area of the read-out anode pixel x3-y6 is populated, while at later times this pixel is also hit when the laser focus points to the closer vicinity of the pixel position. The spatial and temporal extent of $\sim$20 mm and $\sim$4.5 ns of these events fit well with expectations for recoil electrons that are backscattered from the MCP-In before being amplified. The events at even later times could be caused by electrons that are backscattered more than once. Fig.~\ref{fig:recoilE} (top left) also shows a significant population of the regions adjacent to the read-out pixel at 14 - 21 mm, but the timing is the same as for the main laser peak. These can be attributed to charge-sharing crosstalk hits where the charge cloud extends across the pixel boundaries.

\subsection{Afterpulses}
\label{sec:perfAfterpulse}

The DiRICH/TRB3 data acquisition system makes it easy to identify and quantify afterpulse hits. Afterpulses are caused by residual gas ions traveling in the opposite direction to the electrons and hitting the PC of the PMT. They can knock further electrons out of the PC, which are then also amplified in the MCPs, but arrive much later than the photoelectrons from the main trigger. The effect is sketched in the right part of Fig.~\ref{fig:RecoilAP}. As can be seen in Fig.~\ref{fig:trbinfo} (bottom) the arrival times of afterpulse hits range from $\sim$10 ns after the trigger peak at 100 ns to $\sim$1 \textmu{s}. The afterpulse ions also reach the surroundings of the illuminated area and can be detected in adjacent pixels.

\subsubsection{Afterpulse Time Distribution and Fraction}

\begin{figure}[!htbp]
	\centering
	\includegraphics[width=.99\columnwidth]{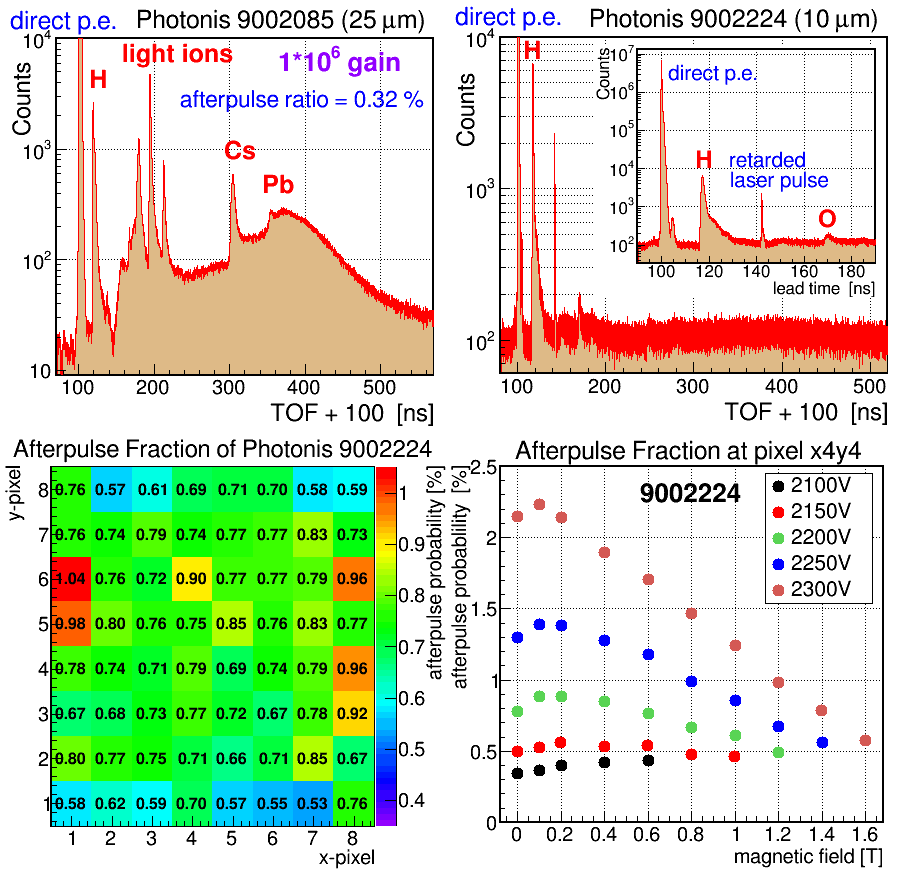}
	\caption{Afterpulse time-of-flight (TOF) distributions of the MCP-PMTs PHOTONIS SN 9002085 (non-ALD, top left) and SN 9002224 (ALD, top right). The trigger peak at 100 ns corresponds to direct photoelectrons (p.e.). Some of the other peaks can be assigned to feedback ions from the residual gas, whereas the weak "retarded laser pulse" in the upper right plot is a laser feature (also visible in (Fig.~\ref{fig:afterpulse_WF})). Bottom left: afterpulse fraction at each pixel for SN 9002224; bottom right: afterpulse probability as a function of voltage and magnetic field.}
	\label{fig:afterpulse}
\end{figure}

In Fig.~\ref{fig:afterpulse} (top) the afterpulse TOF distributions of two different MCP-PMTs are compared. The TOF peaks in these plots can be attributed to light (e.g. H, He, ...) and heavy (e.g. Cs and Pb) residual gas ions most likely generated in the first MCP layer. The broad continuum below these peaks is likely caused by ions scattered within the MCP pores, and the later broad hit accumulation originates from ions generated in the second MCP layer. The overall structure of the afterpulse TOF distribution of a single MCP-PMT is often difficult to interpret. However, it is possible to quantify the afterpulse rate for each MCP-PMT. To do this, we calculate the ratio of hits within the time interval from 105 to 505 ns to those within the trigger window from 99 to 104 ns, corrected for DCR in each case. This results in the fraction of afterpulse hits per anode pixel. The fractions shown in Fig.~\ref{fig:afterpulse} (bottom left) indicate that there is no major difference over the active area. With $<$1\%, the afterpulse fraction is rather moderate in the MCP-PMTs shown, but we have also encountered tubes where the fraction increased to $\gg$10\%. In such PMTs, multiple hydrogen ion peaks are seen at constant time intervals (see Fig.~\ref{fig:afterpulse_WF}), which are caused by afterpulses generated by the previous afterpulse ion. By measuring a large number of MCP-PMTs, we hope to get a better picture of the afterpulse behavior, which could help in the development of new MCP-PMTs with even fewer feedback ions.

It is still unclear whether the afterpulse component depends only on the gain or also on the applied voltage. We were trying to answer this question by taking measurements in a magnetic field. Since the gain is a function of the voltage, we analyzed the afterpulse fraction in a B-field, where the gain decreases at higher fields. The results are shown in Fig.~\ref{fig:afterpulse} (bottom right). For large voltages (and gains), the data is similar to the gain over the magnetic field (Fig.~\ref{fig:gainB_K}). However, while the gain at 1 T has decreased by a factor $\sim$3 compared to 0 T, the afterpulse fraction only drops by a factor $<$2. These data could indicate that the afterpulse probability depends on both the total PMT voltage and the gain. Unfortunately, this argument becomes dubious at lower voltages, as a low gain and comparatively large amplitude afterpulse signal (see Section~\ref{AP_WF}) passes the detection threshold more easily than a low gain and rather small amplitude photoelectron signal. This also has an effect on the measured afterpulse fractions at high magnetic fields.


\subsubsection{Afterpulse Amplitude with Waveform Sampling} \label{AP_WF}

An unpleasant feature of the DiRICH/TRB3 DAQ is that a channel is practically dead for $\sim$30 ns after a hit is triggered. This means that hits arriving up to 30 ns after the main laser pulse are generally not detected and are strongly suppressed. In addition to recoil electrons, this mainly affects H$^+$, H$_2^+$ and He$^+$ ions, which arrive about 15 to 30 ns after the photoelectrons. To further evaluate the extent and impact of these missing afterpulses on the performance of the MCP-PMT, several studies were performed by measuring long waveform samples with an oscilloscope and/or a CAEN digitizer. As a bonus, the waveforms also allow the charge amplitude of afterpulse hits to be determined.

\begin{figure}[!htbp]
	\centering
	\includegraphics[width=.95\columnwidth]{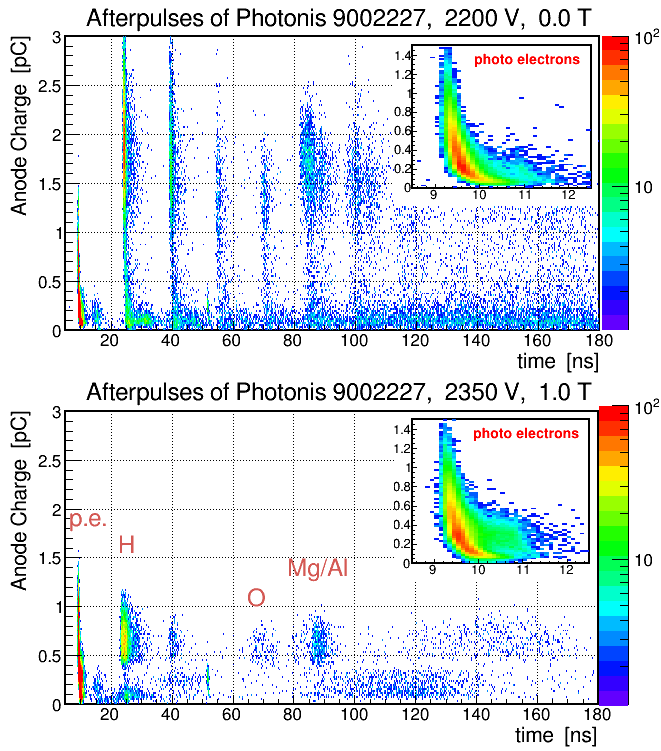}
	\caption{Afterpulse TOF versus charge distributions for the MCP-PMTs PHOTONIS SN 9002227 at the same gain of $\sim$1.5$\cdot$10$^6$; top: 2200 V and 0 Tesla; bottom: 2350 V and 1 Tesla. The insets show the zoomed photoelectron region, which show that the PMT gain is essentially the same in both configurations. The regular time structure, especially in the upper plot, can be attributed to a "cascade effect" of the feedback ions: an H afterpulse generates another H afterpulse, etc. (peaks at 25, 40 and 55 ns), but also an Mg/Al afterpulse ($\sim$85 ns) may induce another H afterpulse at 100 ns. Although the peak labeled as "O" could also stem from CH$_{4}$ \cite{andreotti21}, the abundance of oxygen in the ALD layers makes the latter more likely.}
	\label{fig:afterpulse_WF}
\end{figure}

An example is shown in Fig.~\ref{fig:afterpulse_WF}, where the afterpulse hits are shown in a time versus signal amplitude plot for the lower-quality PHOTONIS MCP-PMT SN 9002227. While the TOF distribution shows a number of discrete structures, it is obvious that the amplitude of the afterpulse ions (e.g. H at 25 ns) is much higher than that for the photoelectrons (also visible in Fig.4 of \cite{vavra2020}). This implies that each ion impact generates several electrons from the PC. An interesting observation is that at the same gain but different magnetic fields, the afterpulse spectrum looks different, both in terms of the time structure and the amplitude height, which is significantly lower at 1 T. For the parameter setting given in Fig.~\ref{fig:afterpulse_WF}, which corresponds to the same gain of $\sim$1.5$\cdot$10$^6$, the afterpulse fraction in the time range shown is 27\% at 0 T and 9\% at 1 T. The different time structure has not yet been clarified, but the lower amplitude indicates that fewer knock-out electrons reach the MCP. This suggests that a high magnetic field reduces the afterpulse fraction and possibly also the damage that each ion does to the PC.

\subsection{Crosstalk}
\label{sec:perfCrosstalk}

Crosstalk between the anode pixels is one of the most disturbing effects of multianode PMTs. In MCP-PMTs, we observe three different types of crosstalk: charge-sharing, electronic, and oscillations. The first two were investigated with the DiRICH/TRB3 DAQ, while the latter was studied with a digital oscilloscope.

\subsubsection{Charge Sharing}

\begin{figure}[H]
	\begin{minipage}{.42\columnwidth}
	\centering
	\begin{tikzpicture}[scale=0.88]
	\hspace{-0.3pc}

		\draw[blue,very thick](2.1,2.02) parabola (2.4,0.4); 
		\node[blue] at (2.0,1.7) {$e^-$};
		
		\fill[green!70!black](4.26,0) -- (4.4,0) -- (4.46,0.4) -- (4.32,0.4) -- cycle;
		\fill[green!70!black](4.06,-0.2) -- (4.2,-0.2) -- (4.26,-0.6) -- (4.12,-0.6) -- cycle;
		\fill[green!70!black](4.26,-0.2) -- (4.4,-0.2) -- (4.46,-0.6) -- (4.32,-0.6) -- cycle;
		\fill[green!70!black](4.46,-0.2) -- (4.6,-0.2) -- (4.66,-0.6) -- (4.52,-0.6) -- cycle;
		
		\fill[blue](2.26,0) -- (2.4,0) -- (2.46,0.4) -- (2.32,0.4) -- cycle; 
		\fill[blue](2.06,-0.2) -- (2.2,-0.2) -- (2.26,-0.6) -- (2.12,-0.6) -- cycle;
		\fill[blue](2.26,-0.2) -- (2.4,-0.2) -- (2.46,-0.6) -- (2.32,-0.6) -- cycle;
		\fill[blue](2.46,-0.2) -- (2.6,-0.2) -- (2.66,-0.6) -- (2.52,-0.6) -- cycle;
		
		\foreach \i in {1,1.2,1.4,...,5}{
			\filldraw[thick,fill=black!30!white] (\i,0) -- (\i+0.06,0) -- (\i+0.12,0.4) -- (\i+0.06,0.4) -- cycle;
		}
		\draw[thick](1,0) rectangle (5.12,0.4);
		\foreach \i in {1,1.2,1.4,...,5}{
			\filldraw[thick,fill=black!30!white] (\i+0.06,-0.6) -- (\i+0.12,-0.6) -- (\i+0.06,-0.2) -- (\i,-0.2) -- cycle;
		}
		\draw[thick](1,-0.6) rectangle (5.12,-0.2);
		
		\draw[green!70!black,very thick] (4.1,2.02) parabola (4.4,0.38); 
		\node[green!70!black] at (4.0,1.7){$e^-$};
		
		\fill[ultra thick,cyan] (1,2) rectangle (5.12,2.08)node[at start, above, black]{PC};
		
		\foreach \l in {4.13,4.21,...,4.54}{
			\draw[green!70!black](4.33,0) parabola (\l,-0.2);
		}
		
		\foreach \k in {3.9,4.02,...,4.52}{
			\draw[green!70!black](4.2,-0.6) parabola (\k,-1.2);
		}
		\foreach \k in {4.1,4.22,...,4.72}{
			\draw[green!70!black](4.4,-0.6) parabola (\k,-1.2);
		}
		\foreach \k in {4.3,4.42,...,4.92}{
			\draw[green!70!black](4.6,-0.6) parabola (\k,-1.2);
		}

		\foreach \l in {2.13,2.21,...,2.54}{
			\draw[blue](2.33,0) parabola (\l,-0.2);
		}
		
		\foreach \k in {1.9,2.02,...,2.52}{
			\draw[blue](2.2,-0.6) parabola (\k,-1.2);
		}
		\foreach \k in {2.1,2.22,...,2.72}{
			\draw[blue](2.4,-0.6) parabola (\k,-1.2);
		}
		\foreach \k in {2.3,2.42,...,2.92}{
			\draw[blue](2.6,-0.6) parabola (\k,-1.2);
		}

		\draw[very thick](1,-1.2) -- (2.24,-1.2);
		\draw[very thick](2.44,-1.2) -- (3.68,-1.2);
		\draw[very thick](3.88,-1.2) -- (5.12,-1.2); 
		
		\foreach \j in {1.62,3.04,4.5}{
			\draw[very thick](\j,-1.2) -- (\j,-1.4);						
		}
		\draw[very thick, ->](5.3,-1.4) -- (5.3,2)node[pos=0.5,right]{$\vv{E}$};
		
	\end{tikzpicture}
\end{minipage}
%
	\hspace{1.3pc}%
	\begin{minipage}{.42\columnwidth}
	\centering
	\begin{tikzpicture}[scale=0.85]
		
		\foreach \j in {1.1,1.25,...,5.1}{
			\draw[thick, ->,>=latex,white!70!black](\j,2.2) -- (\j,-1.5);
		} 
		\node[white!70!black] at (0.8,1.7) {$\vv{B}$};
		
		\draw[blue,thick,decorate,decoration={coil,segment length=12pt}] (2.4,2.02) -- (2.4,0.38); 
		\node[blue,scale=1.2] at (2.0,1.7) {$e^-$};
		
		\fill[green!70!black](4.26,0) -- (4.4,0) -- (4.46,0.4) -- (4.32,0.4) -- cycle;
		\fill[green!70!black](4.26,-0.2) -- (4.4,-0.2) -- (4.46,-0.6) -- (4.32,-0.6) -- cycle;
		
		\fill[blue](2.26,0) -- (2.4,0) -- (2.46,0.4) -- (2.32,0.4) -- cycle; 
		\fill[blue](2.26,-0.2) -- (2.4,-0.2) -- (2.46,-0.6) -- (2.32,-0.6) -- cycle;
		
		\foreach \i in {1,1.2,1.4,...,5}{
			\filldraw[thick,fill=black!30!white] (\i,0) -- (\i+0.06,0) -- (\i+0.12,0.4) -- (\i+0.06,0.4) -- cycle;
		}
		\draw[thick](1,0) rectangle (5.12,0.4);
		\foreach \i in {1,1.2,1.4,...,5}{
			\filldraw[thick,fill=black!30!white] (\i+0.06,-0.6) -- (\i+0.12,-0.6) -- (\i+0.06,-0.2) -- (\i,-0.2) -- cycle;
		}
		\draw[thick](1,-0.6) rectangle (5.12,-0.2);
		
		\draw[green!70!black,thick,decorate,decoration={coil,segment length=12pt}] (4.4,2.02) -- (4.4,0.38); 
		\node[green!70!black, scale=1.2, ultra thick] at (4.1,1.7){$e^-$};
		
		\fill[ultra thick,cyan] (1,2) rectangle (5.12,2.08)node[at start, above, black]{PC};
		
		\foreach \l in {4.26,4.28,...,4.41}{
			\draw[green!70!black](4.33,0) parabola (\l,-0.2);
		}
		
		\foreach \k in {4.26,4.29,...,4.45}{
			\draw[green!70!black](4.35,-0.6) parabola (\k,-1.2);
		}
		\foreach \k in {4.31,4.34,...,4.5}{
			\draw[green!70!black](4.4,-0.6) parabola (\k,-1.2);
		}
		\foreach \k in {4.36,4.39,...,4.55}{
			\draw[green!70!black](4.45,-0.6) parabola (\k,-1.2);
		}
		
		\foreach \l in {2.23,2.25,...,2.42}{
			\draw[blue](2.33,0) parabola (\l,-0.2);
		}
		
		\foreach \k in {2.26,2.29,...,2.45}{
			\draw[blue](2.35,-0.6) parabola (\k,-1.2);
		}
		\foreach \k in {2.31,2.34,...,2.5}{
			\draw[blue](2.4,-0.6) parabola (\k,-1.2);
		}
		\foreach \k in {2.36,2.39,...,2.55}{
			\draw[blue](2.45,-0.6) parabola (\k,-1.2);
		}

		\draw[very thick](1,-1.2) -- (2.24,-1.2);
		\draw[very thick](2.44,-1.2) -- (3.68,-1.2);
		\draw[very thick](3.88,-1.2) -- (5.12,-1.2); 
		
		\foreach \j in {1.62,3.04,4.5}{
			\draw[very thick](\j,-1.2) -- (\j,-1.4);						
		}
		\draw[very thick, ->](5.3,-1.4) -- (5.3,2)node[pos=0.5,right]{$\vv{E}$};

	\end{tikzpicture}
\end{minipage}
	\caption{Schematic comparison of the signal development from PC to anode without (left) and with the presence of a magnetic field (right): events without crosstalk (green), which only hit one anode pad, and with charge-sharing crosstalk, which hits more than one anode pads (blue). In a magnetic field, the charge cloud is more strongly focused.}
\label{fig:crosstalk_scheme}
\end{figure}
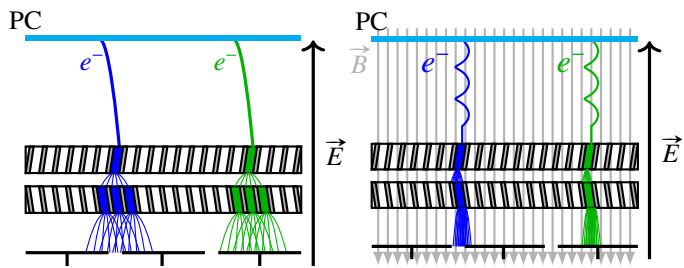

A special feature of MCP-PMTs is that many electrons emerge from the pores of the second MCP in different directions, creating a charge cloud that grows to a certain lateral extent until it reaches the anode plane. Its width depends on the internal structure of the MCP-PMT, in particular the distance between the MCP-Out and the anode. The magnitude of the electric and magnetic fields also has a significant influence on the width (see Section~\ref{Bwidth}). If only one fired pixel per photon (=1 hit) was detected on the entire PMT surface, the full charge cloud was contained in this pixel. With 2 or more hits, the charge cloud is distributed over several anode pixels, being most pronounced between adjacent pixels. This is roughly sketched in Fig.~\ref{fig:crosstalk_scheme}.


\begin{figure}[!htbp]
	\centering
	\includegraphics[width=.99\columnwidth]{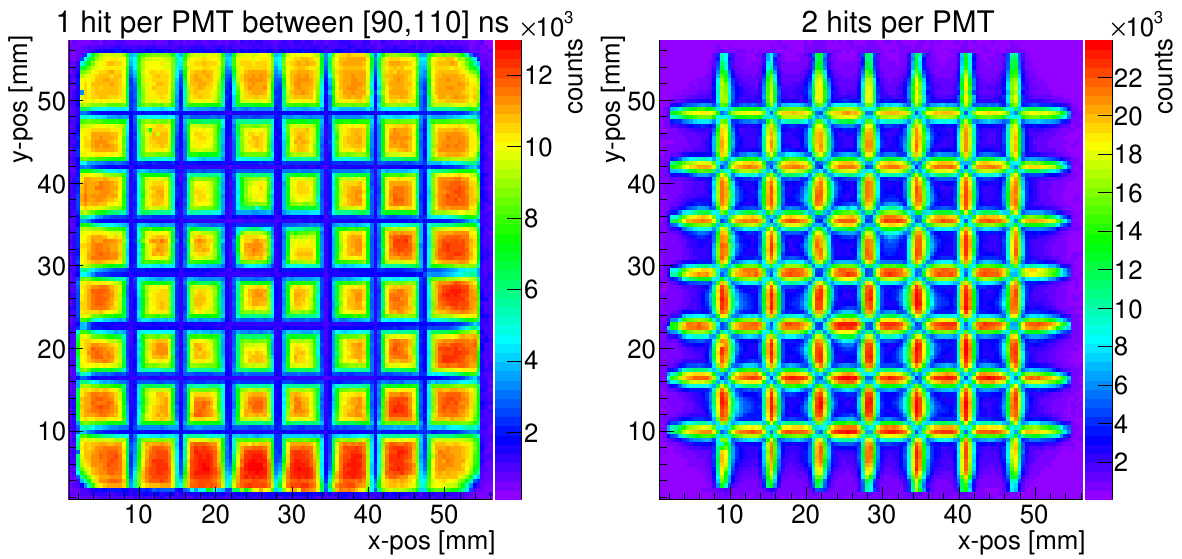}
	\caption{Charge-sharing crosstalk in the xy distribution for hits within 20 ns around the laser trigger for PHOTONIS XP85112 (SN 9002224): left: 1 anode hit; right: 2 anode hits.}
	\label{fig:cross}
\end{figure}

By counting the number of anode pixels hit in a narrow time window around the trigger peak at 100 ns, the charge-sharing crosstalk can be investigated. These hits coincide in time, even if the charge cloud is distributed over two or more anode pads. In Fig.~\ref{fig:cross} (left) we see the 8$\times$8 pixel structure when only one hit is measured in the PMT. With two simultaneous hits in Fig.~\ref{fig:cross} (right), we observe a population consisting mainly of the pixel borders. Events with more than 2 simultaneous hits would be visible in the pixel corners. The plots were created from xy-scans acquired with a $\sim$50 \textmu m diameter laser spot in 0.5 mm steps. Analyzing the distributions of the charge-sharing events also allows estimating the charge cloud width, which varies from $>$1 mm ($\sigma$) in absence of to $\ll$0.5 mm in presence of a B-field.

%

\subsubsection{Electronic}

\begin{figure}[!htbp]
	\centering
	\includegraphics[width=.99\columnwidth]{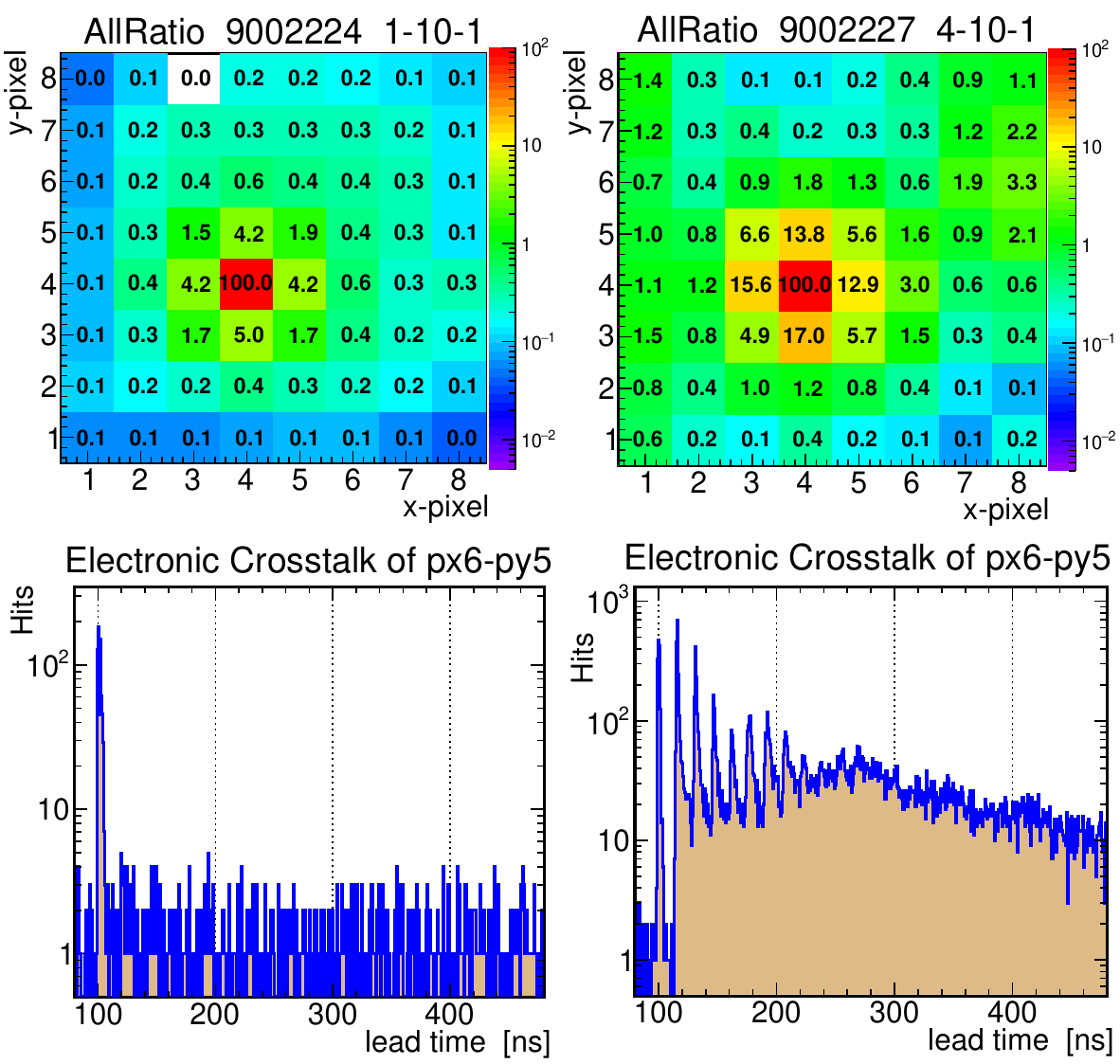}
	\caption{Electronic crosstalk in all other pixels when only the central 2.5$\times$2.5 mm$^2$ of pixel x4-y4 were illuminated. Top: crosstalk maps in \% for the PHOTONIS SN 9002224 (left) and SN 9002227 (right) MCP-PMTs. Bottom: time distributions of hits registered in the more distant pixel x6-y5, which should contain pure electronic crosstalk hits. The regular time structure in the bottom right plot is due to electronic crosstalk induced by the much larger afterpulse signals (compare to Fig.~\ref{fig:afterpulse_WF}).}
	\label{fig:elect_cross}
\end{figure}

The TRB scans also allow a straightforward investigation and quantification of electronic crosstalk effects between the anode pixels. This is unique and easy to accomplish by illuminating a single pixel and then analyzing all other PMT pixels for hits in the time ranges of the main trigger peak (around 100 ns) and the afterpulses ($\le$1000 ns). The data from standard position scans can be used for this, as the illuminated position is saved for each xy step. The results for two PHOTONIS MCP-PMTs (SN 9002224, SN 9002227) are shown in Fig.~\ref{fig:elect_cross}, where a threshold value of $\sim$20\% of the single photon pulse height was selected. The top row shows the fraction of hits in each pixel, normalized to x4-y4, which was illuminated in a central area of 2.5$\times$2.5 mm$^2$. The analyzed time window in the maps includes all hits from 99 to 999 ns. The detected hits in each pixel are corrected by a correspondingly scaled amount of dark events counted in the time window [-10, 0] \textmu{s}. The pixels directly adjacent to x4-y4 may still be populated by hits from recoil and afterpulse signals, but already in the second column/row from the illuminated pixel onwards, no more real hits are to be expected, except for electronic crosstalk.

The maps show that the fraction of electronic crosstalk is typically $\sim$1\% or less, depending on the MCP-PMT and also on the configuration of the voltage divider (1-10-1 versus 4-10-1). In-depth analysis shows that direct photoelectron signals cause only a small amount of electronic crosstalk, while the much larger afterpulse signals lead to significant electronic crosstalk even at the PMT rim. As expected, pulses with a large amplitude tend to induce more electronic crosstalk. The plots in the bottom row of Fig.~\ref{fig:elect_cross} show the temporal distributions of the signals induced by electronic crosstalk, which can vary substantially from tube to tube.

\subsubsection{Oscillations}

\begin{figure}[!htbp]
	\centering
	\includegraphics[width=.99\columnwidth]{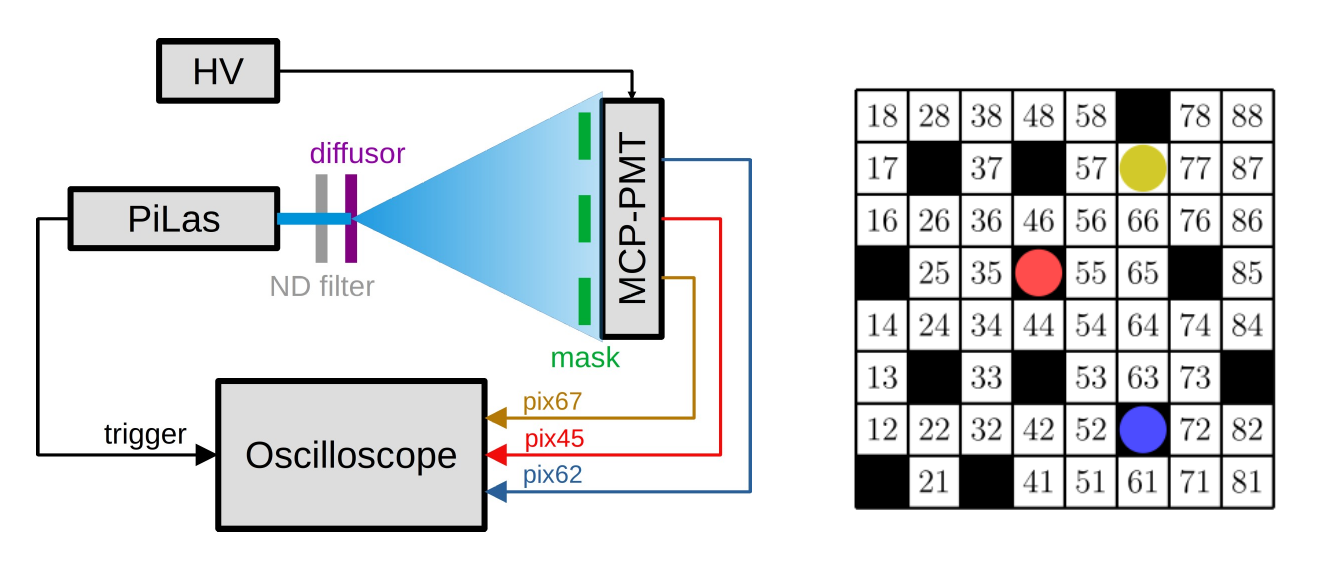}
	\caption{Left: block diagram for oscillation measurements and right: pixel map with illuminated (black) and read-out (colored) anode pixels. Masks with different amounts of open pixels were used.}
	\label{fig:oscillations}
\end{figure}

Another type of crosstalk is the so-called "coherent oscillations". Some years ago, J. Va'vra \cite{vavra2017} pointed out this "coherent oscillation problem" in older PHOTONIS Planacon tubes. The phenomenon occurs when several photons hit the PMT at the same time and can lead to fake anode hits. Inspired by this observation, we investigated the effect in various 8$\times$8-pixel MCP-PMTs. A schematic representation of the setup used can be seen in Fig.~\ref{fig:oscillations}. The active surface of the entire PMT is illuminated with light of different intensities. Using simple masks with several holes across the PC surface exactly at pixel positions, we simulated the arrival of photons with various intensities and at different positions. Three pixels were read out with an oscilloscope.

\begin{figure}[!htbp]
	\centering
	\includegraphics[width=.98\columnwidth]{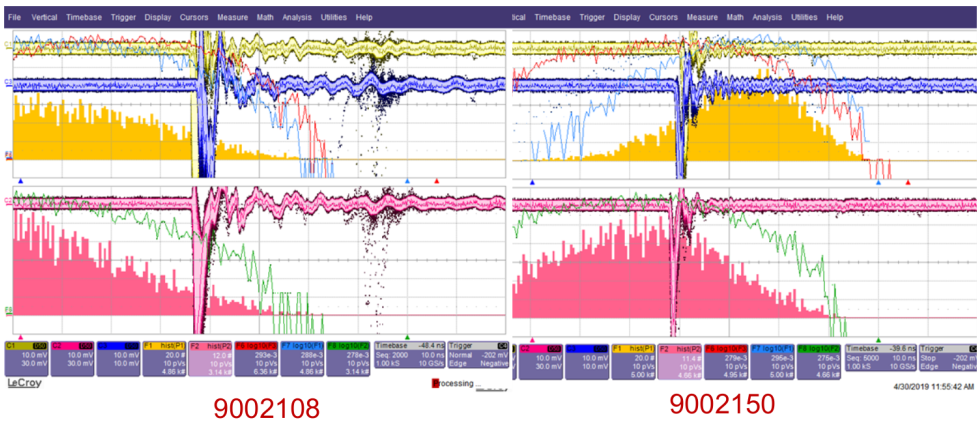}
	\caption{Oscilloscope screenshots of the photon-induced ringing in comparison between PHOTONIS SN 9002108 (standard backplane) and SN 9002150 (modified backplane) \cite{merlin2019}. During the measurements, the entire PC was homogeneously illuminated with $\sim$7 photons per pixel and per laser pulse.}
	\label{fig:oscitraces}
\end{figure}

The results obtained are shown in the form of oscilloscope screenshots in Fig.~\ref{fig:oscitraces}. The left image shows the waveforms for an older Planacon MCP-PMT (SN 9002108) with standard backplane, which shows the coherent oscillations when multiple pixels are hit simultaneously. This is particularly annoying because it could lead to fake hits in a real experiment. The number of possible fake hits was also studied with the TRB3 DAQ. In these measurements, half the area of the MCP-PMT was masked, while the open side was homogeneously illuminated with $\sim$1 detected photon/pixel. Although hits are only expected on the open half of the PMT, fake hits were also observed on the masked half due to crosstalk and/or the oscillations described above. A more detailed discussion of this fake hit problem can be found in ref.~\cite{merlin2019}.

After we informed PHOTONIS about this problem, they designed a new backplane for their MCP-PMTs. The result for a PMT with modified backplane is shown in the screenshot of Fig.~\ref{fig:oscitraces} (right). Obviously, ringing is significantly reduced with the new backplane and is no longer a problem for our application. It should be noted that the 2-inch Hamamatsu MCP-PMTs also suffered from oscillation problems at high photon counts, whereas this effect was much less severe with the Photek PMTs from the beginning.

%
%

\section{xy-Scans in Magnetic Fields}  \label{sec:scans_bfield}

\subsection{Scanning Setup}

\begin{figure}[!htbp]
	\begin{minipage}{.48\columnwidth}
		\centering
		\includegraphics[width=.99\columnwidth]{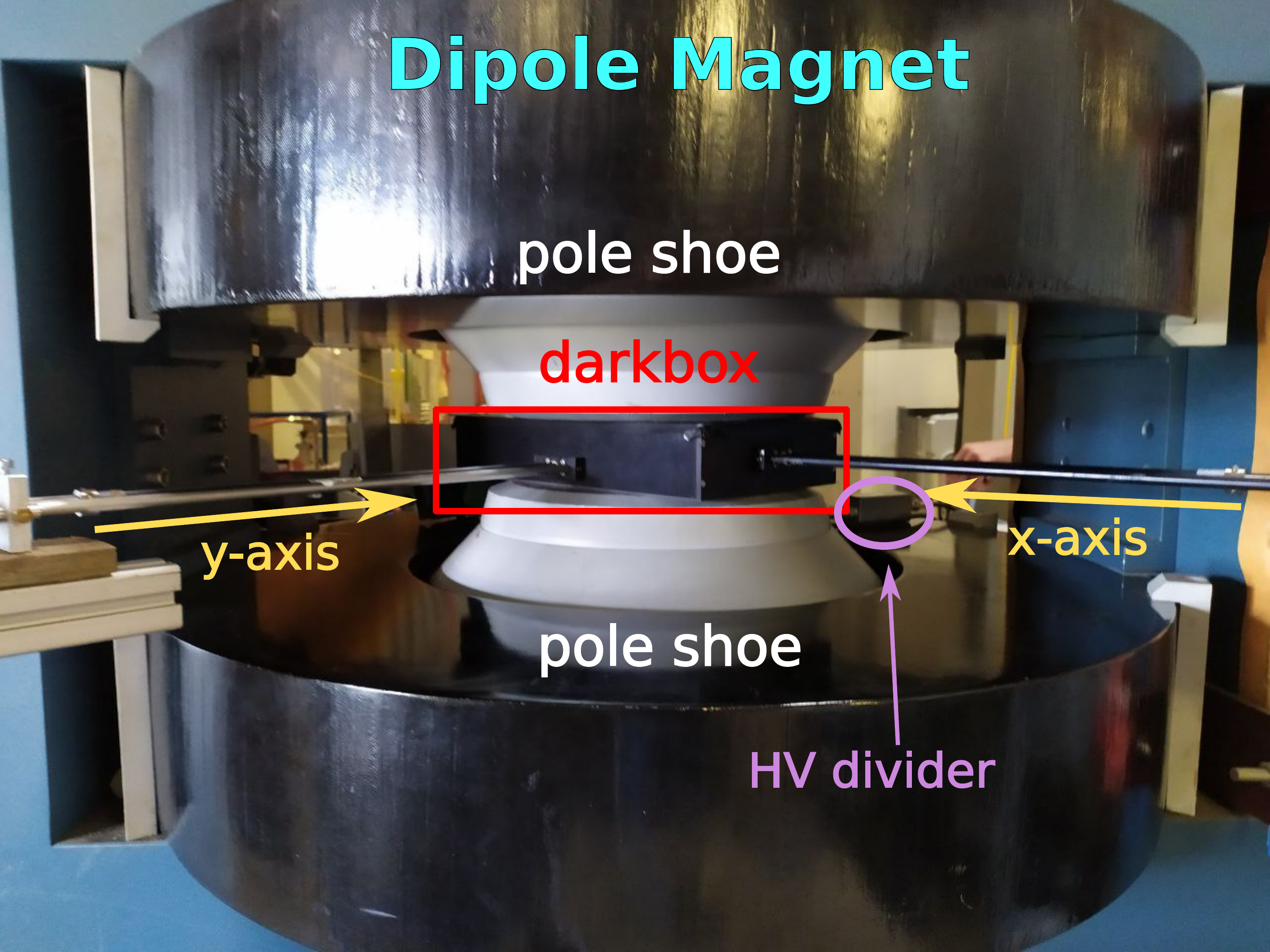}
	\end{minipage}
	\begin{minipage}{.51\columnwidth}
		\centering
		\includegraphics[width=.99\columnwidth]{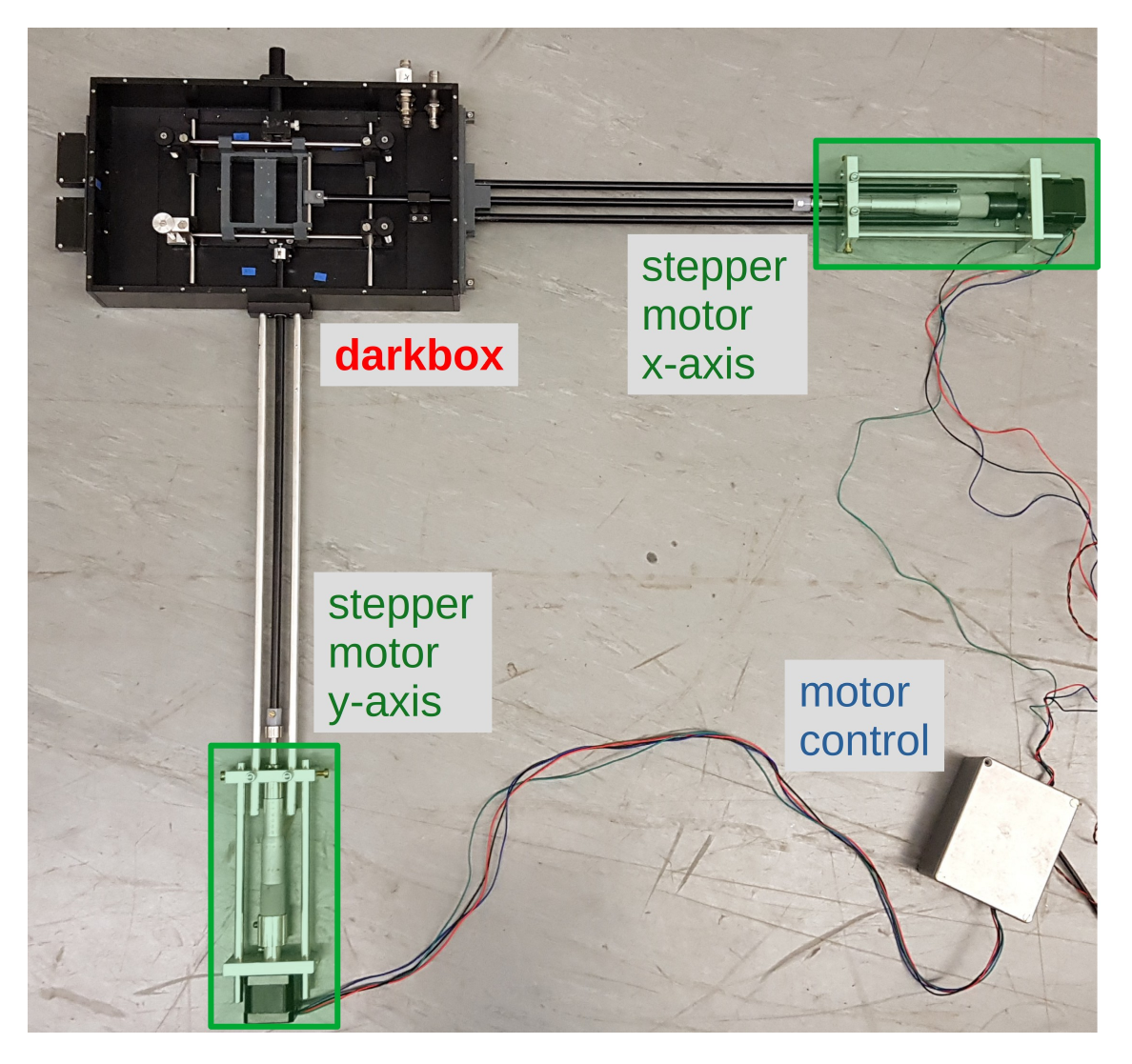}
	\end{minipage}
	\caption{Left: image of the dipole magnet with the dark box inserted. Right: image of the open dark box (same as in Fig.~\ref{fig:Bfield1}) with the extension rails in x and y directions and the stepper motors attached to the end of the rails \cite{steffen}.}
	\label{fig:Bfield}
\end{figure}

The dark box shown in Fig.~\ref{fig:Bfield1} has been modified to also allow for xy-scans in a magnetic field. For these measurements, the micrometer screws are extended by long threaded spindles stabilized by two metal rails that extend well beyond the range of the B-field, as shown in Fig.~\ref{fig:Bfield} (right). The extension rails are connected to the x- and y-axis of the sledge in the dark box on one side and to a stepper motor on the opposite side. This arrangement allows to perform xy-scans in a magnetic field similar to those in the laboratory in absence of magnetic field. The TRB3 data acquisition system is the same as that used in the lab, but the DiRICH boards are replaced by PADIWA frontend cards connected to the MCP-PMT by Samtec high-speed LSHM140 connectors and HLCD coaxial cables and placed outside the B-field. The automatic measurements at many xy positions and the subsequent analyses are controlled via a Python script, which can also be used to set different B-fields and high voltages. This enables very efficient measurements even overnight, for example. The scan range does not cover the full active area due to space limitations within the dark box. Nevertheless, a scan range of almost 25$\times$25 mm$^2$ in x and y direction is possible, which covers about 4$\times$4 pixels of an 8$\times$8 anode MCP-PMT and $\sim$50$\times$1 pixels (in x direction) with a 100$\times$3 anode MCP-PMT.

The laser focus is optimized as described in Section~\ref{positionQE} for the QE setup, and the photocathode current is measured with a Keithley picoammeter for different positions. At optimal focus, the PC current is the lowest due to saturation effects in the PC. For scan measurements, we usually place the lens slightly out of focus to avoid saturation of the anode current.

	
\subsection{Charge Sharing} \label{Bwidth}

\begin{figure}[!htbp]
	\centering
	\includegraphics[width=.98\columnwidth]{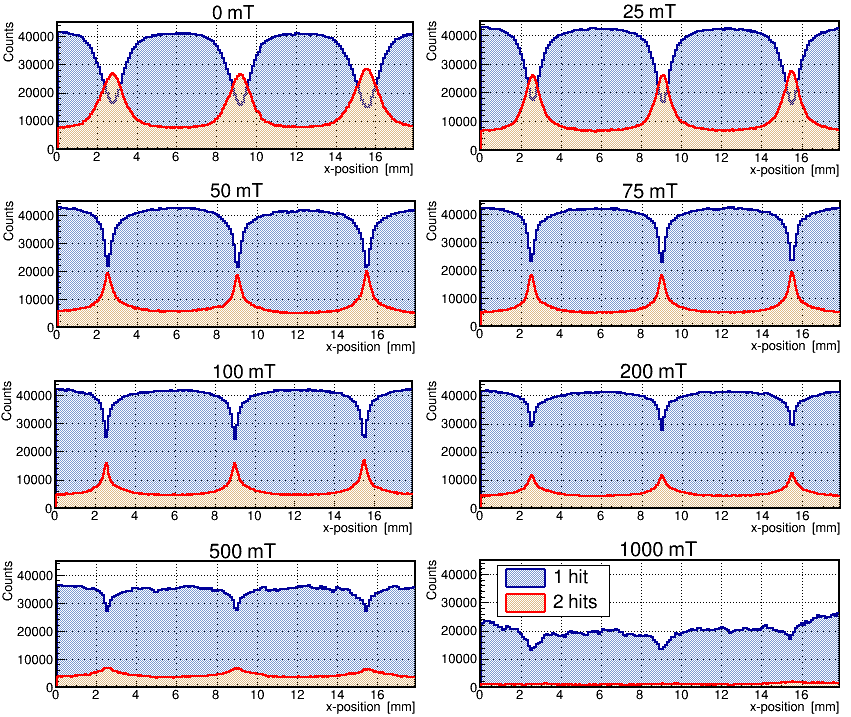}
	\caption{Distributions of charge-sharing crosstalk within different B-fields, determined with a high resolution x-scan for hits within 1 ns around the laser trigger for PHOTONIS XP85112 (SN 9002224).}
	\label{fig:crossb}
\end{figure}

Fig.~\ref{fig:crossb} shows the crosstalk behavior at different B-fields for the PHOTONIS MCP-PMT SN 9002224 with 8×8 anodes. The high-resolution (50 \textmu m steps) x-scan ran over four inner pixels, and all read-out pixels are summed in these plots. In both the 1-hit and 2-hit distributions, one can clearly identify the regions corresponding to the four anode pads. Only hits within a narrow time window of $\pm$0.5 ns around the main trigger peak at 100 ns are selected.

The regions between two adjacent pixels are filled with 2-hit events. This reflects the charge sharing crosstalk, and becomes less pronounced and narrower with increasing magnetic fields. The explanation for this is that the width of the charge cloud reaching the anode gets narrower within a magnetic field. The electrons between MCP-Out and anode are constrained onto helicoidal paths, so that at high B-fields there is almost no charge sharing effect. At 1000 mT the pixel boundaries are barely visible, the remaining 2-hit events are probably caused by electronic crosstalk.

\subsection{Recoil Electrons} \label{Brecoil}

\begin{figure}[!htbp]
	\centering
	\includegraphics[width=.99\columnwidth]{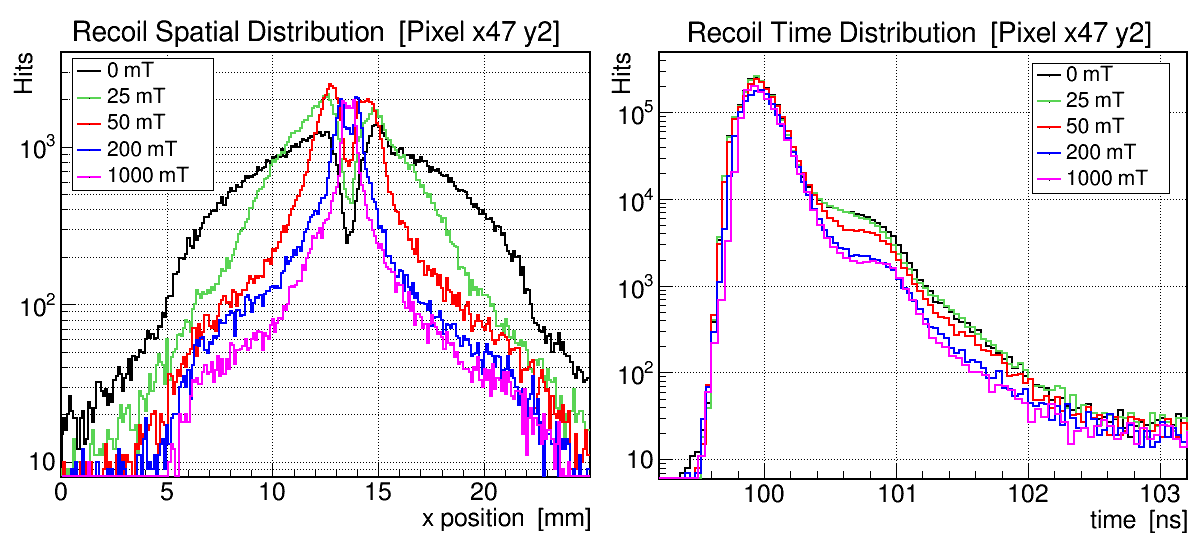}
	\caption{Behavior of recoil hits at different magnetic fields for the PHOTONIS 946P541 with 100$\times$3 anode pixels. The anode pixel x47-y2 was read out while the laser position was changed in steps of 25 \textmu m. Left: spatial distribution of recoil electrons within a time window of [100.4,101.5] ns; right: temporal distributions of recoil electrons.}
	\label{fig:crossb1}
\end{figure}

Similar studies can be carried out with the recoil electrons. The effects are shown for the PHOTONIS 946P541 2-inch MCP-PMT with 100$\times$3 anode pixels, which corresponds to a pixel pitch of $\sim$0.5 mm. Fig.~\ref{fig:crossb1} (left) shows the spatial expansion of recoil electrons with increasing magnetic fields. Obviously, the distribution of recoil electrons becomes much narrower at higher B-fields. This is due to the effect that the electrons are confined within a relatively limited area when a magnetic field is applied. At high fields, the recoiling electron falls back to almost the same position at which it started. If this recoil electron arrives at the same pixel, it can sometimes not be detected due to the $\sim$30 ns dead time of the DAQ channel described above (see section ~\ref{AP_WF}). This results in the holes at the position of the read-out pixel that can be seen in the spatial distributions. In contrast, the temporal distribution of recoil electrons is hardly influenced by a B-field. This is reflected in the distributions of Fig.~\ref{fig:crossb1} (right), where the slope of the tail at later times on the right side of the main laser peak (from 100.5 to 102 ns) shows only minor changes. The lower intensities for higher fields are due to the fewer recoil electrons from adjacent pixels reaching the read-out pixel x47-y2 as these are spatially more localized.

\vspace*{-3mm}
\subsection{Geometric and Lorentz Shift}


The electrons moving in the MCP-PMT are subjected to the Lorentz forces generated by the electric field along the PMT axis (z) and the magnetic field inside the dipole magnet (e.g. y-z plane). This leads to significant electromagnetic focusing of both the photoelectrons (and any recoil electrons) between the PC and the MCP-In and the electron cloud exiting at the MCP-Out. When the PMT axis is tilted with respect to the direction of the B-field, local shifts of the center of gravity of the charge cloud in the anode plane in the x-direction (Lorentz shift) and in the y-direction (geometric shift) are also observed. The non-negligible magnitude of these effects was simulated for different magnetic fields and compared with our measurements \cite{merlindiss}. This will be published in a forthcoming paper.

\section{Summary and Conclusions}  \label{summary}

In this paper, we have presented a systematic approach to the detailed measurement of most performance parameters of MCP-PMTs that can be carried out within a reasonable time frame. In particular, we have focused on the description of the experimental setups and procedures used to measure the different parameters. In parallel, the analyses used and some representative results for each parameter were discussed.

Below is a list of the most important findings and lessons learned from our numerous measurements with MCP-PMTs:

\vspace*{-2mm}
\begin{itemize}
	\item 	It is essential that the anode backplane of the MCP-PMT is very carefully designed to limit electronic crosstalk and coherent oscillations between the anode pixels.
\vspace*{-2mm}
	\item 	Whenever possible and not absolutely necessary, the illumination spots should not be focused too sharply on the PC during test measurements in order to avoid local gain saturation. This applies to QE scans as well, as the PC current can also saturate at high local intensity.
\vspace*{-2mm}
	\item 	QE position scans can be used as a good indicator for vacuum micro leaks in the PMT housing.
\vspace*{-2mm}
	\item 	The CE can be significantly increased to $>$90\% by recapturing the electrons bouncing back from the MCP-In. However, this requires a higher voltage between PC and MCP-in to optimize the RMS time resolution.
\vspace*{-2mm}
	\item 	ALD-coated MCP-PMTs are more susceptible to gain drop in magnetic fields than devices without ALD.
\vspace*{-2mm}
	\item 	At high gain and illumination conditions, some of the MCP-PMTs with 2 ALD layers (MgO as secondary electron emitter and Al$_{2}$O$_{3}$ as resistive layer) started to emit photons in the visible range, which led to "escalation" due to photon feedback with the PC.
\vspace*{-2mm}
	\item 	The typical rate capability of the investigated 2-inch MCP-PMTs is $\lesssim$1 MHz/cm$^{2}$. A few tubes have shown locally different rate capabilities which may indicate inhomogeneities in the resistive ALD layer.
\vspace*{-2mm}
	\item 	The time resolution and the DCR tend to be slightly worse at the rims of the active area of the MCP-PMT.
\vspace*{-2mm}
	\item 	The fraction of afterpulses caused by ion feedback varies from $\ll$1\% to $>$10\%. Normally, the ions identified by the time-of-flight from the MCP to the PC are dominated by H$^{+}$ ions, but in tubes with a high afterpulse rate we also identify O and Mg/Al ions, presumably released by a defective ALD coating. At high afterpulse fractions, we observe further afterpulses triggered by the feedback ions.
	
\vspace*{-2mm}
	\item 	Electromagnetic focusing reduces the width of the charge cloud arriving at the anode plane, which lowers crosstalk due to charge sharing. The spatial expansion of recoil electron events is considerably diminished. The latter can lead to an increased local electron density, particularly at high photon rates, and impair the tube's rate capability.
\vspace*{-2mm}
	\item 	None of the MCP-PMTs with 2 ALD layers in the lifetime setup has significantly lost gain at IAC $\approx$ 10 C/cm$^{2}$.
\end{itemize}

\vspace*{-2mm}
Our hope is that this paper will help to standardize the approaches used to measure the performance parameters of MCP-PMTs. This would facilitate the comparison of results obtained at different sites.

\section*{Acknowledgements}

This work is supported by the German BMBF and GSI Darmstadt.


\biboptions{sort&compress}
\bibliographystyle{plain}

\end{document}